\documentclass[longauth]{aa} 
\usepackage{graphicx}
\usepackage{subcaption} 
\usepackage{float}
\usepackage{placeins}
\usepackage{lscape} 
\usepackage{mathtools}
\usepackage{dblfloatfix}    % To enable figures at the bottom of page
\usepackage{tabularx} %two parts table
\usepackage{url}
\usepackage[normalem]{ulem}
\usepackage{tikz}
\usepackage{siunitx}
\usepackage[switch]{lineno} 
\DeclareMathAlphabet{\mathlib}{OT1}{LinuxLibertineT-OsF}{m}{it}
\DeclareMathAlphabet{\mathbio}{OT1}{LinuxBiolinumT-OsF}{m}{it}

\usepackage{natbib}
\bibpunct{(}{)}{;}{a}{}{,} % to follow the A&A style
\usepackage[colorlinks,allcolors=blue]{hyperref}
\usepackage{rotating}

\def\H0{{\rm \,km\,s^{-1}\,Mpc^{-1}}}

\def\checkmark{\tikz\fill[scale=0.4](0,.35) -- (.25,0) -- (1,.7) -- (.25,.15) -- cycle;} 

\begin{document}

\title{Methanol masers in NGC\,253 with ALCHEMI}

          \author{P. K. Humire \inst{\ref{inst.MPIfR}}
          \and C. Henkel \inst{\ref{inst.MPIfR},\ref{inst.Abdulaziz},\ref{inst.Xinjiang}}
          \and A. Hern\'andez-G\'omez \inst{\ref{inst.MPIfR}}
          \and S. Mart\'in \inst{\ref{inst.ESOChile},\ref{inst.JAO}}
          \and J. Mangum \inst{\ref{inst.NRAOCV}}
          \and N. Harada \inst{\ref{inst.NAOJ},\ref{inst.ASIAA}, \ref{inst.SOKENDAI}}
          \and S. Muller \inst{\ref{inst.ONSALA}}
          \and K. Sakamoto \inst{\ref{inst.ASIAA}}
%----------DATA REDUCTION
          \and K. Tanaka \inst{\ref{inst.KeioUniversity}}
          \and Y. Yoshimura  \inst{\ref{inst.UTokio}}
          \and K. Nakanishi \inst{\ref{inst.NAOJ},\ref{inst.SOKENDAI}}
          \and S. M\"uhle \inst{\ref{inst.UBonn}}
          \and R. Herrero-Illana \inst{\ref{inst.ESOChile},\ref{inst.ICECSIC}}
%----------DATA REDUCTION
          \and D. S. Meier \inst{\ref{inst.NMIMT},\ref{inst.NRAOSocorro}}
          \and E. Caux \inst{\ref{inst.IRAP}}
          \and R. Aladro \inst{\ref{inst.MPIfR}}
          \and R. Mauersberger \inst{\ref{inst.MPIfR}}
          \and S. Viti\inst{\ref{inst.Leiden},\ref{inst.UCL}}
          \and L. Colzi \inst{\ref{inst.CAB-INTA},\ref{inst.Arcetri}}
          \and V. M. Rivilla \inst{\ref{inst.CAB-INTA},\ref{inst.Arcetri}}
          \and M. Gorski \inst{\ref{inst.ONSALA}}
          \and K. M. Menten \inst{\ref{inst.MPIfR}}
          \and K.-Y. Huang\inst{\ref{inst.Leiden}}
          \and S. Aalto \inst{\ref{inst.ONSALA}}
          \and P. P.~van der Werf \inst{\ref{inst.Leiden}}
          \and K. L. Emig \inst{\ref{inst.NRAOCV}} 
          \thanks{Jansky Fellow of the National Radio Astronomy Observatory}
          }

%If there are too many authors, use \authorrunning
%\authorrunning{First Author et al.}

%

\institute{
\label{inst.MPIfR}Max-Planck-Institut f\"ur Radioastronomie, Auf-dem-H\"ugel 69, 53121 Bonn, Germany, \label{email}phumire@mpifr-bonn.mpg.de %\email{phumire@mpifr-bonn.mpg.de} \label{email}phumire@mpifr-bonn.mpg.de   
\and\label{inst.Abdulaziz}Astron. Dept., Faculty of Science, King Abdulaziz University, P.O. Box 80203, Jeddah 21589, Saudi Arabia
\and\label{inst.Xinjiang}Xinjiang Astronomical Observatory, Chinese Academy of Sciences, 830011 Urumqi, China
\and\label{inst.ESOChile}European Southern Observatory, Alonso de C\'ordova, 3107, Vitacura, Santiago 763-0355, Chile  
\and\label{inst.JAO}Joint ALMA Observatory, Alonso de C\'ordova, 3107, Vitacura, Santiago 763-0355, Chile
\and\label{inst.NRAOCV}National Radio Astronomy Observatory, 520 Edgemont Road, Charlottesville, VA 22903-2475, USA
\and\label{inst.NAOJ}National Astronomical Observatory of Japan, 2-21-1 Osawa, Mitaka, Tokyo 181-8588, Japan
\and\label{inst.ASIAA}Institute of Astronomy and Astrophysics, Academia Sinica, 11F of AS/NTU Astronomy-Mathematics Building, No.1, Sec. 4, Roosevelt Rd, Taipei 10617, Taiwan
\and\label{inst.SOKENDAI}Department of Astronomy, School of Science, The Graduate University for Advanced Studies (SOKENDAI), 2-21-1 Osawa, Mitaka, Tokyo, 181-1855 Japan
\and\label{inst.ONSALA}Department of Space, Earth and Environment, Chalmers University of Technology, Onsala Space Observatory, SE-43992 Onsala, Sweden
\and\label{inst.KeioUniversity}Department of Physics, Faculty of Science and Technology, Keio University, 3-14-1 Hiyoshi, Yokohama, Kanagawa 223--8522 Japan
\and\label{inst.UTokio}Institute of Astronomy, Graduate School of Science, The University of Tokyo, 2-21-1 Osawa, Mitaka, Tokyo 181-0015, Japan
\and\label{inst.UBonn}Argelander-Institut f\"ur Astronomie, Universit\"at Bonn, Auf dem H\"ugel 71, D-53121 Bonn, Germany
\and\label{inst.ICECSIC}Institute of Space Sciences (ICE, CSIC), Campus UAB, Carrer de Magrans, E-08193 Barcelona, Spain
\and\label{inst.IRAP}IRAP, Universit\'e de Toulouse, CNRS, UPS, CNES, Toulouse, France
\and\label{inst.Leiden}Leiden Observatory, Leiden University, PO Box 9513, NL - 2300 RA Leiden, The Netherlands
\and\label{inst.UCL}Department of Physics and Astronomy, University College London, Gower Street, London WC1E6BT, UK
\and\label{inst.CAB-INTA}Centro de Astrobiología (CSIC-INTA), Ctra. de Torrej\'on a Ajalvir km 4, 28850, Torrej\'on de Ardoz, Madrid, Spain       
\and\label{inst.Arcetri}INAF Osservatorio Astrofisico di Arcetri, Largo Enrico Fermi 5, I-50125 Firenze, Italy
\and\label{inst.NMIMT}New Mexico Institute of Mining and Technology, 801 Leroy Place, Socorro, NM 87801, USA
\and\label{inst.NRAOSocorro}National Radio Astronomy Observatory, PO Box O, 1003 Lopezville Road, Socorro, NM 87801, USA
}
\date{Received xxx, yyy; accepted xxx, yyy}

\titlerunning{New methanol masers in NGC\,253}
\authorrunning{Humire et al.}

%\abstract {} {} {} {} {Our findings also confirms the previously reported maser at 84\,GHz.} 
%In order to have a deeper insight regarding the presence and the excitation conditions of methanol maser in external galaxies, w

\abstract{Methanol masers of Class\,I (collisionally-pumped) and Class\,II (radiatively-pumped) have been studied in great detail in our Galaxy in a variety of astrophysical environments such as shocks and star-forming regions and are helpful to analyze the properties of the dense interstellar medium. However, the study of methanol masers in external galaxies is still in its infancy.}{Our main goal is to search for methanol masers in the central molecular zone (CMZ; inner 500\,pc) of the nearby starburst galaxy NGC\,253.}{Covering a frequency range between 84 and 373\,GHz ($\lambda$ = 3.6 to 0.8\,mm) at high angular (1\farcs6$\sim$27\,pc) and spectral ($\sim$8--9\,km\,s$^{-1}$) resolution with the ALMA large program ALCHEMI, we have probed different regions across the CMZ of NGC\,253. In order to look for methanol maser candidates, we employed the rotation diagram method and a set of radiative transfer models.} {We detect for the first time masers above 84\,GHz in NGC\,253, covering an ample portion of the $J_{-1}\rightarrow(J-$ 1)$_{0}-E$ line series (at 84, 132, 229, and 278\,GHz) and the $J_{0}\rightarrow(J-$ 1)$_{1}-A$ series (at 95, 146, and 198\,GHz). This confirms the presence of the Class\,I maser line at 84\,GHz, already reported but now being detected in more than one location. For the $J_{-1}\rightarrow(J-$ 1)$_{0}-E$ line series, we observe a lack of Class\,I maser candidates in the central star-forming disk.}{The physical conditions for maser excitation in the $J_{-1}\rightarrow(J-$ 1)$_{0}-E$ line series can be weak shocks and cloud-cloud collisions as suggested by shock tracers (SiO and HNCO) in bi-symmetric shock/active regions located in the outskirts of the CMZ. On the other hand, the presence of photodissociation regions due to a high star-formation rate would be needed to explain the lack of Class\,I masers in the very central regions.}

\maketitle

%________________________________________________________________

\section{Introduction}
\label{sec.introduction}

Methanol (CH$_{3}$OH) is a molecule prone to population inversion under specific excitation conditions in the interstellar medium (ISM) \citep[e.g.][]{Cragg1992}, causing maser emission. In particular, methanol masers are unique tools for studying physical properties of dense gas associated with Young Stellar Objects (YSOs). Given their brightness and compactness \citep{Menten1991b}, their positions can be determined with high precision astrometry (i.e., at milli-arcsecond accuracy with very long baseline interferometry) and over vast distances.

Thousands of such methanol masers have been detected in the Milky Way \citep[][]{Cotton2016,Green2017,Yang2019}. However, in nearby galaxies we can only account for a handful of successful detections \citep[e.g.][]{McCarthy2020}. In particular, the brightest Galactic CH$_3$OH maser transition at 6.7\,GHz \citep[][]{Breen2015}, remains elusive in extragalactic objects outside the Local Group \citep{Ellingsen1994,Darling2003}, with the only exception of NGC\,3079 \citep{Impellizzeri2008}, and possibly Arp220 \citep{Salter2008}, where this line is detected in absorption.

The early discovery that CH$_{3}$OH masers can be divided into two classes, a collisionally-pumped Class\,I and a radiatively-pumped Class\,II \citep{Batrla1987,Menten1991b}, allows us to trace either stellar-induced outflows (Class\,I) or ultra-compact {H\,\sc{ii}} regions (Class\,II). Class\,I methanol masers have been observed toward high and low-mass stars \citep{Kalenskii2006,Kalenskii2010,Rodriguez2017}, while Class\,II masers have been observed only toward high-mass YSOs \citep{Breen2013}. Unlike H$_{2}$O and OH masers, Class\,II methanol masers seem to be exclusively correlated with star-forming regions \citep{Walsh2001,Breen2013}. 

Because Class\,II masers are usually brighter than Class\,I masers in our Galaxy, the former have been studied in great detail, leading to surveys targeting exclusively their relation with the surrounding conditions \citep{Yang2017,Billington2020}. However, outside our Galaxy Class\,II masers were only detected in the Magellanic Clouds \citep{Sinclair1992,Green2008,Ellingsen2010} and the Andromeda galaxy \citep{Sjouwerman2010}, with luminosities not surpassing those in our Galaxy. On the other hand, extragalactic Class\,I masers can be more luminous than those of Class\,II, and have been successfully observed beyond the Local Group, particularly in nearby barred spiral galaxies like NGC\,253, IC\,342, or NGC\,4945 \citep[e.g.][]{Ellingsen2014,McCarthy2017,Gorski2018}. 

There are two types of methanol. For E-type methanol, one of the protons in the hydrogen atoms of the methyl (CH$_{3}$) group has an anti-parallel nuclear spin with respect to the others, analogous to the case of para-NH$_{3}$. In the A-type methanol, the nuclear spins of the three protons in the methyl group are parallel, as in the case of ortho-NH$_{3}$. As the two methanol types have different transition frequencies and may arise in different physical environments, we decided to analyse them separately. 

Hereafter we use the conventional notation for A$^{+}$ and A$^{-}$ introduced by \citet{Lees1968}, related to a combination between the A--CH$_{3}$OH overall-parity and Mulliken symbols\footnote{\url{https://mathworld.wolfram.com/CharacterTable.html}} A$_{1}$ and A$_{2}$. This is done to discriminate between splitted $+K$ and $-K$ levels (doublets), with $K$ being the projection of the angular momentum along the molecular symmetry axis. $+K$ and $-K$ levels are torsionally degenerate for the case of A-type methanol, contrary to the case of E-type methanol (see also \citealt{Cragg1993}).

As shown by \citet{Lees1973}, among E-type methanol transitions (E--CH$_{3}$OH) Class\,I population inversion is favoured in the $K=-$1 relative to the $K=0$ or $K=-$2 ladders. This leads to the prominence of the $J_{-1}\rightarrow(J-$ 1)$_{0}-E$ series \citep[see e.g.][their Fig.\,3]{Leurini2016}. Indeed, the $4_{-1}\rightarrow3_{0}-E$ (36.2\,GHz) and $5_{-1}\rightarrow4_{0}-E$ (84.5\,GHz) lines have been recently discovered to be masing in one extragalactic object, NGC\,253 \citep{Ellingsen2014,McCarthy2018}. For A-type methanol, population inversion in the $K=0$ relative to the $K=1$ ladder is favoured, playing out in the $J_{0}\rightarrow(J-$1)$_{1}-A^{+}$ series, of which the 44.1\,GHz line, at $J=$7, has been also detected in NGC\,253 \citep{Ellingsen2017}. The emission of Galactic Class\,II methanol masers are more compact than those of their Galactic Class\,I cousins \citep{Moscadelli2003,Matsumoto2014}, and that could make Class\,II methanol masers more difficult to detect at extragalactic distances. In addition, Class\,I masers require lower densities and temperatures than Class\,II masers \citep{Menten2012}, making them more numerous. Specifically for the $J_{-1}\rightarrow(J-$ 1)$_{0}-E$ series with $J=$ 4 and 5, their intensities were predicted to be of the order of 50\,mJy in the case of NGC\,253 \citep{Sobolev1993}.

Class\,I methanol masers may be associated with a variety of phenomena, such as supernova remnants \citep{Plambeck1990,Pihlstrom2014}, massive protostellar induced outflows \citep{Cyganowski2018}, and interactions of expanding H\,{\sc ii} regions with surrounding molecular gas \citep{Voronkov2010}, namely regions where shocks compress and heat the gas. In the central molecular zone (CMZ) of our Galaxy (i.e., the inner $\sim$200\,pc in radius; \citealt{Morris1996}), cosmic ray interactions with molecular clouds have been claimed to be an additional source of methanol production. While a high methanol abundance alone is certainly not sufficient to trigger maser emission, there seems to be indeed a clear enhancement of Class\,I masers in this region \citep{Yusef-Zadeh2013,Cotton2016,Ladeyschikov2019}. In particular, extended strong emission showing characteristics of maser action had been found by \citet{HaschickBaan1993} and \citet{Salii2002} in the G1.6$-$0.025 region at the periphery of the CMZ and by \citet{Szczepanski1989} and \citet{LiechtiWilson1996} in a region in which the supernova remnant Sgr A East interacts with a Giant Molecular Cloud (GMC). Noteworthy, CMZ conditions in the Milky Way should provide some guidance to the central regions of starburst galaxies \citep{Belloche2013}.

Class\,I methanol masers at 36, 44, and 84\,GHz have already been reported in NGC\,253 by \citet{Ellingsen2014}, \citet{Ellingsen2017}, and \citet{McCarthy2018}, respectively. The former, at 36\,GHz, has been also observed in NGC\,4945, IC\,342, NGC\,6946, and Maffei\,2 \citep{McCarthy2017,Gorski2018,Humire2020}. A 84\,GHz mega-maser ($\geq$10$^{6}$ times more luminous than typical Galactic masers; \citealt{Lo2005}) has been reported in NGC\,1068 \citep{Wang2014}, but a confirmation would be needed to put this onto a firm basis. Although a tentative detection of methanol mega-maser emission was claimed in Arp\,220 \citep{Chen2015}, later studies ruled it out \citep{Humire2020,McCarthy2020}. Making use of the unprecedented spectral coverage available by the Atacama Large Millimeter/sub-millimeter Array (ALMA) Large Program ALCHEMI \citep{Martin2021}, we have been able to make a comprehensive study toward one of the best candidates to search for extragalactic maser emission, the Sculptor galaxy NGC\,253.

NGC\,253 is a nearby \citep[D $\sim$3.5\,Mpc,][]{Rekola2005} highly inclined ($i\sim70^{\circ}$--79$^\mathrm{\circ}$, \citealt{Pence1980,Iodice2014}) SAB(s)c galaxy \citep{deVaucouleurs1991} with a systemic heliocentric velocity ($v_{\rm sys}$) of $\sim$258.8\,km\,s$^{-1}$ \citep{Meyer2004}. Its large-scale bar feeds the nuclear region producing stars at an approximate rate of $\sim$1.7\,M$_{\odot}$\,yr$^{-1}$ \citep{Bendo2015} within the central 20\arcsec$\times$10\arcsec\,, where 1\arcsec\,corresponds to $\sim$17\,pc. The position angle (PA) of the large-scale bar is 51$^{\circ}$ \citep{Pence1980}. The PA continues till the central $\sim$170 parsecs. Further into the core, the isovelocity contours of the gas change their orientation by about 90$^{\circ}$ due to the existence of a nuclear bar \citep{Cohen2020}. Like our Galaxy, NGC\,253 is characterized by a particularly strong and diverse molecular emission in its central 500\,pc \citep{Sakamoto2006}, which we therefore identify as its CMZ. Previous studies do not suggest that an Active Galactic Nucleus (AGN) is important for the properties of the molecular gas in the CMZ of NGC\,253 \citep{MullerSanchez2010}.

This paper is organized as follows: our observations are described in Sect.\,\ref{Sec.Observations}. In Sect.\,\ref{sec.selected_regions} we show the selected regions to be studied in the CMZ of NGC\,253. In Sect.\,\ref{sec.methanol_maser_emission_ID} we introduce the two methods used to identify methanol masers. In Sect.\,\ref{sec.discussion} we discuss the results and provide a number of conditions to explain maser emission. We finally draw the conclusions of our findings in Sect.\,\ref{sec.summary}.

\section{Observations}
\label{Sec.Observations}

We use ALMA observations of NGC\,253 taken in Cycles~5 and 6 as part of the large program ALCHEMI
(ALMA Comprehensive High-resolution Extragalactic Molecular Inventory, 2017.1.00161.L, followed up by program 2018.1.00162.S, see \citealt{Martin2021}). The central region of the galaxy ($50\arcsec \times 20\arcsec$ = $850\times340$~pc, with a position angle of $65^{\circ}$) was covered in bands~3 to 7 (84.2 -- 373.2\,GHz), with both the 12\,m and 7\,m antenna arrays, achieving a final homogeneous angular resolution of 1\farcs6. The mentioned region was covered with a single pointing in band\,3, with a primary beam ranging between 57$\arcsec$ and 68$\arcsec$, and Nyquist-sampled mosaic patterns of 5 up to 19 pointings for the remaining bands, ensuring a homogeneous sensitivity across the selected region\footnote{ALMA sets up by default a hexagonal Nyquist sampled pattern to ensure homogeneous sensitivity across the mapped region. Furthermore, mosaic images are primary beam corrected, in our case this includes the studied $850\times340$~pc region. For more information about mosaicing, we refer the reader to Sect\,7.7 in the ALMA technical Handbook (\url{https://almascience.eso.org/documents-and-tools/cycle9/alma-technical-handbook})}. The phase center of the observations is $\alpha=00^h47^m33^s.26$, $\delta=-25^\circ17'17\farcs7$ (ICRS\footnote{International Celestial Reference System, equivalent to J2000 within a few milli--arcseconds \citep{Ma1998}}).

The absolute flux calibration accuracy is of the order of 15\% for all ALMA bands, since most percentages are lower across individual bands. The flux density RMS noise ranges from 0.18 to 5.0\,mJy beam$^{-1}$, and the averaged sensitivity is 14.8\,mK. For more details about the data reduction procedures and image processing, see \citet{Martin2021}.

\section{Selected positions}
\label{sec.selected_regions}

\begin{figure}[h]
\includegraphics[width=\linewidth, trim={0 0 0 0},clip]{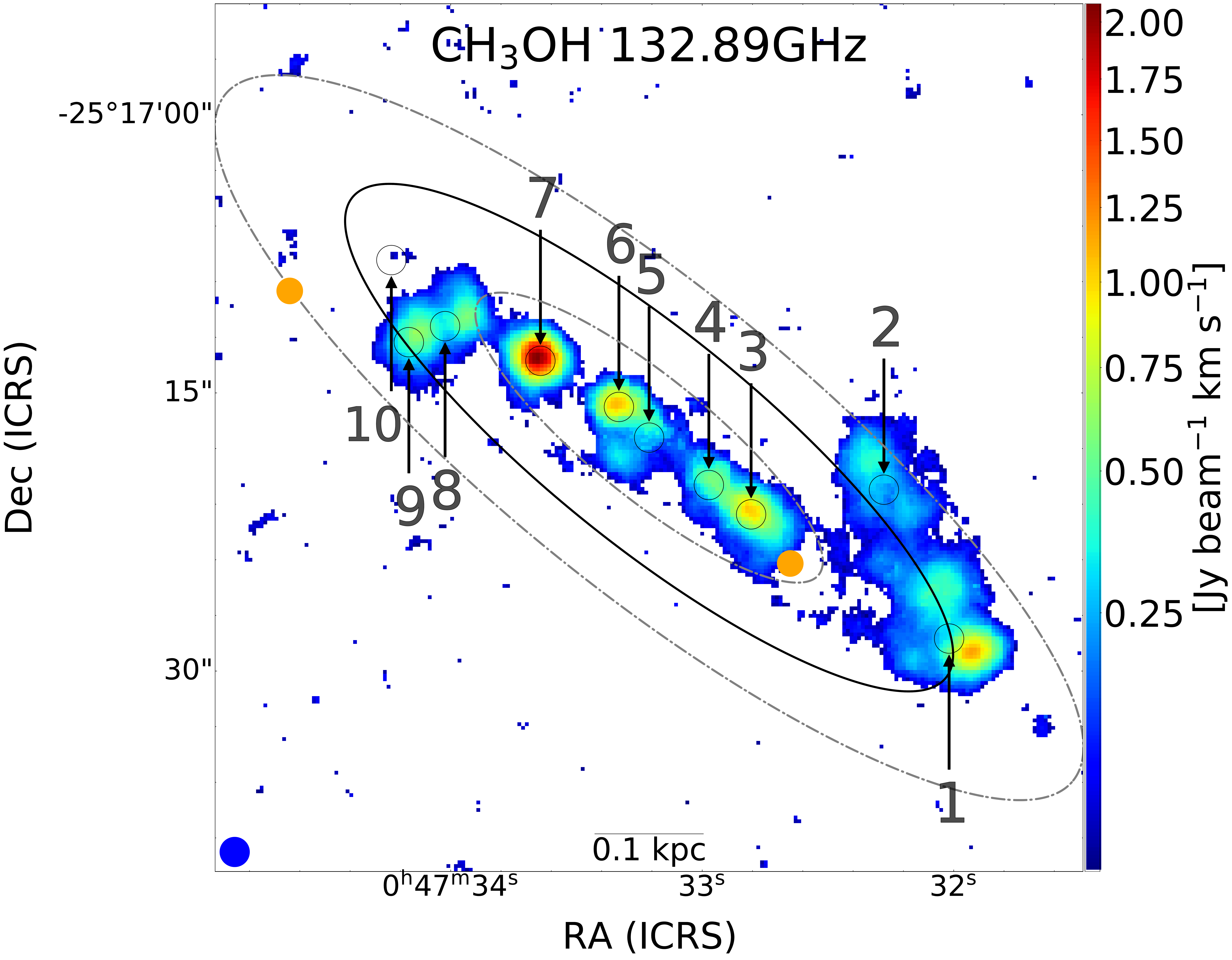}
\caption{Integrated  intensity map for the $6_{-1}\rightarrow5_{0}-E$ methanol line at 132.9\,GHz (within a velocity interval of $v_{\rm sys}\lesssim\pm$200\,km\,s$^{-1}$). Selected positions for spectral extraction (Sect.\,\ref{sec.selected_regions}) are encircled in beam-sized apertures and numbered from 1 to 10. The center and edges of the inner Lindblad resonance from \citet{Iodice2014} are denoted with a black ellipse and dash-dotted grey ellipses, respectively. Superbubbles identified by \citet{Sakamoto2006} are indicated as orange circles. The ALMA beam (1\farcs6) is shown in blue in the bottom left corner.}
\label{fig.integrated_intensity}
\end{figure}

In Fig.\,\ref{fig.integrated_intensity}, we show the integrated intensity map of the 6$_{-1}\rightarrow$5$_{0}-E$ methanol emission (132.9\,GHz rest frequency) using a 3$\sigma$ clip threshold and applying the formalism suggested by \citet[][their Appendix C]{Mangum2015} to obtain the uncertainties.  The selection of this methanol line is motivated by its high signal-to-noise (S/N) ratio, the absence of notable blending candidates, and maser emission in some regions (see below). In NGC\,253, the inner Lindblad resonance (ILR) was found to be co-spatial with its circumnuclear ring (CNR) by \citet{Iodice2014}. The total ILR extension measured by them is 0.3$\pm$0.1--0.4$\pm$0.1\,kpc (deprojected size). We draw concentric ellipses at 0.2 and 0.5\,kpc to denote the ILR limits in Fig.\,\ref{fig.integrated_intensity} (dash-dotted grey ellipses) as well as its center at 0.35\,kpc (solid black ellipse). Along the text, we use the ILR position as a proxy to the x$_{2}$ orbit in NGC\,253 (see Sect.\,\ref{sec.discussion}).

% trim={<left> <lower> <right> <upper>}

\begin{table}[!htp]
\caption{Selected regions.} 
\label{tab.positions}
\begin{center}
\begin{tabular}{llll}
\hline \hline
Region & $\alpha_{\rm ICRS}$ & $\delta_{\rm ICRS}$ & $v_{\rm LSR}$ \\
       & \multicolumn{1}{c}{[00$^{h}$:47$^{m}$:--$^{s}$]}  & \multicolumn{1}{c}{[-25$^{\circ}$:17':--"]}  & \multicolumn{1}{c}{[km\,s$^{-1}$]} \\
       \hline \\
R1  & 32.02 & 28.3 & 304\\ 
R2  & 32.28 & 20.2 & 330\\
R3  & 32.81 & 21.6 & 286\\
R4  & 32.97 & 20.0 & 252 \\
R5  & 33.21 & 17.4 & 231\\
R6  & 33.33 & 15.8 & 180\\
R7  & 33.64 & 13.3 & 174\\
R8  & 34.02 & 11.4 & 205 \\
R9  & 34.17 & 12.3 & 201\\
R10 & 34.24 & 7.8 & 144\\
\hline
\end{tabular}
\tablefoot{The Local Standard of Rest (LSR) velocities were obtained by fitting a local thermodynamic equilibrium model (LTE) to the spectra and averaging between the A-- and E--CH$_{3}$OH symmetry species. Our spectral resolution of $\sim$8--9\,km\,s$^{-1}$ dominates the velocity uncertainty.}
\end{center}
\end{table}

The presence of methanol masers depends on the physical conditions prevailing across the CMZ of NGC\,253.
%in the observed positions.
We decided to select ten distinct positions within the CMZ (see Figure\,\ref{fig.integrated_intensity} and Table\,\ref{tab.positions}). These positions were established based on intensity peaks of the CS \textit{J}=2--1 and H$^{13}$CN \textit{J}=1--0 transitions, applying a GMC identification approach based on \citet{Leroy2015}, i.e., using the CPROPS software \citep{Rosolowsky2006}. Our region coordinates are not exactly the same as those used by \citet{Leroy2015}, even though our numbering is close to theirs. We preferred to use those GMC locations instead of the peak intensities of our CH$_{3}$OH lines at a given frequency, because the peak location varies depending on the chosen transition \citep[see also,][]{Zinchenko2017,McCarthy2018}. We list the coordinates and velocities of all these positions in Table\,\ref{tab.positions}. The velocities were obtained from preliminary LTE radiative transfer models, similar to the ones described in Sect.\,\ref{sec.models}, and averaged over each methanol symmetric type.

For each region, we extracted the full spectrum (84--373\,GHz, see Sect.\,\ref{sec.models}) from the data convolved to a common circular beam of 1\farcs6 diameter, equivalent to the maximum angular resolution of the observations (see Sect.\,\ref{Sec.Observations}). At higher resolution, however, GMCs further divide into molecular clumps with a size range of 0\farcs07--0\farcs25 (1.2--4.3\,pc; \citealt{Leroy2018}). Contrary to the spectrum of the other nine studied regions, we have found a very complex spectrum in region\,5 located next to the dynamical center of the galaxy \citep{MullerSanchez2010}. Radio recombination line observations revealed an S-shaped pattern with complex kinematics including a counter-rotating core in the inner 2$\arcsec$ suggestive of a secondary bar \citep{Anantharamaiah1996} and evidencing a black hole mass of $\sim$10$^{7}$\,M$\odot$ \citep{Cohen2020}. This highly-perturbed environment is likely causing the crowded spectrum observed in Region\,5. In this position, the broad line emission, with a full width at half maximum (FWHM) above 100\,km\,s$^{-1}$, prevents an accurate Gaussian line fitting for the rotation diagram analysis in Sect.~\ref{Sec.rot_diags}, but can be used for the radiative transfer modelling in Sect.\,\ref{sec.models}. In addition, toward region\,5 and mostly in the low frequency range ($\sim$84--163\,GHz, ALMA bands\,3 and 4) we observe absorption components. 

\begin{figure}[!htp]
\includegraphics[width=1.1\linewidth, trim={0cm 1cm 0cm 0cm},clip]{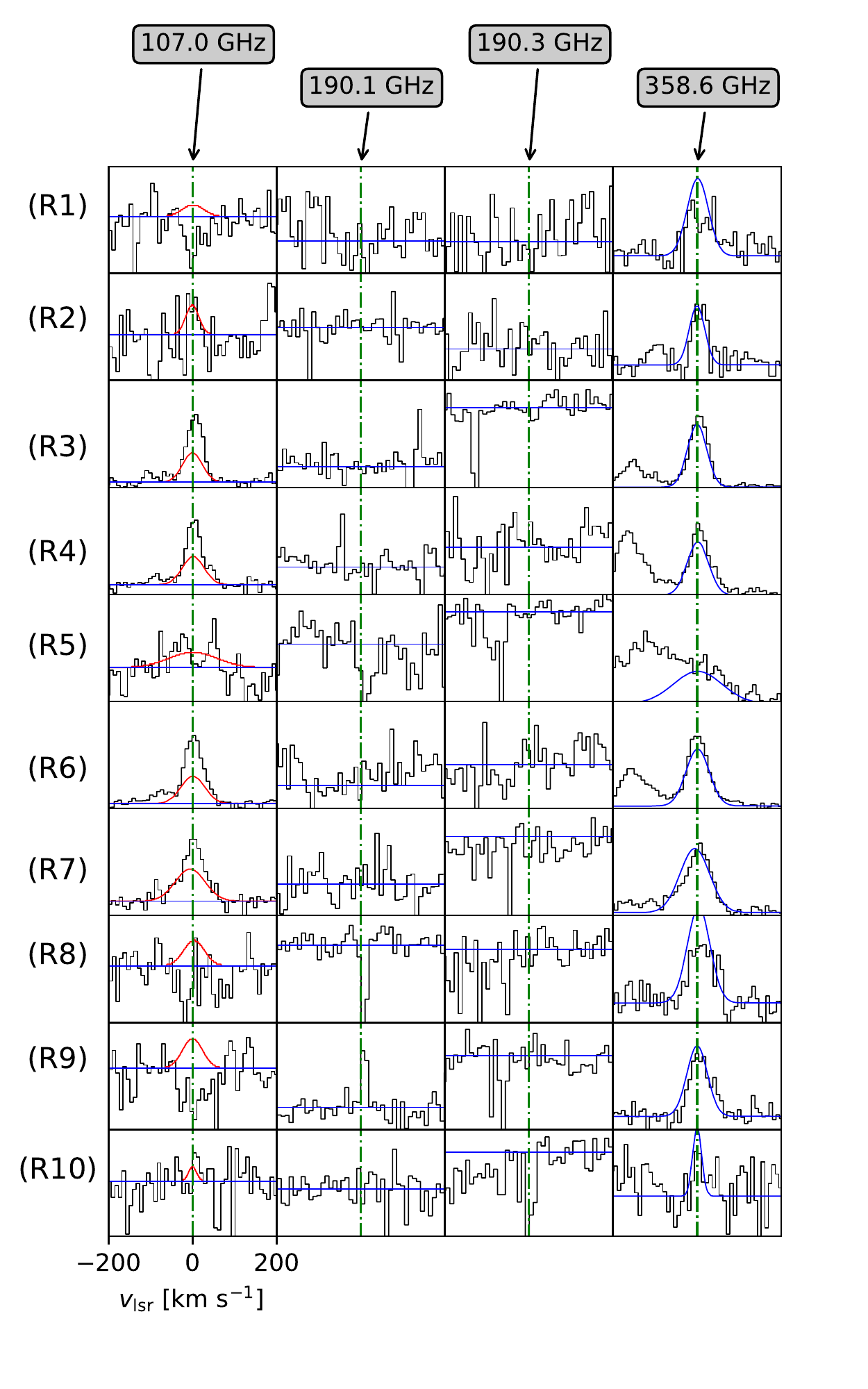}
\caption{Methanol lines observed in absorption or self-absorption in one or more of the selected regions (Table\,\ref{tab.positions}). Red and blue lines are synthetic spectra for methanol A- and E- type, respectively (see\,Sect.\,\ref{sec.models}). Line frequencies are labeled in the top of the Figure, while the velocity range (in km\,s$^{-1}$, after applying the radio convention\protect\footnotemark \,and subtracting the $v_{\rm LSR}$ velocities from Table\,\ref{tab.positions}) is indicated in the bottom-left corner. Regions 1 to 10 are ordered from top to bottom as indicated in the leftmost panel (y-axis: R from "Region" plus the corresponding number). For the location of the regions, see Fig.\,\ref{fig.integrated_intensity} and Table\,\ref{tab.positions}.}
\label{fig.abs_lines}
\end{figure}

Methanol lines in absorption or self-absorption (in the case of the line at 358.6\,GHz) along the entire ALCHEMI spectral coverage are observed in at least one region for the following four transitions: the 3$_{1}\rightarrow 4_{0}-A^{+}$ line at 107.0\,GHz, the 4$_{-2}\rightarrow 4_{-0}-E$ line at 190.1\,GHz, the 5$_{-2}\rightarrow 5_{-0}-E$ line at 190.3\,GHz (in absorption in region\,10 only), and the 4$_{1}\rightarrow 3_{0}-E$ line at 358.6\,GHz. They are shown in Figure\,\ref{fig.abs_lines}.

\footnotetext{\url{https://web-archives.iram.fr/IRAMFR/ARN/may95/node4.html}}

\section{Methanol maser emission identification}
\label{sec.methanol_maser_emission_ID}

In some cases, it can be difficult to assess whether a given line is excited under thermal or maser conditions. Methanol maser lines in Galactic star-forming regions are usually very bright (in the sense of brightness temperature) and narrow. However, in the case of extragalactic objects, beam dilution makes them apparently weaker and less distinguishable from thermal emission due to line blending from surrounding gas. In the spectral dimension this also complicates a proper line identification since the contribution of other species might affect the true brightness of the putative maser lines (see the last column of Table\,\ref{tab.methanol_lines}).

In this section we explore two methods to identify methanol emission lines outside local thermodynamic equilibrium (LTE) as an indicator of potential maser emission, namely rotation diagram analysis and comparisons with synthetic spectra from radiative transfer modelling. For both methods, we constrained the upper energy above the ground level ($E_{\rm up}$/k) of the lines to $\leq$150\,K, since the entirety of lines above that limit are blended with other methanol transitions below that limit, and are likely not contributing significantly to the overall emission (see the analysis in Sect.\,\ref{sec.models}). 

In this study, we consider as LTE conditions those reached when a single excitation temperature, $T_{\rm ex}$, characterizing the energy level population according to the Boltzmann distribution, is sufficient to explain the line emission of methanol along all its observed transitions. Namely, an equivalency between kinetic ($T_{\rm kin}$) and excitation temperatures is not strictly required.

\subsection{Rotation diagram}
\label{Sec.rot_diags}

As a first approach, we use the rotation diagram method to compare the relative intensities of methanol lines in NGC\,253. The details of this method can be found in \citet{Goldsmith1999} and are also summarized in our Appendix\,\ref{sect.apen.rot_diag}.

Since this method assumes that the gas is under LTE conditions and that the lines are optically thin, any transition deviating from a straight-line fit in the rotation diagram indicates that, for values below the fitted line, opacities are not low and that, for values above the fitted line, maser emission may be a suitable explanation. 

If a given line is affected by blending, we could overestimate its integrated intensity, which would lead to a wrongly classified maser line. Therefore, after identifying all the methanol transitions in our observations and before performing the rotation diagrams, we have initially searched both in the CDMS \citep{Mueller2005} and JPL \citep{Pickett1998} databases for potentially contaminating lines from other molecular species (last column in Table\,\ref{tab.methanol_lines}).
Special care was taken on previously reported molecules in NGC\,253 \citep[e.g.][]{Martin2006,Meier2015,Ando2017}, as well as preliminary line identification performed on the 1\farcs6 resolution ALCHEMI data (Mart\'in et al. in prep.). Blended non-methanol line candidates are required to fall inside the FWHM of the given methanol line, which varies for each symmetric type and region. 

To produce the rotation diagrams we have used the CASSIS\footnote{\url{http://cassis.irap.omp.eu}} software. In particular, we have used the spectroscopic VASTEL database, which comes from the JPL catalog\footnote{\url{http://cassis.irap.omp.eu/?page=catalogs-vastel}}, since it distinguishes between A and E methanol forms. Covering transitions with $E_{\rm up}$/k$<$150\,K, we did not consider methanol lines separated from each other by less than their FWHM. Because of that, we did not include in our analysis the $J_{K}\rightarrow (J-1)_{K}$ line series at $\sim$96.7\,GHz ($J=2$), $\sim$145.1\,GHz ($J=$3), $\sim$193.5\,GHz ($J=$4), $\sim$241.8\,GHz ($J=$5), and $\sim$290.1\,GHz ($J=$6).

%A summary of the considered transitions for each region is shown in Table\,\ref{tab.methanol_lines}, where line parameters were taken from the CDMS database (the same holds for the other tables throughout this work). For each of the selected lines, we have obtained the integrated line intensity $\int T_{\rm{mb}}$d$v$ by fitting a single Gaussian profile. Then, for each region we constructed the rotation diagrams for A-- and E--CH$_{3}$OH symmetry species separately (see Fig.\,\ref{fig.rotationdiagrams}). For the rotation diagrams, we conservatively assumed a flux calibration uncertainty of 15\% for the integrated intensities (see\,Sect.\,\ref{Sec.Observations}), in agreement with previous works using ALCHEMI data \citep[see e.g.][]{Martin2021}. This uncertainty plus Gaussian fitting uncertainties were added in quadrature for each line. 

\begin{figure*}[!htp]
\includegraphics[width=0.93\linewidth, trim={3cm 12cm 3cm 8.1cm},clip]{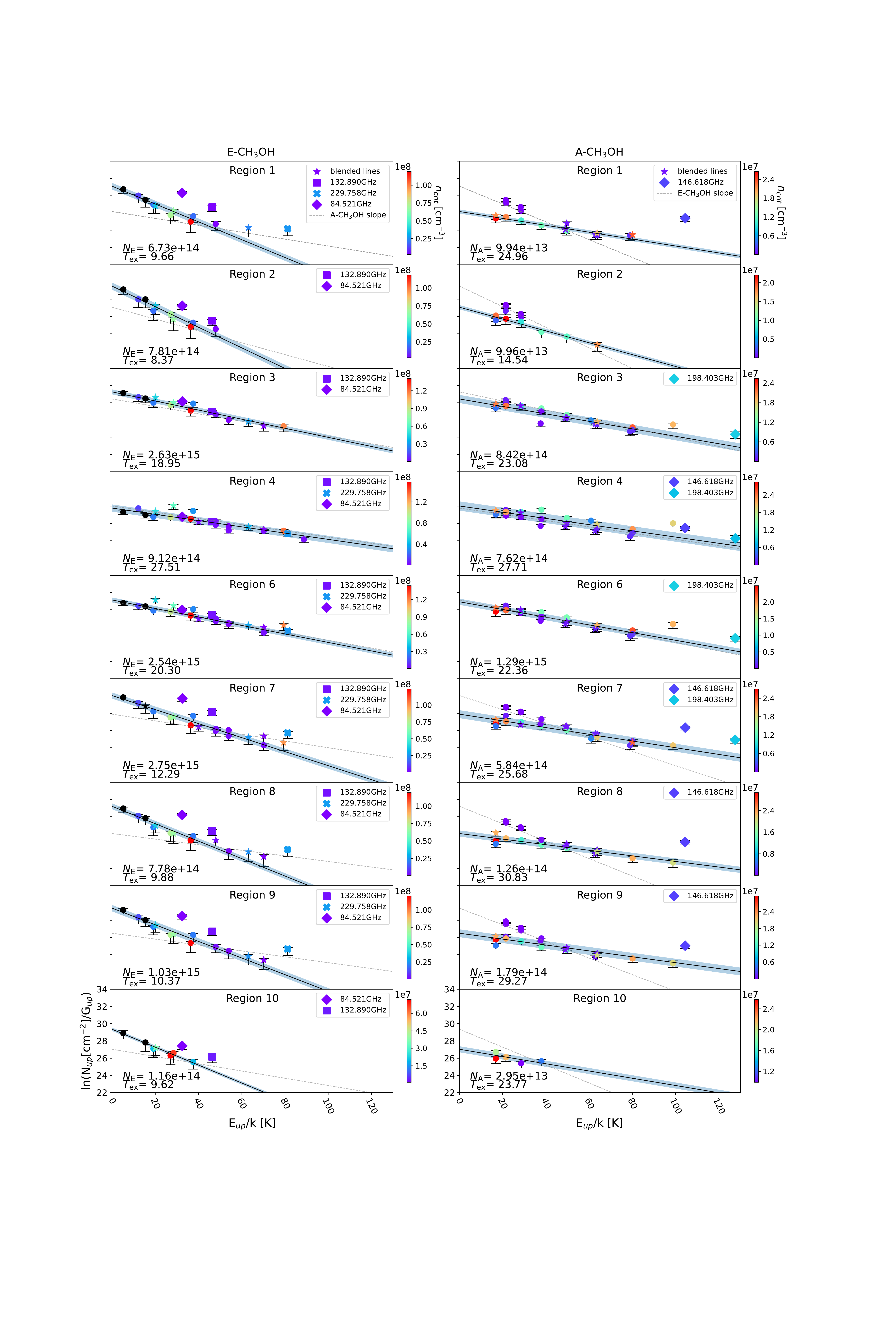}
\caption{Rotation diagrams based on our selected methanol transitions (Table\,\ref{tab.methanol_lines}) and separated by methanol symmetric type (E--CH$_{3}$OH and A--CH$_{3}$OH). Straight black lines represent our best fits to the data considering only methanol lines not blended with other methanol lines (see Sect.\,\ref{Sec.rot_diags}) and following LTE conditions. Dotted grey lines are the slopes of the complementary methanol symmetric type in the same region. Maser candidates (see Sect.\,\ref{Sec.rot_diags}) are labelled with symbols. Points are color-coded depending on their critical densities (in units of cm$^{-3}$), as indicated by the color bar to their right (transitions with unavailable critical densities are in black). In the bottom left corner of each panel, the column density ($N_{\rm E-CH_{3}OH}$ and $N_{\rm A-CH_{3}OH}$, in cm$^{-2}$) and the excitation temperature ($T_{\rm ex}$, in Kelvin) derived from the fit are indicated. Non-blended lines in LTE are marked by circles, while methanol lines blended with other methanol lines (at $E_{\rm up}$/k$>$150\,K) are assigned by stars. Shaded areas represent 1$\sigma$ uncertainties (see Sect.\,\ref{Sec.rot_diags}).}
\label{fig.rotationdiagrams}
\end{figure*}

A summary of the considered transitions for each region is shown in Table\,\ref{tab.methanol_lines}, where line parameters were taken from the CDMS database (the same holds for the other tables throughout this work). For each of the selected lines, we have obtained the integrated line intensity $\int T_{\rm{mb}}$d$v$ by fitting a single Gaussian profile. The resulting rotation diagrams are presented in Figure\,\ref{fig.rotationdiagrams}, for A-- and E--CH$_{3}$OH symmetry species separately and for each studied region (Table\,\ref{tab.positions}), where we conservatively assumed a flux calibration uncertainty of 15\%, as recommended by \citet{Martin2021}. This uncertainty plus Gaussian fitting uncertainties were added in quadrature for each line.

The rotation diagram method assumes the Rayleigh–Jeans (RJ) approximation (see Appendix\,\ref{sect.apen.rot_diag}). In our case, given the low $T_{\rm ex}$ measured by the rotation diagrams in certain regions (see Table\,\ref{tab.apen.RD_params}), the RJ approximation is not a good assumption for high-frequency transitions, leading to an overestimation of $\sim$40\% for the upper level population of some transitions in the worst case scenario. As explained in Appendix\,\ref{sect.apen.rot_diag}, applying a correction factor to all our transitions, as an attempt to go beyond the RJ approximation, we obtain slight changes in $T_{\rm ex}$ but total column densities decrease by up to $\sim$40\% in extreme cases. The value that would be derived from the Planck function for the corresponding transition assuming the same $T_{\rm ex}$ originally derived with the RJ approximation for each region and methanol symmetric type, is shown inside the error bars in Fig.\,\ref{fig.rotationdiagrams}. The application of the Planck formula increases the lower-limit uncertainty as a function of the $T_{\rm ex}$ (see Eq.\,\ref{eq.CF}) found for each region through RJ and the individual frequency of each transition and therefore mostly affects the regions where $T_{\rm ex}$ is lower than 12\,K, i.e., regions\,1, 2, and 8 to 10 in E--CH$_{3}$OH. The parameters derived after applying this correction are also included in the uncertainties presented in Table\,\ref{tab.apen.RD_params}.

On Figure\,\ref{fig.rotationdiagrams} we also indicate the critical densities of the different transitions shown in the rotation diagrams. To this end, we have used the most recently available collisional rates (C$_{\rm{ij}}$), interpolating their values, given in steps of 10\,K by \citet{Rabli2010}, by the corresponding $T_{\rm ex}$ obtained through the rotation diagrams in each region and for each methanol symmetric type. The critical densities thus obtained, in units of cm$^{-3}$ are then indicated by colors in the rotation diagrams (Fig.\,\ref{fig.rotationdiagrams}; transitions with unavailable C$_{\rm{ij}}$ are in black). 

We started including all the considered transitions (Table\,\ref{tab.methanol_lines}) below a certain $E_{\rm up}$/k (see below) to perform our rotation diagrams and obtain the best linear fit. However, we soon realized that some transitions did not fall on the linear fit line derived from the ensemble of CH$_{3}$OH integrated intensities. We excluded these lines that do not follow LTE conditions but are not likely to correspond to masers, see Appendix\,\ref{sec.apen.rotdiag_outliers}, to perform, in a second iteration, a linear fit to the remaining data in order to estimate the column density and $T_{\rm ex}$. Our fit results are shown at the bottom of each rotation diagram in Fig.\,\ref{fig.rotationdiagrams} with outliers indicated in the legends when they correspond to maser candidates (see below). Column densities and $T_{\rm ex}$ with their uncertainties are listed in Table\,\ref{tab.apen.RD_params}. In general, the distribution of the data points can be well described by a single $T_{\rm ex}$. 

Without considering the outliers, the standard deviations (1$\sigma$) of the residuals of the data points from the fitted lines  are of a difference (ln($N_{\rm {up,\,data\,points}}$) - ln($N_{\rm{up,\,best fit}})$) between 0.2 (E--CH$_3$OH in region\,10) and 0.5 (A--CH$_3$OH in region\,4). These values translate into a factor ($N_{\rm{up,\,data\,points}}/N_{\rm{up,\,best fit}}$) between $\exp(0.2)=$1.2 and $\exp(0.5)=$1.7. 

We conservatively rounded up the average standard deviation ($\sigma=$0.35 to 0.4) and obtained a factor of 3.3 scatter, in the exponential scale of our rotation diagrams (exp(3$\sigma$)). We then classify as outliers all methanol transitions surpassing by more than a factor of 3.3 the expected value from LTE conditions. 

These outliers are listed in Table\,\ref{tab.RD_outliers}, where their transitional quantum numbers, frequencies, symmetry types, maser classes, $E_{\rm up}$/k, A$_{\rm{ij}}$, and optical depths are indicated in columns 1 to 7, respectively. In this table we also provide information about possible line blending with other methanol transitions falling within the FWHM of the main transition; they possess higher $E_{\rm up}$/k and lower A$_{\rm{ij}}$, implying a lower contribution to the observed spectrum due to the need for more extreme conditions to be emitted. In Table\,\ref{tab.RD_outliers}, the information given for these lines is the transition, frequency, symmetric type, $E_{\rm up}$/k, and A$_{\rm{ij}}$ along columns 8 to 12, respectively. 

Deviations from LTE are expected in the ISM, in particular for complex level diagrams and if radiative and collisional excitation and de-excitation compete. But in general, given the success in the fitting, it can be inferred that potential blending lines in Table\,\ref{tab.methanol_lines} are, in most cases, not significantly contaminating.

We wish to emphasize that a sophisticated model, including hundreds or thousands of transitions from many molecules, would be needed to determine methanol line blending with transitions of other species in a thorough way, providing a percentage of contamination to methanol made from other species. Since this is clearly beyond the scope of this paper (but we will sometimes quote initial results from region\,5, obtained from Martin et al., in preparation), we refer in the following to the case of blending of methanol with other methanol lines, unless contamination with other species is explicitly mentioned.

Based on the line width, CASSIS is able to discriminate between unblended methanol lines and those which are contaminated by adjacent lines of CH$_{3}$OH. In Fig.\,\ref{fig.rotationdiagrams} we represent unblended lines with circles and blended lines with stars. Since line widths change per region, the same transition may be denoted by circles or stars depending on the region. Besides of that, we have found that some methanol lines presented in Fig.\,\ref{fig.rotationdiagrams} are blended with other highly energetic methanol lines ($E_{\rm up}$/k$\geq$1000\,K, see Table\,\ref{tab.RD_outliers}) and therefore significant blending is highly unlikely. However, we decided to show them also as blended lines (stars) in order to be impartial along all the regions and transitions. In addition, we note that lines with $E_{\rm up}$/k$\leq$150\,K, being blended with CH$_{3}$OH transitions with $E_{\rm up}$/k$>$150\,K, are plotted with the lower $E_{\rm up}$/k value in Fig.\,\ref{fig.rotationdiagrams}. Finally, the LTE slope is adjusted exclusively considering unblended and non-masing methanol lines (for a definition of the latter, see below). 

%In addition to the uncertainties for each transition, the 1$\sigma$ scatter of the ensemble of non-maser/non-blended transitions per region and methanol symmetric type are indicated by shaded blue areas around the best fit slopes (black solid lines).

% trim={<left> <lower> <right> <upper>}

\begin{table*}[!t]
\caption{Outliers in the rotation diagrams.} 
\label{tab.RD_outliers}
\scriptsize
\renewcommand{\tabcolsep}{0.1cm}
\begin{center}
\begin{tabular}{llllllrllllllll}
\hline \hline 
\multicolumn{2}{l}{Maser candidate}                         &          &                      &                 &                      &   $\tau$\tablefootmark{a}       & blended\tablefootmark{b}      &     &         &       &    & maser\tablefootmark{c}  & location\tablefootmark{d} \\
Transition                  & $\nu$ [GHz] & A/E      & Class                & $E_{\rm up}$/k[K] & A$_{ij}$[s$^{-1}$]   &   [$\times$10$^{-4}$]          & Transition   & $\nu$ [GHz] & A/E     & $E_{\rm up}$/k[K] & A$_{ij}$[s$^{-1}$]     & [K]      & \\
\hline \\
5$_{-1}\rightarrow$4$_{0}$  & 84.521172   & E        & I                    & 32.49      & 1.97$\times$10$^{-6}$  &$-$236.6 & 12$_{10}\rightarrow$12$_{11}$  & 84.531123   & E       & 1085.97    & 7.72$\times$10$^{-7}$  & yes  & 1,2,7--10\\ 
2$_{1}\rightarrow$1$_{1}$   & 95.914310   & A$^{+}$  & --                   & 21.44      & 2.49$\times$10$^{-6}$  &12.5    &not-blended  & --  & --      & --    & --  & no &\\ 
2$_{1}\rightarrow$1$_{1}$   & 97.582798   & A$^{-}$  & --                   & 21.56      & 2.63$\times$10$^{-6}$  &3.8   &not-blended  & --  & --      & --    & --  & no  &\\ 
6$_{-1}\rightarrow$5$_{0}$  & 132.890759  & E        & I                    & 46.41      & 7.74$\times$10$^{-6}$  &$-$67.7 &not-blended  & --  & --      & --    & --  & yes & 1,2,7--10\\ 
3$_{1}\rightarrow$2$_{1}$   & 143.865795  & A$^{+}$  & --                   & 28.35      & 1.07$\times$10$^{-5}$  &7.8   &not-blended  & --  & --      & --    & --  & no  &\\ 
3$_{1}\rightarrow$2$_{1}$   & 146.368328  & A$^{-}$  & --                   & 28.59      & 1.13$\times$10$^{-5}$  &6.5 &not-blended  & --  & --      & --    & --  & no  &\\ 
9$_{0}\rightarrow$8$_{1}$   & 146.618697  & A$^{+}$  & I                    & 104.41     & 8.03$\times$10$^{-6}$  &$-$0.1 &14$_{1}\rightarrow$13$_{2}$    & 146.617429  & A$^{+}$ & 256.02     & 6.6$\times$10$^{-6}$   & yes  & 1,7--9\\
2$_{1}\rightarrow$3$_{0}$   & 156.602395  & A$^{+}$  & --                   & 21.44      & 1.78$\times$10$^{-5}$  &162.3  &not-blended  & --  & --      & --    & --  & no  &\\ 
10$_{0}\rightarrow$9$_{1}$  & 198.403067  & A$^{+}$  & I                    & 127.60     & 4.10$\times$10$^{-5}$  &$-$0.01 &not-blended  & --  & --      & --    & --  & yes  & 3,6,7\\ 
8$_{-1}\rightarrow$7$_{0}$  & 229.758756  & E        & I                    & 81.20      & 4.19$\times$10$^{-5}$  &$-$2.1 &not-blended  & --  & --      & --    & --  & yes  & 2,7--9\\ 
2$_{-1}\rightarrow$1$_{-0}$   & 261.805675  & E  & --                       & 28.01      & 5.57$\times$10$^{-5}$  &402.3 &20$_{5}\rightarrow$21$_{3}$      & 261.800288  & E    & 611.06     & 9.85$\times$10$^{-9}$  & no   &\\
6$_{2}\rightarrow$5$_{1}$   & 315.266861  & E        & --                   & 63.10      & 1.18$\times$10$^{-4}$  &6.7 &37$_{-6}\rightarrow$36$_{-7}$  & 315.237555  & E       & 1808.32     & 5.73$\times$10$^{-5}$ & no   &\\
8$_{1}\rightarrow$8$_{0}$   & 318.318919  & A$^{-+}$ & II\tablefootmark{e}  & 98.82      & 1.79$\times$10$^{-4}$  &1.5 &not-blended  & --  & --      & --    & --  & no   &\\
7$_{-0}\rightarrow$6$_{-0}$   & 338.124488 & E        & --         & 78.08      & 1.70$\times$10$^{-4}$  &10.9 &38$_{12}\rightarrow$37$_{10}$  & 338.117653  & E       & 2425.29     & 7.38$\times$10$^{-8}$ & no   &\\
4$_{-0}\rightarrow$3$_{1}$  & 350.687662  & E  & --                & 36.33      & 8.67$\times$10$^{-5}$  &1205.0 &18$_{-3}\rightarrow$18$_{-2}$      & 350.723878  & E & 812.46     & 4.17$\times$10$^{-5}$  & no   &\\
4$_{1}\rightarrow$3$_{0}$   & 358.605799  & E        & --                   & 44.26      & 1.32$\times$10$^{-4}$    &224.1  &not-blended  & --  & --      & --    & --  & no &\\
7$_{2}\rightarrow$6$_{1}$   & 363.739868  & E        & --                   & 79.36      & 1.70$\times$10$^{-4}$  &1.7 & 23$_{6}\rightarrow$22$_{9}$    & 363.680652  & E       & 1327.80    & 1.78$\times$10$^{-7}$  & no  & \\

 \hline

\end{tabular}
\tablefoot{\tablefoottext{a}{Optical depth according to our non-LTE model of region\,8 (see Appendix\,\ref{sec.apen.detailed_modelling}}). \tablefoottext{b}{Methanol lines next to the frequency given in column (2). Their frequency is given in column (9)}. \tablefoottext{c}{Maser candidates based on the discussion in Sect.\,\ref{sec.maser_line_candidates_RD}}. \tablefoottext{d}{Regions where the transition is masing}.\tablefoottext{e}{According to the classification of the $J_{1}-J_{0} A^{-+}$ line series in \citet{Zinchenko2017}. The dash (--) symbol indicates no information/detection. For more information about the columns see Sect.\,\ref{Sec.rot_diags}.}}

\end{center}
\end{table*}

\subsubsection{Rotation diagram results}
\label{sec.maser_line_candidates_RD}

%To distinguish maser candidates among the outliers presented in Table\,\ref{tab.RD_outliers}, that is, lines located beyond a 3$\sigma$ scatter; the respective line must belong to known methanol maser line series already discovered in NGC\,253 at lower frequencies. This is because maser action is more prominent at lower frequencies than the ones covered in this work: as A$_{\rm ij}$ is proportional to the cube of the frequency, lower A$_{\rm ij}$ leads to a longer time lapse to accumulate inverted populations. In addition, we also require that the candidate line shows maser behaviour in more than one region. 

To distinguish maser candidates among outliers, that is, lines located beyond a 3$\sigma$ scatter; the respective line must belong to known methanol maser line series already discovered in NGC\,253 at lower frequencies. This is because maser action is more prominent at lower frequencies than the ones covered in this work: as A$_{\rm ij}$ is proportional to the cube of the frequency, lower A$_{\rm ij}$ leads to a longer time lapse to accumulate inverted populations. In addition, we also require that the candidate line shows maser behaviour in more than one region.

In the following we measure the departure from LTE of the maser lines as the ratio between their nominal upper level column densities in the rotation diagrams ($N_{\rm up, maser}$) over the expected upper level column density in LTE ($N_{\rm up, LTE}$). In the computation of $N_{\rm up}$ one generally assumes a proportional relation with the integrated intensity (see Eq.\,\ref{eq.N_up}). For maser lines, however, intensities and abundances are not related, as their negative opacity amplifies radiation from the background. Therefore this $N_{\rm up, maser}$/$N_{\rm up, LTE}$ factor must be taken as an intensity difference instead of abundance difference.

\subsubsection{E--type methanol masers}
\label{subsec.Emasers}

Among the E--CH$_{3}$OH maser line candidates belonging to the $J_{-1}\rightarrow(J-$ 1)$_{0}-E$ series (at 84.5, 132.9, and 229.8\,GHz), none of them were detected out of LTE in regions\,3 to 6 (see Table\,\ref{tab.RD_outliers}), which are therefore excluded from the following analysis. 

\begin{figure*}[!htp]
\includegraphics[width=\linewidth, trim={0 0 0 3cm},clip]{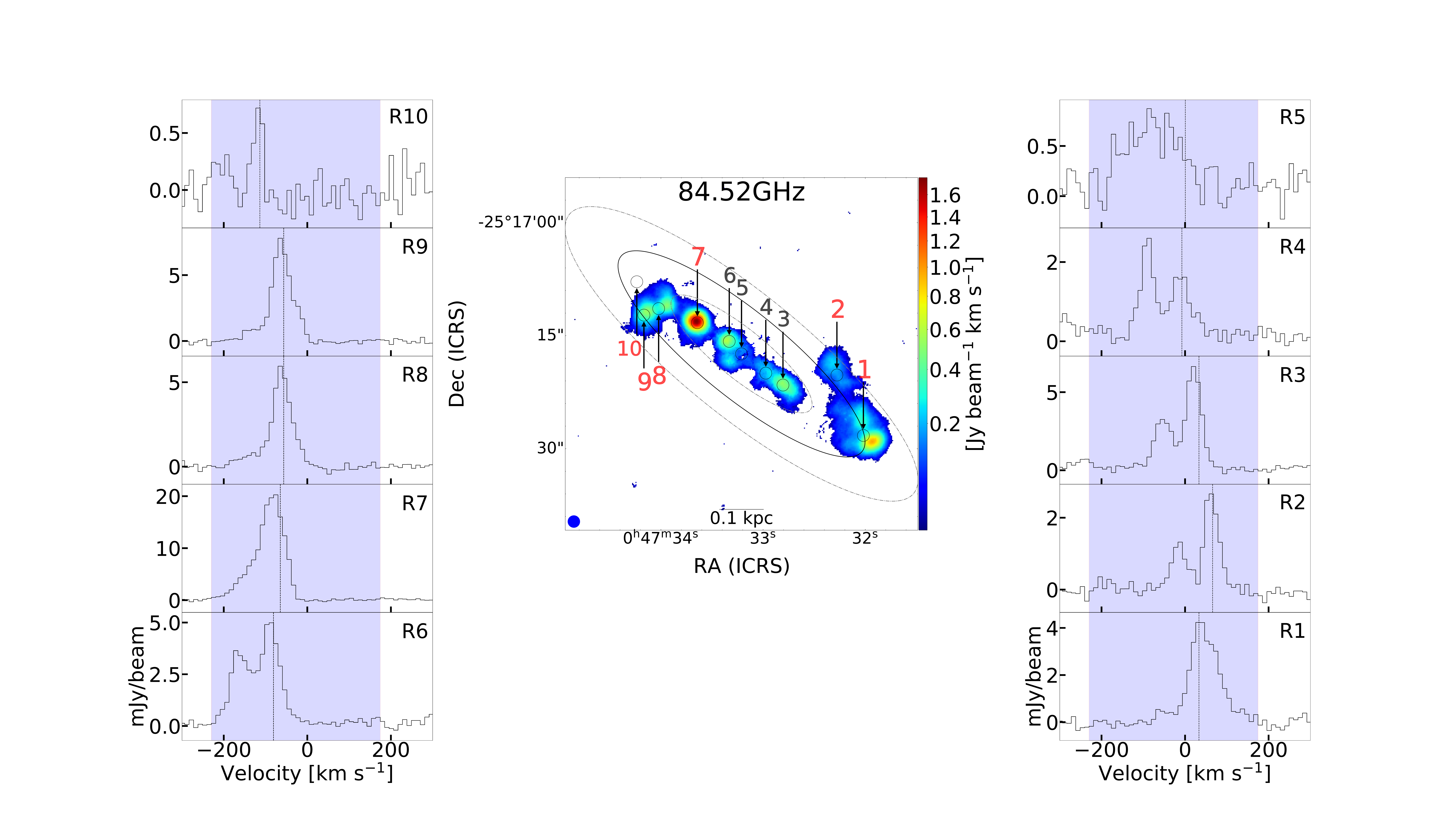}
%\caption{Middle panel: Velocity ($v_{\rm sys}$ $\leq\pm$220\,km\,s$^{-1}$) integrated intensity
\caption[]{Central panel: Velocity integrated intensity of the $5_{-1}\rightarrow4_{0}-E$ methanol line at 84.5\,GHz obtained by integrating the channels inside the colored areas shown in the side panels. Regions where we detect Class\,I maser emission in the $J_{-1}\rightarrow (J-1)_{0}-E$ series (see Sect.\,\ref{subsec.Emasers}) are labeled in red and with slightly larger numbers. Side panels: methanol spectra of the different regions as labeled in the top right corner of each panel (a $v_{\rm sys}$ of 258.8\,km\,s$^{-1}$ was subtracted). Vertical lines indicate the peak velocity of the $^{13}$CO $J=$1--0 line (110.2\,GHz) extracted in each region, as a reference for the systemic velocity at that position. Coordinate system, beam size and other parameters are the same as in Fig.\,\ref{fig.integrated_intensity}.}
\label{fig.m0_and_spectra_84}
\end{figure*}

\begin{figure*}
\includegraphics[width=\linewidth, trim={0 0 0 3cm},clip]{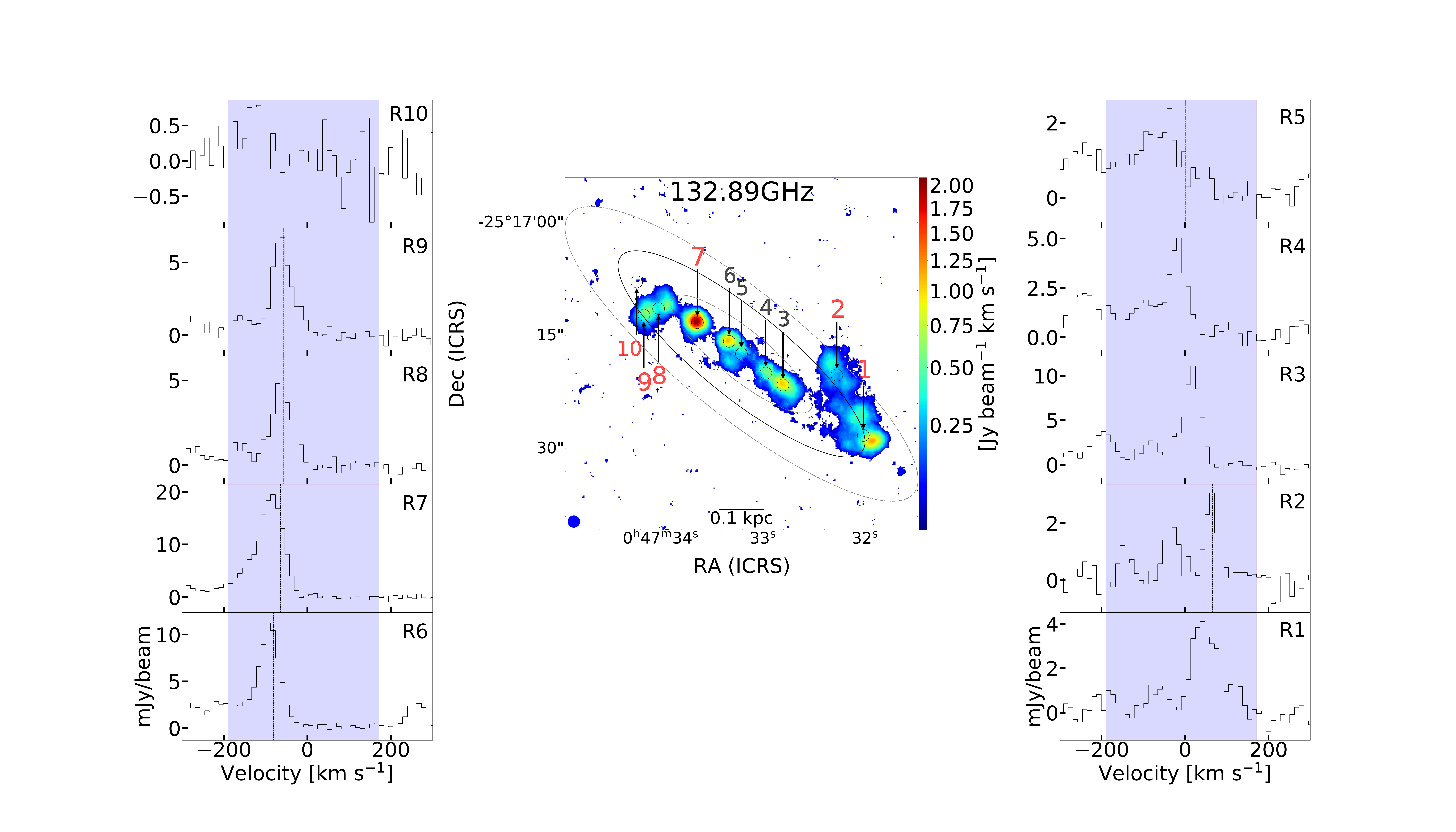}
\caption{Same as for Figure\,\ref{fig.m0_and_spectra_84} but for the $6_{-1}\rightarrow5_{0}-E$ methanol line at 132.9\,GHz.}
\label{fig.m0_and_spectra_132}
\end{figure*}

We present the spectra and velocity integrated intensity maps of the two transitions with the highest S/N ratios ($5_{-1}\rightarrow 4_{0}-E$ at 84.5\,GHz and $6_{-1}\rightarrow 5_{0}-E$ at 132.9\,GHz) in Figures\,\ref{fig.m0_and_spectra_84} and \ref{fig.m0_and_spectra_132}.

\textbf{The} $\mathbf{5_{-1}\rightarrow 4_{0}-E}$ line at 84.5\,GHz departs from LTE by factors ($N_{\rm up, maser}$/$N_{\rm up, LTE}$) ranging from 4.2$\pm$1.3 to 13.2$\pm$1.2 in regions\,10 and 1, respectively (see Fig.\,\ref{fig.rotationdiagrams}). This is the first detection of the 84.5\,GHz maser line in more than one position, after its first discovery by \citet{McCarthy2018}. The higher $J$ masers in the $J_{-1}\rightarrow(J-1)_{0}-E$ series are usually observed to originate in the same regions, suggesting similar distributions and pumping mechanism.

\textbf{The} $\mathbf{6_{-1}\rightarrow 5_{0}-E}$ line at 132.9\,GHz departs from its expected integrated intensity in LTE by factors ranging from 4.6$\pm$1.2 to 10.6$\pm$1.2, in regions\,2 and 1, respectively. Contrary to the previous transition at 84.5\,GHz, it is not blended with other methanol lines.

\textbf{The} $\mathbf{8_{-1}\rightarrow 7_{0}-E}$ line at 229.8\,GHz is observed to mase in the same regions as the ones at 84.5 and 132.9\,GHz except in regions\,2 and 10, which display the lowest LTE fitted $T_{ex}$ in the E-type methanol species and also the lowest S/N ratios. In these couple of regions, the emission is too weak to be classified as a maser. The 8$_{-1}\rightarrow 7_{0}-E$ line is the highest frequency transition detected as a maser by the rotation diagram method; its intensity departs from LTE by 9.8$\pm$1.2, in region\,7 and up to 32.0$\pm$1.2 in region\,1.

% trim={<left> <lower> <right> <upper>}

% trim={<left> <lower> <right> <upper>}

\subsubsection{A--type methanol masers}
\label{subsec.Amasers}
With respect to A--CH$_{3}$OH, maser line candidates are part of the $J_{0}\rightarrow (J-1)_{1}-A^{+}$ series (9$_{0}\rightarrow 8_{1}-A^{+}$ and 10$_{0}\rightarrow 9_{1}-A^{+}$ transitions). We note that a lower $J$ (and lower frequency) transition in this series is the 7$_{0}\rightarrow 6_{1}-A^{+}$ line at 44.1\,GHz, the strongest Galactic Class\,I maser.

\textbf{The} $\mathbf{9_{0}\rightarrow 8_{1}-A^{+}}$ line at 146.618\,GHz departs from LTE by factors in the range of 8.4$\pm$1.2 (in region\,8) to 29.8$\pm$1.2 (in region\,1), excluding region\,4 because of line blending (see Appendix\,\ref{sec.apen.rotdiag_outliers}). This maser transition has been suggested to be part of the Class\,I family of methanol masers \citep[e.g.][and references therein]{Yang2020}.

\textbf{The} $\mathbf{10_{0}\rightarrow 9_{1}-A^{+}}$ line at 198.4\,GHz departs from LTE in regions\,3, 6, and 7, being the only masing line candidate for regions\,3 and 6. In those regions, this line departs from LTE by factors of 4.3$\pm$1.3, 3.6$\pm$1.2, and 8.5$\pm$1.2, respectively. 

\begin{figure}[!t]
\includegraphics[width=\linewidth, trim={1cm 1cm 2.2cm 3cm},clip]{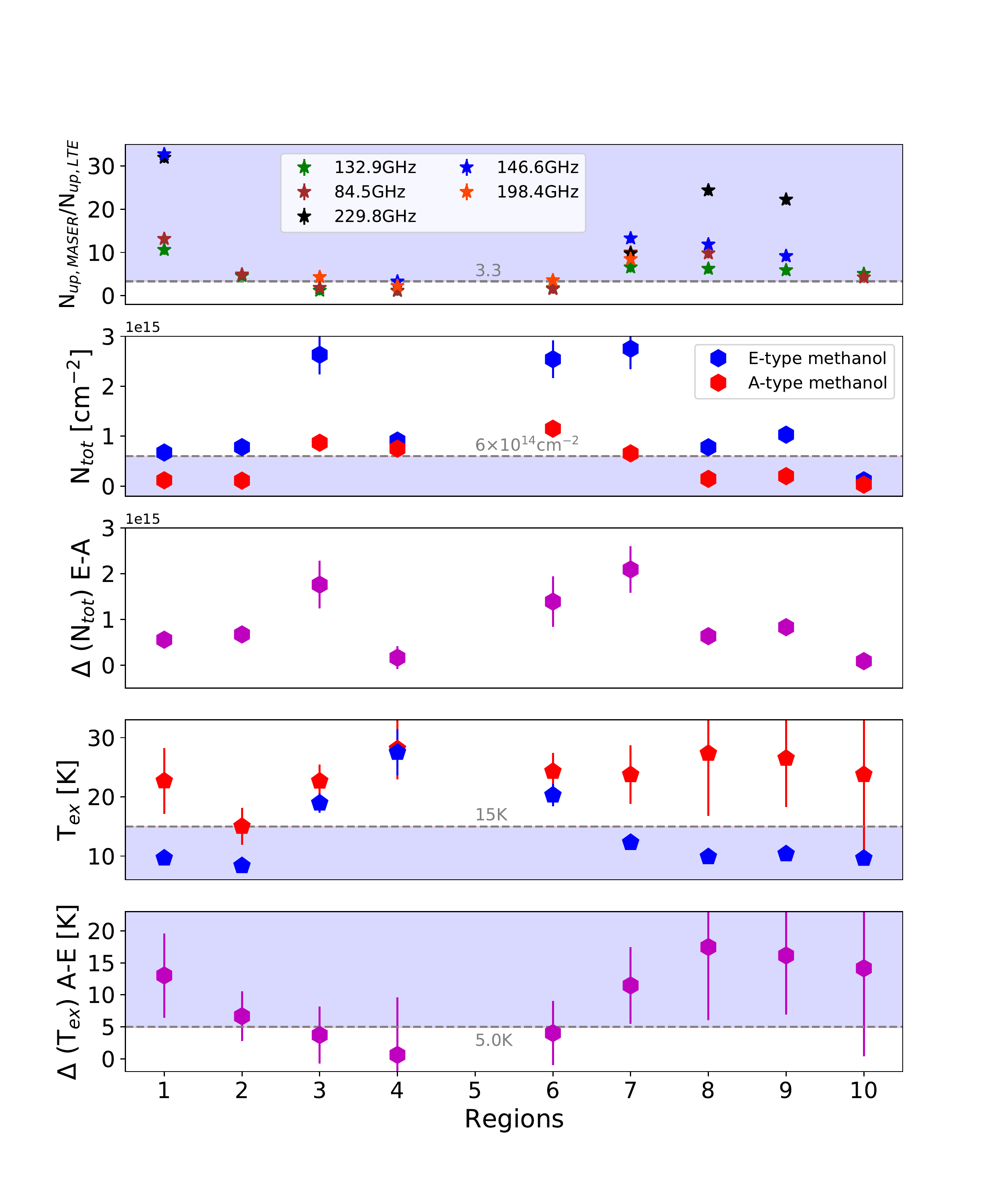}
\caption{Physical conditions for the methanol maser candidates (see Table\,\ref{tab.RD_outliers}) derived from our rotation diagrams. The x--axis indicates the number of the region. The y--axis represents, from top to bottom, (1) the ratio between the nominal upper level column densities of the lines $N_{\rm up, MASER}$ and the expected values from our best fit to LTE conditions $N_{\rm up}, LTE$, (2) the total column density, (3) the difference between total E-- and A--type column densities, (4) the LTE excitation temperatures and (5) the difference between the A-- and E--type excitation temperatures. Shaded areas highlight regions were maser emission in the $J_{-1}\rightarrow(J-$ 1)$_{0}-E$ line series (at 84, 132, and 229\,GHz), and the $J_{0}\rightarrow (J-1)_{1}-A^{+}$ line series (at 146 and 198.4\,GHz), are observed, according to the first panel.}
\label{fig.delta}
\end{figure}

We have found that all our maser candidates, except the $10_{0}\rightarrow 9_{1}-A^{+}$ line at 198.4\,GHz, are masers at the outskirts of the CMZ of NGC\,253 (see Table\,\ref{tab.RD_outliers}). The physical conditions in the regions giving rise to those masers are characterized (see Fig.\,\ref{fig.delta}) by total column densities for A--CH$_{3}$OH lower than 6$\times$10$^{14}$\,cm$^{-2}$, an excitation temperature lower than 15\,K in E--CH$_{3}$OH, and differences between A-- and E--CH$_{3}$OH excitation temperatures larger than 5.0\,K. Conversely, we do not see a correlation between the presence of masers and total column densities in E-type methanol, the temperature described by A--CH$_{3}$OH, or differences between column densities of the two methanol types. 

With the exception of the line at 198.4\,GHz, it is worth noting that all the regions where we detect Class\,I maser emission lie inside the Lindblad resonances (see e.g. Fig.\,\ref{fig.m0_and_spectra_84}), with the only exception of region\,7. We will further explore this point in Sect.\,\ref{sec.maser_emission_distribution}.

Based on LTE modelling, we have also found other transitions that are possibly experiencing maser activity. They will be described below.

\subsection{Radiative transfer modelling}
\label{sec.models}

In comparison with the rotation diagram method, synthetic spectra offer a number of improvements, like the possibility to reproduce line profiles both in LTE and out of LTE, which may also subtly indicate the number of gas components present in the observed source (see e.g. Fig.\,\ref{fig.LTE_R4}). Also, as the lines are observed on a linear scale, the effect of slight changes in the assumed gas conditions are noticed in more detail compared with the logarithmic scale used in the rotation diagram method. Apart from those advantages, radiative transfer modelling allows us to fit not only single methanol lines, but also blended ones, as their spectra are summed up when trying to obtain the desired intensities. Alternatively, this provides an estimate for the degree of contamination caused by other species. This allows us to include many more lines, increasing the sample size. In fact, as we will see below, there are 57 lines used for the fitting in our radiative transfer models, in contrast to the 39 finally used in the rotation diagrams (see Sect.\,\ref{sec.maser_line_candidates_RD}).

% trim={<left> <lower> <right> <upper>}

%Blue areas bathe regions were maser emission in the $J_{-1}\rightarrow(J-$ 1)$_{0}-E$ line series (at 84, 132, and 229\,GHz), and the $J_{0}\rightarrow (J-1)_{1}-A^{+}$ line series (at 146 and 198.4\,GHz), are observed.

The possibility to perform non-LTE modelling helps us to discard maser line candidates in case negative optical depths are not required to reproduce their emission. Finally, the nature of blended methanol lines with other methanol lines at known maser frequencies, such as the transitions at 95.169 and 278.305\,GHz, could not be unveiled through the rotation diagram method. As mentioned in Sect.\,\ref{Sec.rot_diags}, we use the term blending to refer to contamination of methanol lines with other methanol transitions, unless contamination by other species is explicitly mentioned.

We have computed synthetic spectra for each of the 10 selected regions in NGC\,253, covering the entire ALCHEMI frequency range ($\sim$84--373\,GHz, see Sect.\,\ref{Sec.Observations}), by using the CASSIS software, capable of producing LTE and non-LTE spectral modelling. For the non-LTE case, CASSIS is used as a wrapper of RADEX \citep{vanderTak2007}, a one-dimensional non-LTE radiative transfer code based on the escape probability formulation. In CASSIS it is possible to create a physical model defined by 6 parameters: column density of the species ($N(Sp)$), excitation ($T_{ex}$, LTE case) or kinetic ($T_{kin}$, non-LTE case) temperature, full width at half maximum (FWHM) of the lines, velocity of the source in the Local Standard of Rest ($V_{\rm LSR}$) system, size of the source in arcseconds, and the H$_2$ volumetric density ($n_{\rm H2}$) in the case of non-LTE modelling. For simplicity, in this section we have considered that only a single physical component for each methanol type is responsible for the observed emission; for a more detailed analysis, see Appendix\,\ref{sec.apen.detailed_modelling}. CASSIS makes use of the Monte Carlo Markov Chain (MCMC) method \citep{Hastings1970} to explore a user-predefined range of values for each of the parameters previously mentioned. By means of the $\chi^{2}$ minimization method, CASSIS is able to find the best ensemble of solutions. Computing the synthetic spectra we assumed a beam filling factor of one, by selecting a source size of 1\farcs6 ($\sim$27\,pc), and a slab geometry \citep[appropriate for shocks, e.g.][]{Leurini2016}.

Model solutions could be strongly influenced by the initial conditions. Therefore we start by selecting unblended methanol lines to be fitted and then add blended lines whose total line profile is successfully reproduced with the starting models. This was achieved by exploring the success of CASSIS in reproducing the lines throughout our selected regions.

After our initial attempts to fit methanol lines along the entire ALCHEMI spectral coverage, assuming either LTE or non-LTE conditions, it became clear from our 10 regions that it is impossible to properly fit all the lines with a single physical set of parameters, even after separating between A-- and E--CH$_{3}$OH flavours. This is especially important at frequencies below $\sim$156\,GHz, and it is possibly due to a couple of factors: the presence of a series of lines out of LTE, and the higher number of maser candidates.

Using our LTE model, we found that the $J_{K}\rightarrow (J-1)_{K}$ transitions (for both methanol symmetric types), which were avoided in Sect\,\ref{Sec.rot_diags} due to line blending, are not following LTE conditions. Within the ALCHEMI frequency coverage, these line series have the following frequencies: $\sim$96.7\,GHz ($J=$2), $\sim$145.1\,GHz ($J=$3), $\sim$193.5\,GHz ($J=$4), $\sim$241.8\,GHz ($J=$5), and $\sim$290.1\,GHz ($J=$6). Performing a non-LTE model in region\,8, they are satisfactorily reproduced, as can be seen in Fig.\,\ref{fig.apen.RADEX_R8} of Appendix\,\ref{sec.apen.detailed_modelling} without invoking the presence of masers.

At frequencies below $\sim$156\,GHz we also cover four maser candidates, three of them reported in Subsect.\,\ref{subsec.Emasers} plus an additional one that we will see below (Subsect.\,\ref{sec.LTE_outliers}). Therefore, at lower frequencies our models fail to reproduce an important quantity of available, not blended, methanol lines.

We discard possible software issues by doing a sanity check with another radiative-transfer code capable of producing LTE models, MADCUBA \citep{Martin2019b}. It shows similar results including convergence primarily toward lines above 156\,GHz. 

%Another issue that can prevent the reproduction of some methanol lines below 156\,GHz is that our 1\farcs6 data are based on a combination of both the 12-m meter ALMA and the 7-m ACA array. Therefore, we are also sensitive to relatively extended cool gas that may predominantly affect the lower frequency transitions. This may impede the reproduction of methanol lines with a single set of parameters.

%We explored the possibility of having an extended and colder outer layer that could be responsible for the emission at lower frequencies in the spectra since our radiative transfer models were unable to reproduce them satisfactorily. However, this was not possible because the predicted emission (both for LTE and out of LTE conditions) of the modelled lines at lower frequencies largely surpasses the line profiles at frequencies higher than 200\,GHz.

%In the following we discard from the fit the families of lines near 96.7 and 145.1\,GHz, as well as the maser candidates, in order to reduce the numbers of parameters and gas layers. This restriction also allows for a better agreement between synthetic models and rotation diagrams (which require much less assumptions), since for our diagrams we have measured a filling factor of 1 by considering a beam-sized region (1\farcs6) and we observe only two main sets of parameters per region, one per methanol species (see Table\,\ref{tab.LTE_model_params}), without an extra, cold and/or extended, component.

Having said the above, we do not include the mentioned $J_{K}\rightarrow (J-1)_{K}$ transition series nor the maser candidates in our LTE modelling. Even with this restriction, we observe that these series are well reproduced ($>$50\%) in region\,4 by merely fitting the other transitions, and this can be improved with a two-component LTE model (see Appendix\,\ref{sec.apen.detailed_modelling}). 

%When performing a non-LTE model for region\,8 (see Appendix\,\ref{sec.apen.detailed_modelling}), both series of lines are also better reproduced, although we cannot be completely confident about that non-LTE model due to an overestimation of the line series at 157.3\,GHz (see Appendix\,\ref{sec.apen.detailed_modelling} for a two-component modelling of regions\,4 and 8).

When inspecting and comparing the results for all regions between the LTE and non-LTE models, considering a single component for each methanol symmetric type, we observe that they are in agreement within an uncertainty of about 15\% (although this agreement is not observed when we consider a two-component non-LTE model, see Appendix\,\ref{sec.apen.detailed_modelling}). This is not surprising since high H$_{2}$ densities of $>$10$^{7}$\,cm$^{-3}$ are needed in the non-LTE models to reproduce most line profiles. At such densities collisions play a major role, tending to constrain the spread in excitation temperatures between different lines. Therefore, we conclude that the simpler LTE conditions are sufficient to represent the observed spectra in NGC\,253. From now on, we will therefore mainly refer to our LTE modelling, except in a few exceptional cases where this is explicitly mentioned. 

As described in Sect.\,\ref{sec.methanol_maser_emission_ID}, we set an upper $E_{\rm up}$/k threshold of 150\,K for the synthetic spectra. This was determined through comparing observations to model fits with and without the higher energy levels ($E_{\rm up}$/k$>$\,150\,K). A total of 600 LTE models for each individual spectrum, one per selected region (see Table\,\ref{sec.selected_regions}), were computed, reaching a convergence after 300--400 iterations. We find that we only require models that include $E_{\rm up}$/k$<$\,150\,K to fit the observed spectra. Thermal line emission with $E_{\rm up}$/k$>$\,150\,K may be there, but is too faint to affect model fits or to be separated from line blends. In the Galaxy, methanol masers with levels around 150\,K above the ground state are scarce. Within the ALCHEMI frequency coverage we can mention the case of the $11_{-1}\rightarrow 10_{-2}-E$ line at 104.3\,GHz \citep{Leurini2018}. However, this Class\,I maser is rarely seen \citep{Voronkov2012}. We do not obtain any strong emission at this frequency in NGC\,253.

Best fit parameters determined from our LTE models for each region and methanol species are listed in Table\,\ref{tab.LTE_model_params}. Differences between A-- and E--CH$_{3}$OH symmetry species are present in terms of temperature and density. In velocity they differ by a few km\,s$^{-1}$, although this discrepancy never exceeds our spectral resolution of $\sim$8--9\,km\,s$^{-1}$. Thus, the averaged $V_{\rm LSR}$ of these two species is adopted as the velocity of the region, in the same way as it has been previously established in Table\,\ref{tab.positions}.

%\setlength\tabcolsep{0.3pt}  % default value: 6pt
%%%TABLA MOFIFICADA CON VALORES PARA REGION 7 , 8 9 Y 10
\begin{table*}[!t]
\caption{Best fit parameters from our LTE modelling.} \label{tab.LTE_model_params}
\scriptsize
\begin{center}
\setlength\tabcolsep{4pt}  % default value: 6pt
\begin{tabular}{llllllllllll}
\hline \hline
       & \multicolumn{4}{c}{E--CH$_{3}$OH}  & \multicolumn{4}{|c}{A--CH$_{3}$OH} &\\
Region & $N$(Sp) & $T_{ex}$ & FWHM & V$\rm{_{LSR}}$ & \multicolumn{1}{|l}{$N$(Sp)} & $T_{ex}$ & FWHM & V$\rm{_{LSR}}$ & mean V$\rm{_{LSR}}$& \\
       & \multicolumn{1}{l}{[$\times$10$^{14}$ cm$^{-3}$]} & \multicolumn{1}{l}{[K]} & \multicolumn{1}{l}{[km\,s$^{-1}$]} & \multicolumn{1}{l}{[km\,s$^{-1}$]}  & \multicolumn{1}{|l}{[$\times$10$^{14}$ cm$^{-3}$]} & \multicolumn{1}{l}{[K]} & \multicolumn{1}{l}{[km\,s$^{-1}$]} & \multicolumn{1}{l}{[km\,s$^{-1}$]} & \multicolumn{1}{l}{[km\,s$^{-1}$]}
       \\ \hline \\
R1     & 3.7$_{-0.1}^{+0.1}$  & 14.8$_{-0.6}^{+0.4}$ & 59.7$_{-2.5}^{+1.0}$  &305.4$_{-1.0}^{+0.8}$ & 0.76$_{-0.04}^{+0.05}$ & 18.8$_{-0.7}^{+0.8}$ & 61.3$_{-0.4}^{+1.7}$ & 302.9$_{-0.8}^{+1.4}$ & 304$\pm$9 \\
R2     & 5.0$_{-0.2}^{+0.3}$  & 9.3$_{-0.4}^{+0.4}$  & 43.6$_{-1.5}^{+2.5}$  &328.1$_{-0.7}^{+0.2}$ & 1.0$_{-0.1}^{+0.1}$ & 14.8$_{-0.9}^{+0.5}$ & 37.4$_{-2.9}^{+2.6}$ & 327.2$_{-1.1}^{+1.6}$ & 328$\pm$9 \\
R3     & 19.5$_{-0.5}^{+0.6}$ & 21.4$_{-0.5}^{+0.4}$ & 53.8$_{-0.9}^{+0.6}$  &283.2$_{-0.4}^{+0.4}$ & 9.3$_{-0.2}^{+0.1}$ & 26.6$_{-0.6}^{+0.3}$ & 55.0$_{-0.3}^{+0.9}$ & 284.7$_{-0.5}^{+0.3}$ & 283$\pm$9 \\
R4     & 10.16$_{-0.03}^{+0.04}$ & 31.0$_{-1.2}^{+0.6}$ & 60.2$_{-1.5}^{+1.7}$  &252.0$_{-0.6}^{+0.5}$ & 9.0$_{-0.2}^{+0.2}$ & 34.2$_{-0.9}^{+0.8}$ & 60.1$_{-1.3}^{+0.7}$ & 251.5$_{-0.6}^{+0.2}$ & 252$\pm$9 \\
R5     & 6.2$_{-0.4}^{+0.5}$  & 32.2$_{-1.7}^{+1.8}$ & 136.0$_{-7.9}^{+8.4}$ &217.2$_{-3.3}^{+4.5}$ & 3.2$_{-0.2}^{+0.3}$ & 32.2$_{-1.7}^{+1.8}$ & 136.0$_{-7.9}^{+8.4}$ &217.2$_{-3.3}^{+4.5}$ & 217$\pm$10 \\
R6     & 22.32$_{-0.03}^{+0.07}$ & 24.8$_{-0.4}^{+0.5}$ & 65.3$_{-0.4}^{+0.9}$  &179.9$_{-0.5}^{+0.4}$ & 14.1$_{-0.2}^{+0.2}$& 31.0$_{-0.6}^{+0.4}$ & 66.0$_{-0.9}^{+0.7}$ & 182.1$_{-0.3}^{+0.7}$ & 181$\pm$9 \\
R7     & 17.4$_{-0.4}^{+0.6}$ & 19.5$_{-0.5}^{+0.4}$ & 82.8$_{-1.4}^{+1.3}$  &169.3$_{-0.9}^{+0.6}$ & 5.5$_{-0.1}^{+0.2}$ & 22.9$_{-0.5}^{+0.4}$ & 83.2$_{-2.9}^{+3.0}$ & 173.6$_{-1.0}^{+1.2}$ & 171$\pm$9 \\
R8     & 5.9$_{-0.1}^{+0.2}$  & 14.8$_{-0.4}^{+0.3}$ & 63.7$_{-1.8}^{+0.9}$  &204.2$_{-1.9}^{+1.3}$ & 1.2$_{-0.1}^{+0.1}$ & 19.5$_{-0.7}^{+0.4}$ & 59.1$_{-3.3}^{+3.7}$ & 205.2$_{-1.6}^{+1.3}$ & 205$\pm$9\\
R9     & 5.6$_{-0.1}^{+0.3}$  & 13.9$_{-0.4}^{+0.5}$ & 57.4$_{-1.3}^{+2.2}$  &202.7$_{-0.9}^{+1.1}$ & 1.5$_{-0.1}^{+0.1}$ & 19.2$_{-0.4}^{+0.6}$ & 56.3$_{-1.9}^{+2.6}$ & 201.4$_{-0.7}^{+1.1}$ & 202$\pm$9\\
R10    & 1.0$_{-0.1}^{+0.1}$ & 12.3$_{-1.1}^{+0.7}$  & 24.9$_{-1.3}^{+1.8}$  &143.9$_{-0.8}^{+1.0}$ & 0.27$_{-0.02}^{+0.03}$ & 27.36$_{-3.1}^{+3.1}$ & 22.8$_{-3.3}^{+3.9}$ & 146.4$_{-1.5}^{+1.1}$ & 145$\pm$9\\
\hline \\
\end{tabular}
\tablefoot{From left to right columns are: Selected region of the CMZ of NGC\,253 (see Table\,\ref{sec.selected_regions}), E-type methanol parameters: column density of the species ($N$(Sp)), excitation temperature ($T_{\rm ex}$), FWHM of the modeled methanol lines, and the local standard of rest velocity of the region $V_{\rm LSR}$. Columns 6--9 provide the same gas parameters for A--CH$_{3}$OH. Uncertainties are 1$\sigma$. An extra uncertainty of 15\% should be added to $N$(Sp) (due to the flux calibration accuracy, see Sect.\,\ref{Sec.Observations}). In the last column we present the average velocity of the two methanol symmetric types. In this case our 1$\sigma$ uncertainty is summed in quadrature with our velocity resolution of $\sim$8--9\,km\,s$^{-1}$, which dominates.}
\end{center}
\end{table*}

A comparison between synthetic and observed spectra allows us to perform a deep scan of the methanol lines, highlighting some lines that slightly deviate from LTE conditions in the rotation diagrams or that are blended with other methanol lines and that were previously not discussed. The result of this inspection is henceforth described.

\subsubsection{LTE modelling results}
%\subsubsection{LTE outliers}
\label{sec.LTE_outliers}

\begin{sidewaysfigure*}
%\begin{figure*}[h]
\begin{center}
\includegraphics[width=0.95\linewidth, trim={3.5cm 1cm 1.5cm 1.9cm},clip]{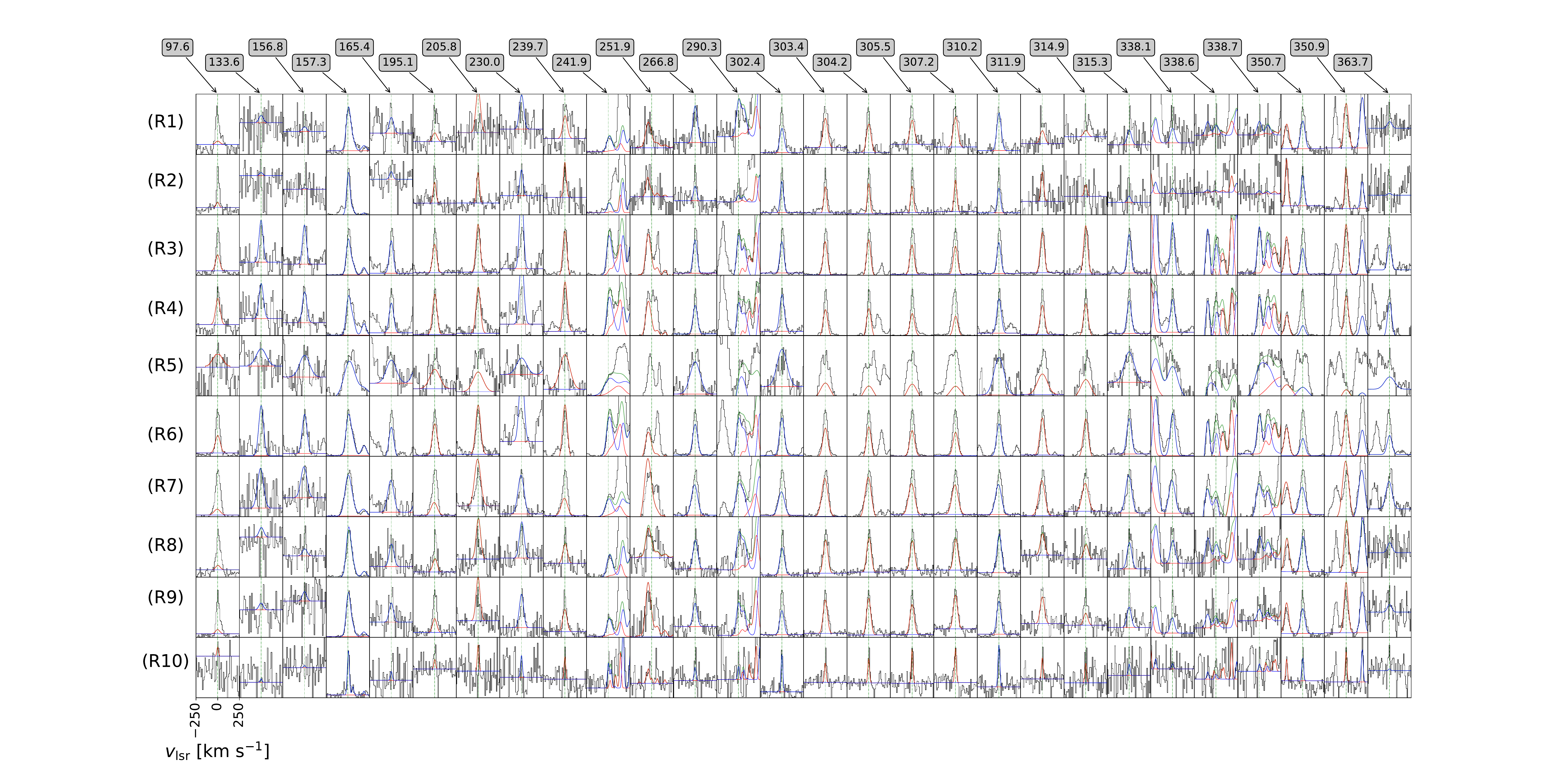}
\caption{Zoom into successfully fitted lines for each region, which can be considered to follow LTE conditions. Red and blue colors indicate methanol A- and E- type, respectively, while the superposition of the two types, the final fit, is indicated in green. Line frequencies (in GHz units) are labeled at the top of the figure and are also indicated as a green dashed vertical line inside panels, while the velocity range (in km\,s$^{-1}$ with respect to the systemic velocity of the individual regions, see Fig.\,\ref{fig.m0_and_spectra_84}) is indicated in the bottom-left corner. Regions 1 to 10 are ordered from top to bottom as indicated in the leftmost panel (R from "Region" plus the corresponding number).}
\label{fig.fitted_lines}
%\end{figure*}
\end{center}
\end{sidewaysfigure*}

As can be seen in our Fig.\,\ref{fig.fitted_lines}, our synthetic spectra fit reasonably well the observed spectra in most of the regions. Region\,5 is the most difficult to reproduce due to the large FWHM ($\sim$140\,km\,s$^{-1}$) of the lines, with its spectrum almost reaching the confusion limit. Fortunately, based on the remaining regions, we selected a large number of lines that are reproduced (see Tables\,\ref{tab.methanol_lines_model_E} and \ref{tab.methanol_lines_model_A}, and Figure\,\ref{fig.fitted_lines}). We attempted to fit the same lines in region\,5, preventing in this way false line identifications and subsequent erroneous fitting attempts. 

The best solution for Region\,5 was reached by fitting both methanol symmetric types simultaneously. When this is done in CASSIS, the whole set of parameters between A- and E-type methanol  is forced to be equal, allowing to change only the ratio between their column densities. As can be seen in Table\,\ref{tab.LTE_model_params}, the column density of E--CH$_{3}$OH is 2.1 times higher than that of A--CH$_{3}$OH. A lower column density E/A ratio leads us always to a worse fitting and is therefore not attempted to reach. We have also discarded from the initial fit the A--CH$_{3}$OH transitions between 303.3 and 309.2\,GHz, as their inclusion always leads to an overestimation of a number of lines (e.g. transitions at 239.7, 241.9, 338.6, and 350.9 GHz, see Tables\,\ref{tab.methanol_lines_model_E} and \ref{tab.methanol_lines_model_A}).

All the proposed maser candidates listed in Table\,\ref{tab.RD_outliers} that were initially unveiled through the rotation diagrams were confirmed by our models. 

In addition to the outliers previously detected (see Table\,\ref{tab.RD_outliers}), we found several transitions with intensities not reproduced by our synthetic spectra. Among those outliers, listed in Table\,\ref{tab.methanol_maser_candidates_LTE_model}, there are a couple of maser candidates. We describe them below.

\subsubsection{Maser line candidates}
\label{sec.maser_line_candidates_model}

Contrary to the remaining outliers found through the radiative transfer modelling (shown in Fig.\,\ref{fig.unknown_maser_lines} and described at some level in Appendix\,\ref{sec.apen.modelling_outliers}), our maser candidates have negative optical depths, depart significantly from LTE (see below), and belong to the same transition series as the maser candidates detected previously through the rotation diagram method (Subsect.\,\ref{subsec.Amasers}), being Class\,I methanol masers.

\textbf{The} $\mathbf{8_{0}\rightarrow 7_{1}-A^{+}}$ transition line at 95.2\,GHz shows intensities $>$12.8 times stronger than predicted by the LTE modelling in all the regions but region\,4, where it is three times stronger. This large departure in the inner regions of the CMZ remind us of the case of the $10_{0}\rightarrow 9_{1}-A^{+}$ methanol line at 198.4\,GHz, which shows emission 3.3 times larger than expected in regions\,3 and 6.
%We therefore include the 95.2\,GHz as maser candidate in Table\,\ref{tab.methanol_maser_candidates_LTE_model}. 

\begin{table}[h]
\caption{Outliers in the LTE model.} \label{tab.methanol_maser_candidates_LTE_model}
\scriptsize
\begin{center}
\renewcommand{\tabcolsep}{0.03cm}
\begin{tabular}{lllllll}
\hline \hline 
Transition                  &$\nu$ [GHz]  & A/E                   & $E_{\rm up}$/k[K]   & A$_{ij}$[s$^{-1}$]  & $\tau$\tablefootmark{d} [$\times$10$^{-5}$] &maser\tablefootmark{e}        \\
\hline \\
8$_{0}\rightarrow$7$_{1}$   & 95.169391   & A$^{+}$                        & 83.54       & 2.13$\times$10$^{-6}$  & $-$6.5 & I\\
3$_{1}\rightarrow$4$_{0}$   & 107.013831\tablefootmark{a}  & A$^{+}$      & 28.35      & 3.06$\times$10$^{-6}$  & 1021 & no\\
0$_{0}\rightarrow$1$_{-1}$ & 108.893945   & E                              & 13.21     & 1.47$\times$10$^{-6}$  & 16590 & no\\
1$_{1}\rightarrow$1$_{0}$   & 165.050175  & E                             & 15.47       & 2.35$\times$10$^{-5}$ & 3423  & no\\
2$_{1}\rightarrow$2$_{0}$   & 165.061130  & E                               & 20.11       & 2.35$\times$10$^{-5}$ & 3801 & no\\
3$_{1}\rightarrow$3$_{0}$   & 165.099240  & E                                & 27.08       & 2.35$\times$10$^{-5}$ & 2579 & no\\
5$_{3}\rightarrow$5$_{2}$   & 251.811956  & A$^{-+}$      & 37.96      & 1.65$\times$10$^{-4}$ & $-$0.1 & no\\
2$_{0}\rightarrow$1$_{-1}$  & 254.015377  & E                              & 12.19      & 1.90$\times$10$^{-5}$ & 9522  & no\\
9$_{-1}\rightarrow$8$_{0}$  & 278.304512\tablefootmark{b}  & E              & 102.07     & 7.67$\times$10$^{-5}$  & $-$2.5 & I\\
7$_{-1}\rightarrow$6$_{-1}$ & 338.344588  & E                                & 62.65       & 1.66$\times$10$^{-4}$ & 1301 & no\\
$J_{K}\rightarrow (J-1)_{K}$\tablefootmark{c} & 97--290 & A,E                          & --          & --                    & $>$0.0 & no \\ \hline
\end{tabular}
\tablefoot{List of lines with intensities that exceed our LTE model. Their line intensities cannot be reproduced without a non-LTE model and there are no other lines subject to produce an important contribution if blended. \tablefoottext{a}{Blended with the methanol 34$_{5}\rightarrow$35$_{3}-A^{-}$ transition at 107.015820\,GHz ($E_{\rm up}$/k$=$1515.904\,K, A$_{ij}=$1.00$\times$10$^{-9}$\,s$^{-1}$)}. \tablefoottext{b}{Blended with the methanol 2$_{-2}\rightarrow$3$_{-1}-E$ transition at 278.342261\,GHz ($E_{\rm up}$/k$=$24.96\,K, A$_{ij}=$1.65$\times$10$^{-5}$\,s$^{-1}$)}. \tablefoottext{c}{The $J_{K}\rightarrow (J-1)_{K}$ line series is described in Appendix\,\ref{sec.apen.detailed_modelling} and \ref{sec.apen.modelling_outliers}}. \tablefoottext{d}{Optical depth according to our non-LTE model in region\,8 (see Appendix\,\ref{sec.apen.detailed_modelling})}. \tablefoottext{e}{Maser candidates based on the discussion in Sect.\,\ref{sec.maser_line_candidates_model}. If it is a candidate, then we add the classification Class\,I or II, according to \citet{Menten1991b}.}}

\end{center}
\end{table}

% trim={<left> <lower> <right> <upper>}

\begin{figure*}[h]
\begin{center}
\includegraphics[width=0.95\linewidth, trim={1.5cm 0.5cm 1.5cm 2cm},clip]{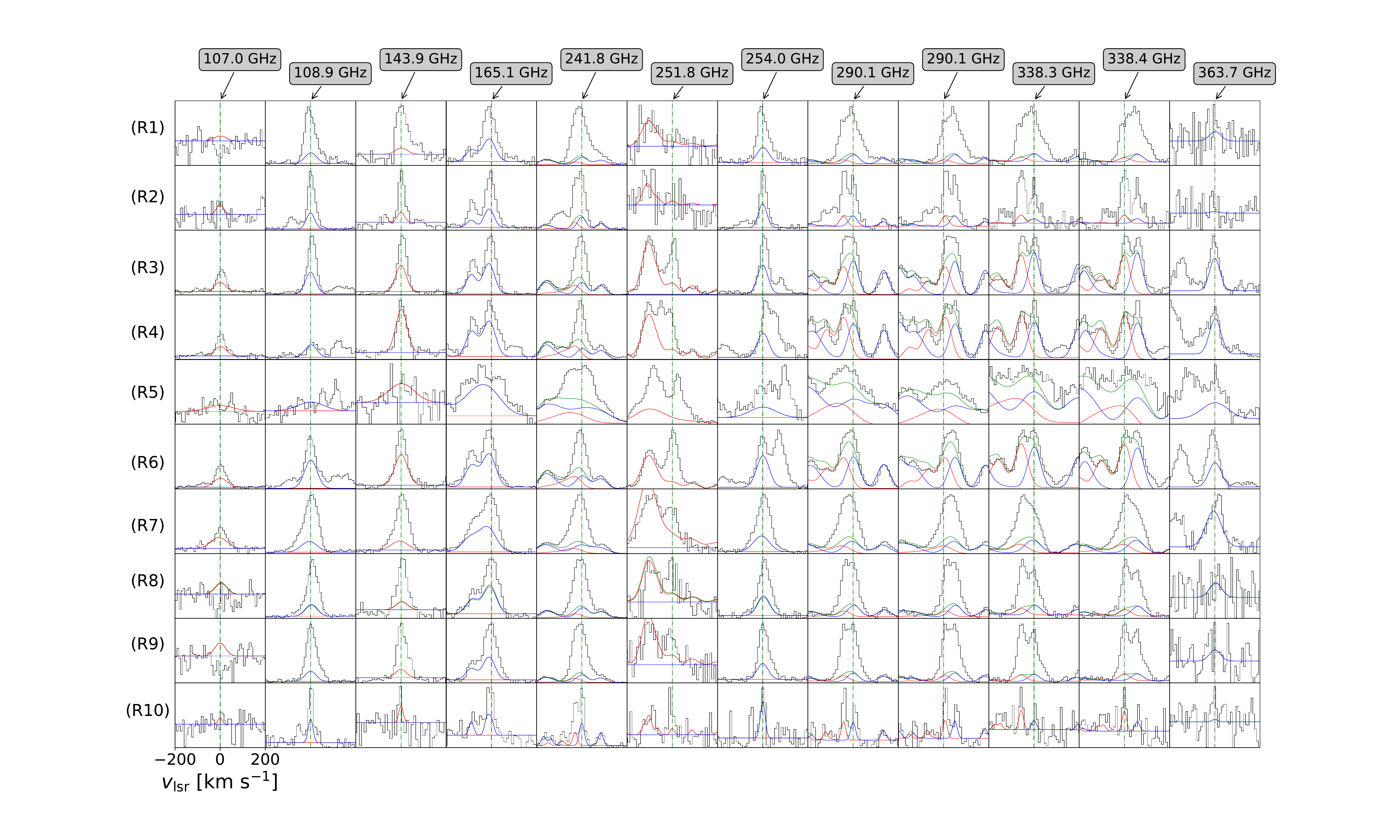}
\caption{Outliers to our LTE modelling (see Appendix\,\ref{sec.apen.modelling_outliers}) with positive optical depths. The only exceptions are the $J_{1}\rightarrow J_{0} A^{-+}$ line series, recently proposed to present methanol maser Class\,II activity by \citet{Zinchenko2017}. Line frequencies are labeled at the top of the Figure. Most of the lines are unclassified so far. Labels and colors are the same as for Figure\,\ref{fig.fitted_lines}.}
\label{fig.unknown_maser_lines}
\end{center}
\end{figure*}

\begin{figure*}[h]
\begin{center}
\includegraphics[width=0.9\linewidth, trim={1.2cm 0.7cm 1.5cm 0cm},clip]{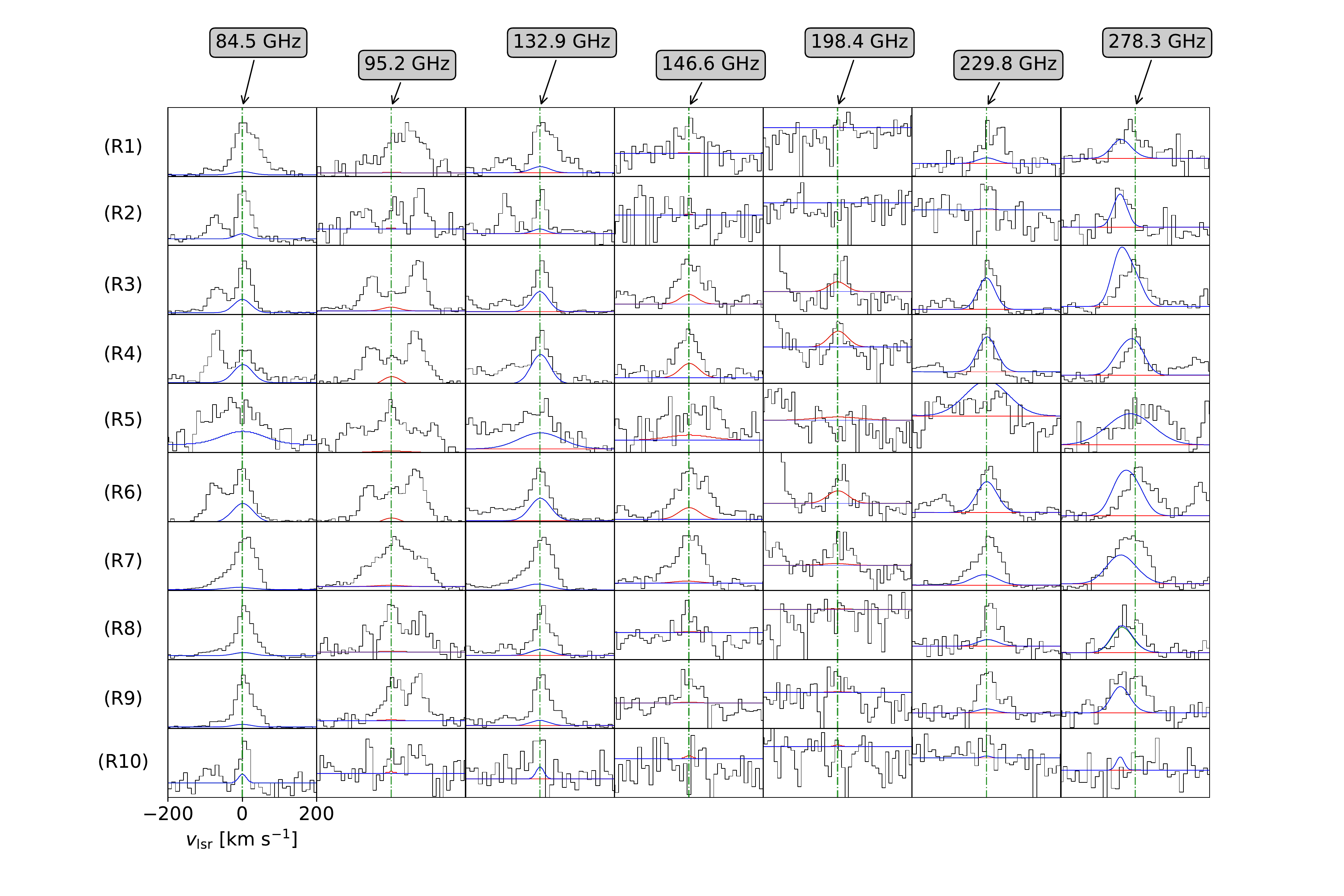}
\caption{Our proposed methanol masers belonging to previously known maser transitions. Labels and colors are the same that for Figure\,\ref{fig.fitted_lines}.}
\label{fig.known_maser_lines}
\end{center}
\end{figure*}

\begin{figure*}[ht!]
\begin{center}
\includegraphics[width=0.9\textwidth]{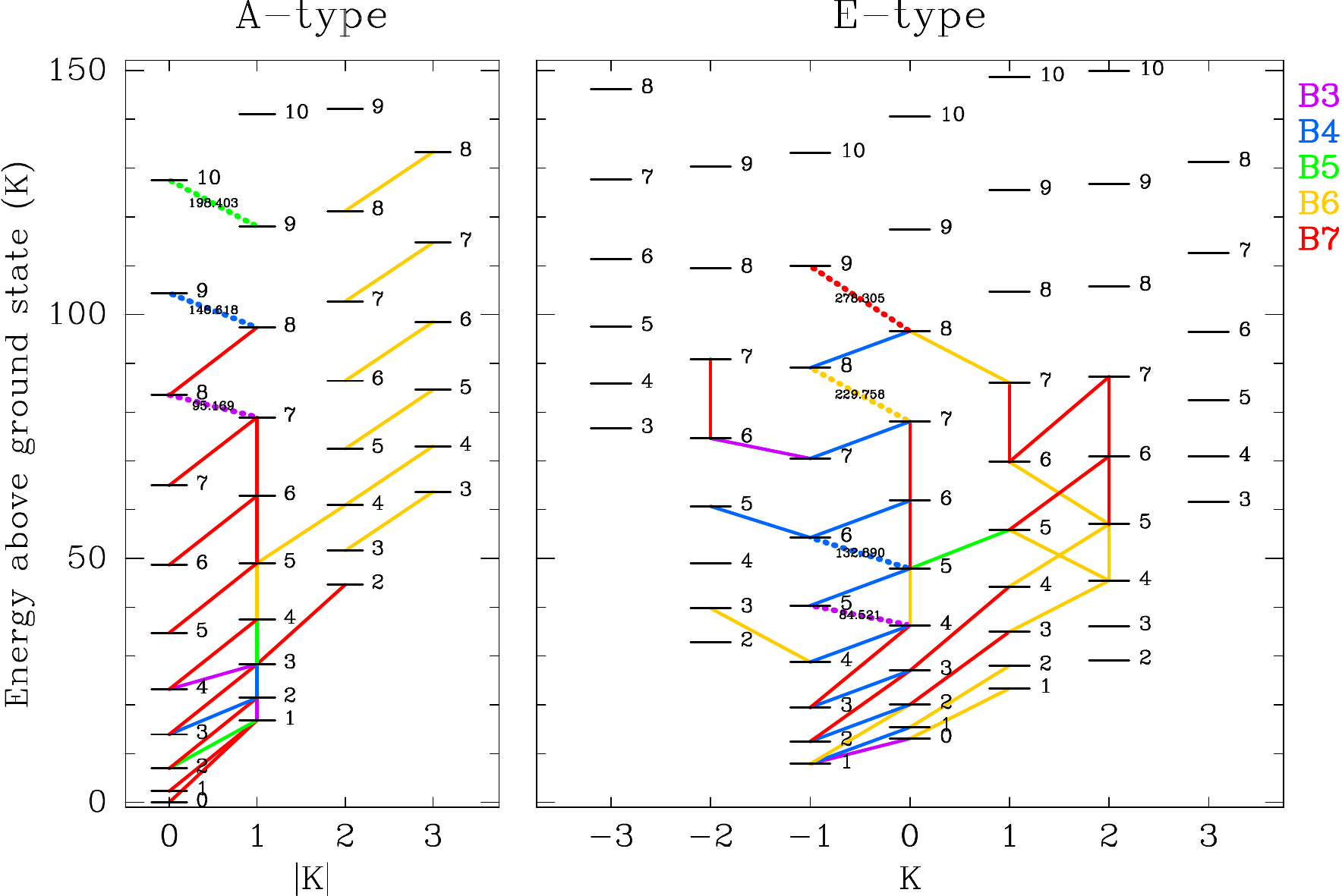}
\caption{Diagram of the energy levels for A-type (left) and E-type (right) methanol transitions covered by ALCHEMI. The x-axis gives the quantum number K, while the number for each level gives J. 
The transitions used in the analysis are indicated with a color code corresponding to ALMA bands, as given in the top right corner. The maser candidate lines are presented as dashed lines, with their frequency in GHz. Note the energy difference of 7.9~K between the ground state of A and E-type.}
\label{fig.EnergyDiagram}
\end{center}
\end{figure*}

\textbf{The} $\mathbf{9_{-1}\rightarrow 8_{0}-E}$ line at 278.305\,GHz is the last methanol transition in the $J_{-1}\rightarrow(J-$ 1)$_{0}-E$ line series covered in this work. As previously mentioned in Sect.\,\ref{subsec.Emasers}, it is strongly contaminated by the 2$_{-2}\rightarrow 3_{-1}-E$ line at 278.342\,GHz and was therefore excluded from the analysis with rotation diagrams. In regions\,1, 7, 8, and 9, however, its line profile is clearly distinguished from the contaminating line when being checked by the synthetic models. According to our radiative transfer modelling, the companion line at 278.342\,GHz is the only one that should be observed (while the masing line intensity should be negligible under LTE, see Fig.\,\ref{fig.known_maser_lines}), with its peak velocity about 40 km\,s$^{-1}$ lower than the maser line in region\,9, a difference four times larger than our spectral resolution. 

%We then include the 9$_{-1}\rightarrow 8_{0}-E$ transition line as a maser candidate in Table\,\ref{tab.methanol_maser_candidates_LTE_model}.

In summary, all the outliers in the LTE models have intensities above the expected one under LTE conditions, but most of them are likely able to be reproduced under non-LTE conditions if the results obtained for region\,8 (Appendix\,\ref{sec.apen.detailed_modelling}) are maintained for the other regions. The only clear maser candidates are the lines at 95.2 and 278.3\,GHz.

%, which belong to line series where maser emission was detected previously through the rotation diagram method (Subsect.\,\ref{subsec.Amasers}).

Given the spatial resolution and software limitations \citep[see][their Sect. 3.6]{vanderTak2007}, we cannot account for a proper model to our maser candidates. Instead, we reproduce their line profiles and check whether negative optical depths are required or not. When no other results are plausible, we can assure that the transitions are effectively experiencing a population inversion. Unfortunately, as Class\,I maser emission arises from spot sizes of the order of 10$^{-5}$--10$^{-3}$\,pc \citep{Voronkov2014,Matsumoto2014}, any attempt to model our observations will be devoid of a real physical meaning. Negative optical depths are determined in the lines belonging to the $J_{-1}\rightarrow(J-$ 1)$_{0}-E$ and $J_{0}\rightarrow(J-$ 1)$_{1}-A^{+}$ series (see Fig.\,\ref{fig.known_maser_lines}) and they constitute our maser candidates. All of them are Class\,I methanol masers and, for the case of those belonging to the $J_{-1}\rightarrow(J-$ 1)$_{0}-E$ line series, depart from LTE at the outskirts of the CMZ of NGC\,253.

We summarize the observed methanol transitions along the entire ALCHEMI coverage in Fig.\,\ref{fig.EnergyDiagram}, where LTE lines are labeled in straight lines and maser candidates are in dashed lines. We indicate with colors the ALMA band for each transition. From Fig.\,\ref{fig.EnergyDiagram} it becomes clear that the 7$_{-1}\rightarrow6_{0}-E$ line at 181.295\,GHz should be pumped by the same conditions than the other maser candidates in the $J_{-1}\rightarrow(J-$ 1)$_{0}-E$ series. Our non-LTE models performed in region\,8 yield a negative optical depth (-2.4) for this transition suggesting maser emission. Unfortunately, our spectral and angular resolution is not sufficient to discriminate between the 181.295\,GHz line and the $J=$2--1 HNC line at 181.324\,GHz.
\section{Discussion}
\label{sec.discussion}

\subsection{Conditions for maser emission}
\label{subsect.conditions_for_maser_emission}
Among all the lines proposed to be masers, the ones detected initially by our rotation diagrams are the most plausible ones. Here we will focus on the Class\,I methanol masers in the $J_{-1}\rightarrow(J-$ 1)$_{0}-E$ series (at 84, 132, 229, and 278\,GHz), as they describe a very clear difference in maser occurrence: a complete LTE behaviour in the central regions (3 to 6) and strong maser activity at the outskirts of the CMZ of NGC\,253, where important $T_{ex}$ differences between A- and E- type methanol take place (see\,Fig.\,\ref{fig.delta}), and also where Lindblad resonances are located (with the only exception of region\,7).

Unfortunately, based on the rotation diagram method we do not have enough evidence to account for the presence of Class\,II masers. The only exception might be the 3$_{1}\rightarrow4_{0}-A^{+}$ transition line at 107\,GHz ($E_{\rm up}$/k$=$28\,K), that follows LTE conditions in regions\,2, 5, and 10 ($\sim$1.5, $\sim$1.5, and 2$\sigma$, respectively). It is slightly above the fit in regions\,3, 4, 6, and 7, and is present in absorption in regions\,1 and 9 (3 and 2$\sigma$, respectively), and maybe also in region\,8 (2$\sigma$); although in this region the line seems to exhibit emission in the middle of the absorption feature.

Maser activity by the line at 107\,GHz was first discovered by \citet{Valtts1995}. They found this line either in absorption or showing quasi-thermal emission. In the first case, this line is spatially correlated with the 5$_{1}\rightarrow 6_{0}-A^{+}$ Class\,II maser line at 6.7\,GHz, that belongs to the same family of lines. Other lines in the $J_{1}\rightarrow (J+1)_{0}-A^{+}$ series are the transitions at 57\,GHz ($J=$4), 156.6\,GHz ($J=$2), and 205.8\,GHz ($J=$1). By means of the rotation diagram method, we note that the transition at 156.6\,GHz is above the LTE trend in the same regions where the line at 107\,GHz surpasses the LTE modelling (regions\,3, 4, 6, and 7). However, as indicated in Appendix\,\ref{sec.apen.rotdiag_outliers}, the line at 156.6\,GHz may also follow the LTE conditions of A--CH$_{3}$OH. 

In NGC\,253, absorption against the continuum in region\,5 is noticeably seen in the rotational ground state transitions of dense gas tracers such as the formyl cation (H$^{13}$CO$^{+}$; \citealt{Harada2021}). Absorption features are also very prominent in the $J$=2--1 (e.g. \citet{Meier2015}) and $J$=3--2 transitions of silicon monoxide (SiO). All those features are uniquely observed in region\,5, contrary to the case of the 3$_{1} \rightarrow 4_{0}-A^{+}$ transition line at 107\,GHz, that presents absorption mostly in regions\,1 and 9. 

Therefore, the edges of the CMZ, namely regions 1, 2, 7, 8, 9, and 10, appear to provide suitable conditions for Class\,I maser emission in the $J_{-1}\rightarrow(J-$ 1)$_{0}-E$ line series. We are aware that these conclusions only refer to the average conditions in our selected regions, as our linear resolution is of the order of 27\,pc, larger than typical clump sizes by a factor of 6--23 \citep[see e.g.][for estimations of clump sizes in the CMZ of NGC\,253]{Leroy2018} but usually small enough to resolve GMCs. 

%About the conditions for maser activity, specifically for galaxies like NGC\,253, the Lindblad resonances (e.g. ellipses in Fig.\,\ref{fig.SiO_HNCO}) seem to be a good candidate for large-scale low-velocity shocks \citep{GarciaBurillo2000, Ellingsen2018}, since such resonances mark the positions where the gas entrained by the spiral arms looses an important amount of angular momentum in order to fall into the galaxy nucleus along the bar \citep[e.g.][]{Combes1993,Knapen2004, Krumholz2015}. In the case of NGC\,253, resonances could even mark the formation of a counter rotating inner bar \citep{Anantharamaiah1996,Cohen2020}. 

\begin{figure*}[!htp]
\includegraphics[width=\linewidth, trim={0 0cm 0 0cm},clip]{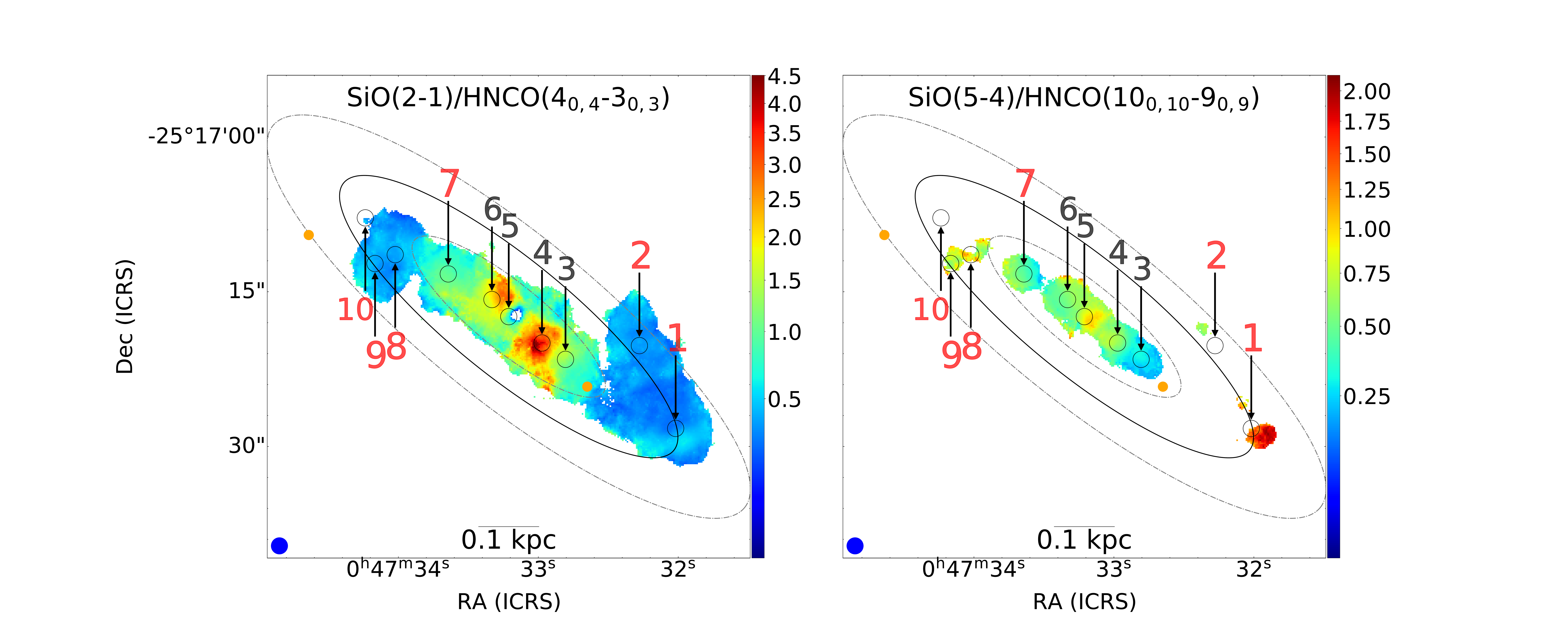}
\caption{Strong over weak shock tracers as accounted by SiO\,$J$=2--1/HNCO\,(4$_{0,4}$--3$_{0,3}$) (left) and SiO\,$J$=5--4/HNCO\,(10$_{0,10}$--9$_{0,9}$) (right) line ratios obtained from ALCHEMI data \citep{Martin2021}. A sigma clip of 3.0 was applied for both SiO and HNCO before obtaining the line ratios. Regions where we detect Class\,I maser emission in the $J_{-1}\rightarrow (J-1)_{0}-E$ series (see Sect.\,\ref{subsec.Emasers}) are labeled in red and with slightly larger numbers. The center and edges of the inner Lindblad resonance from \citet{Iodice2014} are denoted with a black ellipse and dash-dotted grey ellipses, respectively. A square root stretch has been applied for an easy visualization.}
\label{fig.SiO_HNCO}
\end{figure*}

Although Lindblad resonances (e.g. ellipses in Fig.\,\ref{fig.SiO_HNCO}) have been claimed as a good candidate for shocks \citep{GarciaBurillo2000, Ellingsen2018}, for galaxies with a strong bar there is theoretical support against the existence of Lindblad resonances \citep{Regan2003} or their relationship with the circumnuclear ring (CNR) position, which is more accurately defined by x$_{2}$ orbits \citep{Kim2012,Li2015,Schmidt2019}. The interplay between the nuclear dust/gas lanes and the CNR looks as a more likely explanation for the production of bi-symmetric shock/active regions: at the outskirts of the CMZ, where we find higher levels of HNCO as compared to SiO (regions\,1,2, and 7 to 10; see Fig.\,\ref{fig.SiO_HNCO}), in a similar way as in M\,83 \citep{Harada2019}, leading to the appearance of methanol masers in the $J_{-1}\rightarrow(J-$ 1)$_{0}-E$ line series. We should see maser action encircling the CMZ if they were caused by the ILR.

%%%%%

As an attempt to account for gas disturbances in the CMZ of NGC\,253, we obtained line intensity ratios from two well-known shock tracers, SiO and HNCO. Their moment\,0 maps, along with data reduction and a thorough analysis will be described in a forthcoming paper by Huang et al. (in prep.). Here our only interest is to relate the presence of shocks to methanol maser emission in the $J_{-1}\rightarrow(J-$ 1)$_{0}-E$ line series, which is mostly observed toward intersecting areas between the circumnuclear ring and leading-edge dust/gas lanes of the bar.

We selected transitions in the same ALMA bands (3 and 6) in order to reduce instrumental uncertainties. Based on previous studies, here we consider SiO as a tracer of strong shocks and HNCO as a tracer of weak shocks \citep[e.g.][]{Meier2015,Yu2018}, and therefore the SiO/HNCO ratio may be used as an indicator of the shock strength \citep{Kelly2017}.

We note that HNCO\,(4$_{0,4}$--3$_{0,3}$), at 87.9\,GHz, is enhanced in practically the same regions where we detect Class\,I maser activity in the $J_{-1}\rightarrow(J-$ 1)$_{0}-E$ line series (at 84, 132, 229, and 278\,GHz); this is also true for the previously detected methanol transition in this series, at 36\,GHz, if we dismiss the weak maser candidate at the very center shown in \citet{Gorski2018}. 
%Medium-intensity HNCO is observed in region\,6, that present either quasi-thermal emission or a lesser amount of maser spots in the mentioned transitions. 
The resulting SiO\,$J=$2--1/HNCO\,(4$_{0,4}$--3$_{0,3}$) ratios (see Fig.\,\ref{fig.SiO_HNCO}, left panel) are enhanced in the central regions, especially in regions\,4 and 6, where no maser activity, in the $J_{-1}\rightarrow(J-$ 1)$_{0}-E$ line series, is detected.

Considering the SiO and HNCO transitions at frequencies $\sim$220\,GHz (ALMA band\,6), the largest SiO\,$J=$5--4/HNCO\,(10$_{0,10}$--9$_{0,9}$) line ratios come from region\,1 (see Fig.\,\ref{fig.SiO_HNCO}, right panel), the region where we see the largest departure between maser Class\,I emission in the $J_{-1}\rightarrow(J-$ 1)$_{0}-E$ line series (and also in the A-type methanol line at 146.6\,GHz) and LTE conditions (see Fig.\,\ref{fig.delta}). On the other hand, the lowest SiO\,$J=$5--4/HNCO\,(10$_{0,10}$--9$_{0,9}$) ratios are observed in region\,3 and this coincides with the lowest LTE departures of potential Class\,I masers in E--CH$_{3}$OH lines (see Fig.\,\ref{fig.delta}). 

One, however, also needs to take into consideration the varying gas properties likely causing the highly varying trend of these SiO/HNCO ratios. A more complete investigation and further discussion upon the gas properties traced by HNCO and SiO is covered in a forthcoming paper (Huang et al. in prep. ). In addition, there is something missing in describing the intermediate regions. For example, regions\,4 and 6, where we do not see Class\,I maser emission in E--CH$_{3}$OH, present SiO\,$J=$5--4/HNCO\,(10$_{0,10}$--9$_{0,9}$) ratios similar to regions\,1, 2, 7, 8, and 9, where we do observe Class\,I maser emission in E--CH$_{3}$OH.

% trim={<left> <lower> <right> <upper>}

% trim={<left> <lower> <right> <upper>}
\begin{figure*}[!htp]
\includegraphics[width=1\linewidth, trim={0cm 0cm 0cm 0cm},clip]{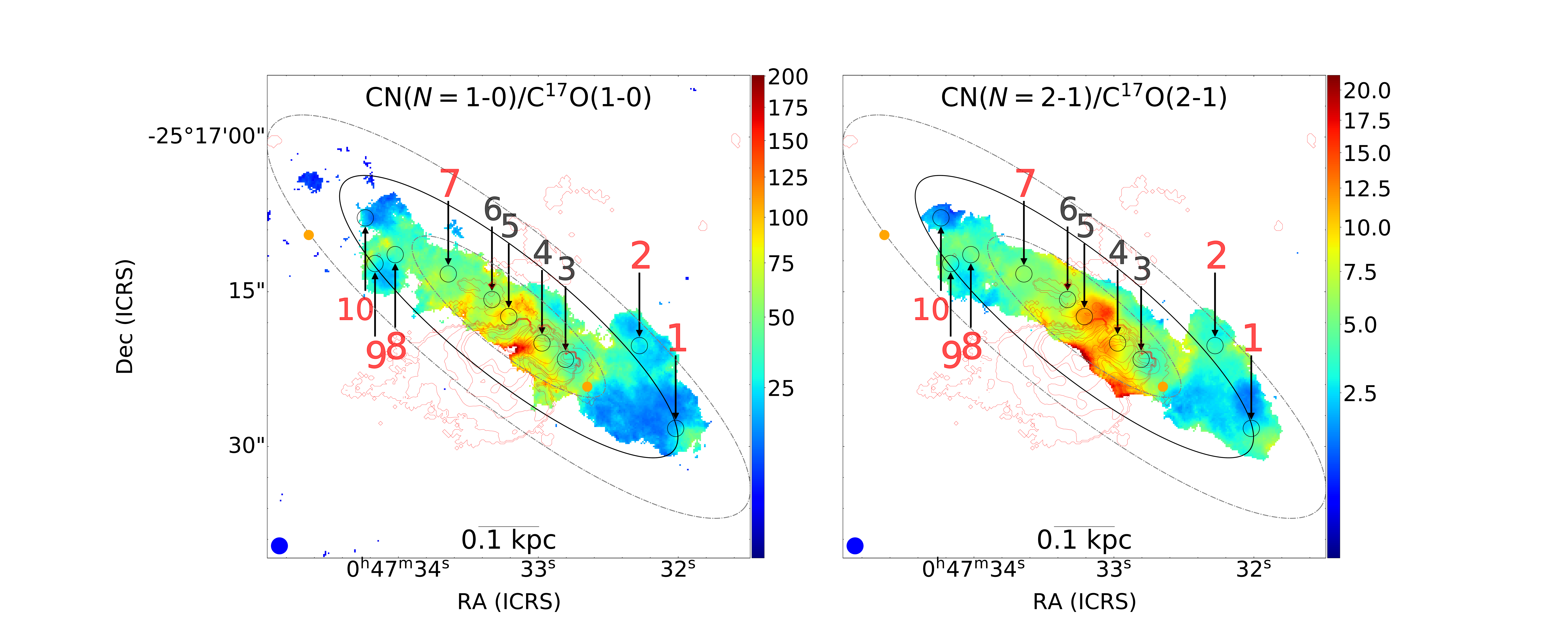}
\caption{CN\,($N=$1--0)/C$^{17}$O\,($J=$1--0) (left) and CN\,($N=$2--1)/C$^{17}$O\,($J=$2--1) (right) line ratios obtained from ALCHEMI data \citep{Martin2021}, indicating the presence of photodissociation regions. A 3$\sigma$ clipping was applied. Red contours indicate the H$_{\alpha}$ emission in logarithmic scale from MUSE archival data (ID:0102.B-0078(A), PI:Laura Zschaechner). Regions where we detect Class\,I maser emission in the $J_{-1}\rightarrow (J-1)_{0}-E$ series (see Sect.\,\ref{subsec.Emasers}) are labeled in red and with slightly larger numbers. The center and edges of the inner Lindblad resonance from \citet{Iodice2014} are denoted with a black ellipse and dash-dotted grey ellipses, respectively. Higher ratios indicate a higher photodissociation level. A square root stretch has been applied for an easy visualization.}
\label{fig.PDR_tracers}
\end{figure*}

The missing element to be considered for the presence or absence of methanol masers can be the occurrence of photodissociation regions (PDRs), since photodissociation is the main mechanism for methanol destruction \citep[see e.g.][]{Hartquist1995}. We can use  PDR tracers such as CN, whose abundance is found to be high in PDR regions \citep[e.g.][]{Fuente1993,Kim2020}, to investigate whether PDRs may be dominant. As mentioned by \citet{Meier2015}, the high rate of star production in the central regions of NGC\,253 can be identified by means of e.g. CN/C$^{17}$O line ratios, where C$^{17}$O traces dense molecular gas \citep[e.g.][]{Thomas2008}.
 
We obtained the CN/C$^{17}$O line ratios by taking the ratio of moment\,0 maps of the chosen molecules. A 3$\sigma$ clipping was applied in the creation of those moment\,0 and the resulting line ratio maps (see Fig.\,\ref{fig.PDR_tracers}).

In Fig.\,\ref{fig.PDR_tracers} we also plot the H$\alpha$ emission from recent MUSE\footnote{Multi Unit Spectroscopic Explorer \citep[MUSE;][]{Bacon2017}} observations obtained from archival data (ID:0102.B-0078(A), PI:Laura Zschaechner) after continuum subtraction by means of the STATCONT software \citep{SanchezMonge2018}. The stronger H$\alpha$ emission is related to the starburst induced outflow emerging from the nuclear bar, its morphology resembles the inner parts of previously observed outflows in X-rays \citep{Strickland2000} as well as the OH plume to the north-west \citep{Turner1985b}. We created six red contours, on a linear scale, covering emission in the range (1--7)$\times$10$^{-14}$~erg\,s$^{-1}$\,cm$^{-2}$. The CN/C$^{17}$O line ratios were obtained from ALMA band\,3 and 6, namely CN\,($N=$1--0)/C$^{17}$O\,($J=$1--0) and CN\,($N=$2--1)/C$^{17}$O\,($J=$2--1) ratios, respectively. These ratios are plotted in Fig.\,\ref{fig.PDR_tracers}, where high values correspond to a high rate of photodissociation.
 
It can be noted from Fig.\,\ref{fig.PDR_tracers} that high values of CN/C$^{17}$O line intensity ratios are closely following the root of the large-scale outflowing gas. Overall, we note the highest level of photodissociation at the central regions of the CMZ (regions 4, 5, and 6), coinciding with regions where we observe a lack of methanol masers in the $J_{-1}\rightarrow (J-1)_{0}-E$ line series, and also the A-type transition at 146.6\,GHz, belonging to the $J_{1}\rightarrow (J+1)_{0}-A^{+}$ line series. The only exception involves the comparison between regions\,3 and 7. Masers are observed in region\,7, even though the CN/C$^{17}$O ratio is  slightly higher than in region\,3, where no maser emission is encountered. 

Although region\,7 is located away from the nuclear ring, it is associated with a hot core cluster indicating an active star formation activity, as deduced from an increment of Complex Organic Molecules (COMs) such as CH$_{3}$COOH \citep{Ando2017}. As tracers of YSO outflows, Class\,I methanol masers are expected to be detected in environments like our region\,7. Additionally, region\,7 is not showing a high level of photodissociation as compared to regions 4 or 5, where the star formation activity is further confirmed by the presence of the H40$\alpha$ line \citep{Bendo2015}.

In conclusion, the PDRs at some level can destroy methanol molecules in the central regions of NGC\,253. This is one of the scenarios suggested by \citet{Ellingsen2017}. Either starburst induced outflows and/or shocks where the leading-edge dust/gas lanes on the bar are connected to the circumnuclear ring, are causing the prevalence of Class\,I methanol masers in the $J_{-1}\rightarrow(J-$ 1)$_{0}-E$ line series (at 84, 132, 229, and 278\,GHz) to regions farther away from the core (regions 1, 2, 7, 8, 9, and 10). Region\,7, although away from the resonances, presents a strong star formation but a rather moderate photodissociation level, giving way to the formation of methanol masers both in the $J_{-1}\rightarrow(J-$ 1)$_{0}-E$ and the $J_{0}\rightarrow (J-1)_{1}-A^{+}$ line series. 

\subsection{Maser emission distribution}
\label{sec.maser_emission_distribution}

\begin{figure*}[!htp]
\includegraphics[width=\linewidth, trim={0 0cm 0 0cm},clip]{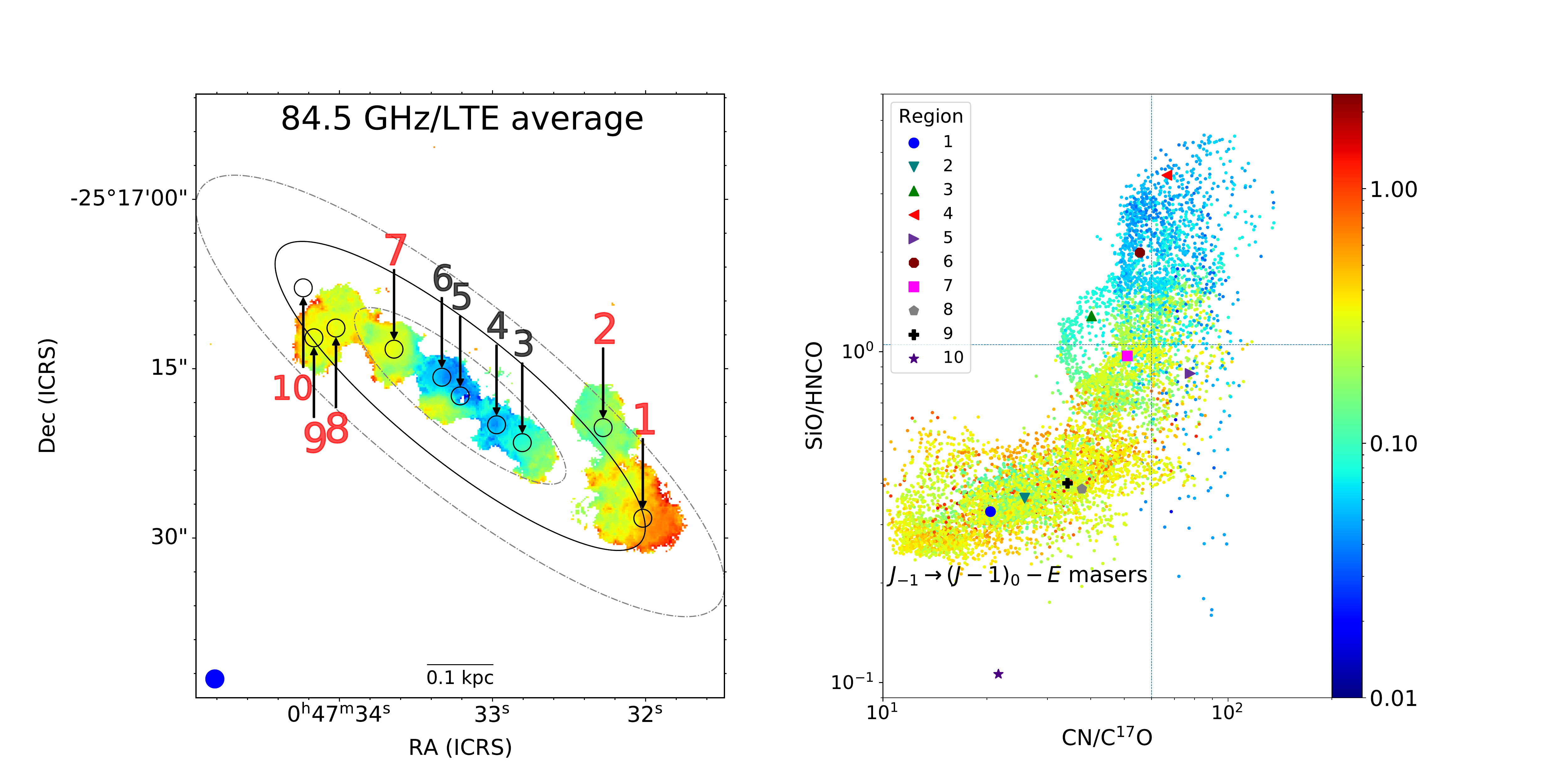}

\caption{Left: Maser emission distribution per spaxel as observed by dividing integrated intensities of the maser methanol line at 84.5\,GHz by the mean intensity value of the surrounding (in terms of $E_{\rm up}$/k) methanol lines in LTE at 218.4 and 310.1\,GHz (see Sect.\,\ref{sec.maser_emission_distribution}). A 3$\sigma$ clipping was applied in all the transitions involved to produce the figure. The center and edges of the inner Lindblad resonance from \citet{Iodice2014} are denoted with a black ellipse and dash-dotted grey ellipses, respectively. Regions where we observe methanol maser emission in the $J_{-1}\rightarrow(J-$ 1)$_{0}-E$ line series are labeled in red. Right: Same spaxels (with their values as colors) as in the figure to the left, but this time distributed according to their ratios in the SiO/HNCO (y-axis) and CN/C$^{17}$O line ratio maps in ALMA band\,3. A logarithmic stretch has been applied in both panels for an easy visualization. Colors are in common for both panels. Our threshold of 3.3 above the rotation diagram fit given in Fig.\,\ref{fig.delta} (first panel) corresponds to an intensity ratio of >0.1244 (see Appendix\,\ref{Sec.apen.equivalency}).}
\label{fig.maser_distribution}
\end{figure*}

Based on our rotation diagrams, we selected a couple of E--CH$_{3}$OH lines that follow LTE conditions in all the studied regions and compare their averaged integrated intensities with the ones observed in the maser line at 84.5\,GHz ($E_{\rm up}$/k$=$32.5\,K) in each spaxel. The selected lines in LTE are the 4$_{2}\rightarrow 3_{1}-E$ and 3$_{1}\rightarrow 2_{0}-E$ transitions at 218.4 ($E_{\rm up}$/k$=$37.6\,K) and 310.1\,GHz ($E_{\rm up}$/k$=$27.1\,K), respectively (see\,Table\,\ref{tab.methanol_lines}). They are not blended with other methanol transitions and do not appear to be significantly contaminated by blending lines from other species (Table\,\ref{tab.methanol_lines}), given their good fit in the rotation diagrams. Additionally, these two lines have $E_{\rm up}$/k values around the one of the maser line at 84.5\,GHz. The result of dividing the integrated intensity of the 84.5\,GHz maser line by the averaged intensity of the two lines proposed to be in LTE (at 218.4 and 310.2\,GHz) is presented in the left panel of Fig.\,\ref{fig.maser_distribution}. In this Figure the spaxels with higher intensity ratios are in line with regions where we previously identified maser emission through the rotation diagram method. The distribution of maser emission we found is also in good agreement with latest detections of methanol maser emission at 36.2\,GHz in NGC\,253 \citep[see][their Fig.\,2]{McCarthy2020}. We have further checked an equivalency between the intensity ratios and upper level column density ratios in our Appendix\,\ref{Sec.apen.equivalency}. We find that a line intensity ratio of >0.1244 corresponds to regions with maser emission in the $J_{1}\rightarrow (J-1)_{0}-E$ line series (upper level column density ratios >3.3 with respect to LTE (Fig.\,\ref{fig.delta}, first panel)). This intensity ratio may appear low, but it is only because the spatially widespread thermal lines have higher intensities than the likely much less widespread non-thermal methanol transition at 84.5\,GHz. Furthermore, it is important to emphasize that the maser's column densities are nominal and do not reflect column densities in a linear way due to amplification effects.

Taking into account the previous subsection\,\ref{subsect.conditions_for_maser_emission}, we noticed that, in order to explain the lack of methanol maser emission in the $J_{-1}\rightarrow(J-$ 1)$_{0}-E$ line series (Class\,I) in the central regions of NGC\,253, it is necessary to invoke a combination of strong photodissociation in region\,5 as well as strong rather than weak shocks in regions\,3, 4, and 6. In other words, suitable conditions of Class\,I maser emission in these line series are weak shocks and low rates of photodissociation, the latter preventing methanol destruction. In the right panel of Fig.\,\ref{fig.maser_distribution}, we empirically determine a threshold for the two conditions previously mentioned as SiO/HNCO ratios lower than 1.05 and CN/C$^{17}$O ratios lower than 60 in ALMA band\,3 data. 

In the right panel of Fig.\,\ref{fig.maser_distribution}, we also show the average values of the 10 selected regions (Table\,\ref{sec.selected_regions}) derived from band\,3 and they are distributed as expected: central regions fall where no maser emission is expected. We performed a similar diagram for ALMA band\,6 data (not shown); in this case region\,7 is difficult to separate from locations where LTE conditions are predominant, and regions\,1, 2, and 8 to 10, are located where SiO/HNCO ratios are in the 0.4 to 1.2 range and CN/C$^{17}$O ratios are lower than 4.75 (both ratios obtained from ALMA band\,6 data). 

A similar distribution for Class\,I methanol maser emission was previously found by \citet{Ellingsen2017} and \citet{Gorski2017,Gorski2019} for the 36.2\,GHz line, which also belongs to the $J_{-1}\rightarrow(J-$ 1)$_{0}-E$ line series.

%\section{Conclusions}
\section{Summary and Conclusions}
\label{sec.summary}
Searching for methanol (CH$_{3}$OH) transitions, we have performed a spectral survey toward the archetypical starburst galaxy NGC\,253 with the ALMA interferometer covering a frequency range between 84--373\,GHz. We focus on ten regions inside the CMZ of this galaxy; those regions are centered at the position of giant molecular clouds. 

After limiting our search to methanol lines with $E_{\rm up}$/k$<$150\,K, since above that limit lines are too weak to be detected, we identified all the methanol transitions of each region separately and used the rotation diagram method in order to determine which transitions deviate from LTE conditions. Assuming LTE conditions, we found that E--CH$_{3}$OH is more abundant at lower $E_{\rm up}$/k and follows predominantly lower excitation temperatures than A--CH$_{3}$OH. We also find that including methanol lines with $E_{\rm up}$/k$<$150\,K is sufficient to fit the observed spectra.

We also performed LTE and non-LTE model calculations with CASSIS-RADEX to find maser candidates. A moderate difference of only about 15\% in intensities between the LTE and non-LTE predictions is found in our data when we consider a single component for each methanol symmetric type (but this difference becomes important when we consider two components, see Appendix\,\ref{sec.apen.detailed_modelling}). 

Although most of the observed methanol lines can be reproduced by our LTE models, we found a number of outliers, of which 7 show maser properties. We have confidently identified a total of 3 A-type and 4 E-type methanol maser transitions, all of them classified as Class\,I.

We have also performed a more detailed non-LTE model in region\,8 (Appendix\,\ref{sec.apen.detailed_modelling}), where the $J_{K}\rightarrow (J-1)_{K}$ line series is better reproduced. From this model we obtained optical depths for the covered methanol transitions and these support the presence of masers indicating negative opacities for our best candidates.

Using rotation diagrams, we have confidently detected all but the last of the covered methanol lines in the $J_{-1}\rightarrow(J-1)_0-$E series to be masing, namely, methanol lines at 84.5, 132.9, and 229.8\,GHz. The last available line in this series, at 278.3\,GHz, is detected to depart from LTE when we compare the observed spectrum with the LTE synthetic spectrum, showing maser characteristics at the edges of the CMZ in NGC\,253. Maser action in the 84.5\,GHz line, previously reported by \citet{McCarthy2018}, is now detected in more than one location for the first time.

An absorption line at 107\,GHz was identified in the spectra of regions\,1, 8 and 9, which are located at the outskirts of the CMZ of NGC\,253. Interestingly, the position of these regions coincides with the position of the circumnuclear ring (co-spatial with the Lindblad resonances according to \citet{Iodice2014}, but see Sect.\,\ref{sec.discussion}) and with Class\,I methanol maser emission in the $J_{-1}\rightarrow(J-$ 1)$_{0}-E$ line series. 

The increment of weak shocks in the interplay between dust/gas lanes and the circumnuclear ring is proposed to harbor favourable conditions to produce Class\,I masers in the $J_{-1}\rightarrow(J-1)_0-$E line series. This happens in regions 1, 2 and 8--10. These resonances are expected to create density waves and locally increase the amount of shocks where they are located, possibly creating favourable conditions to produce methanol masers. On the other hand, considering the Class\,I masers belonging to the $J_{0}\rightarrow(J-$ 1)$_{1}-A$ line series (at 95, 146, and 198\,GHz), we found the first of them ($J=$8, at 95\,GHz) departing at least by a factor of 3 (in region\,4) from LTE in all the regions, and above a factor of 12 without considering region\,4. The levels giving rise to the transition at 146\,GHz are less populated, showing maser emission in regions\,1, 7, 8 and 9. The last in the $J_{0}\rightarrow(J-$ 1)$_{1}-A$ series, at $J=$10, departs from LTE by a factor higher than 3.3 in regions\,3, 6, and 7, being the only one masing in the nuclear parts of the CMZ detected through the rotation diagram method.

We computed the SiO\,$J=$5--4/HNCO\,(10$_{0,10}$--9$_{0,9}$) intensity ratios as a tracer of strong over weak shocks for all regions and found that region 1 and 3 have the highest and lowest values, respectively. When looking at the $J_{-1}\rightarrow(J-1)_{0}-E$ series for those particular regions, we find that in region 1 maser lines deviate the most from the expected LTE behaviour in the rotation diagram. This indicates that the transitions in region\,1 are likely masing due to collisional excitation by shocks. For region 3, the emission seems to be completely in LTE since the transitions align almost perfectly in the rotation diagrams. When considering the SiO\,$J=$2--1/HNCO\,(4$_{0,4}$--3$_{0,3}$) ratios, which cover much more spaxels and trace colder gas, we find an opposite picture, where HNCO is stronger than SiO at the outskirts of the CMZ, exactly where we observe methanol maser emission in the $J_{-1}\rightarrow(J-1)_{0}-E$ line series, with the only exception of region\,7, whose pumping mechanism might be dominated by star-forming processes, in agreement with latest findings in NGC\,253 targeting the methanol 4$_{-1}\rightarrow$3$_{0}-E$ transition at 36\,GHz \citep{Gorski2019}

For the regions in the center of the CMZ of NGC\,253, we compared CN/C$^{17}$O ratios as a PDR tracer. We found out that regions 3, 4, and 5 show the highest levels of photodissociation. The implied strong ultraviolet radiation may lead to the destruction of methanol molecules in the central regions, preventing the appearance of methanol masers.

Although several scenarios are proposed to explain the presence or absence of methanol masers in all the regions in NGC\,253, we cannot conclude which of them is dominant over all the regions along the CMZ. Higher angular resolution observations are needed to spatially resolve smaller regions and to differentiate between the proposed mechanisms that give place to the formation of methanol masers.

\begin{acknowledgements}
PH is a member of and received financial support for this research from the International Max Planck Research School (IMPRS) for Astronomy and Astrophysics at the Universities of Bonn and Cologne. PH is grateful to Arnaud Belloche, Dario Colombo, Yu Gao, and Sudeep Neupane, for their constructive advice and fruitful discussions along a variety of aspects covered in this work. VMR is funded by the Comunidad de Madrid through the Atracci\'on de Talento Investigador (Doctores con experiencia) Grant (COOL: Cosmic Origins Of Life; 2019--T1/TIC--15379), and from the Agencia Estatal de Investigaci\'on (AEI) through the Ram\'on y Cajal programme (grant  RYC2020-029387-I). L.C. has received partial support from the Spanish State Research Agency (AEI; project number PID2019-105552RB-C41). KS thanks MOST grant 109-2112-M-001-020. N.H. acknowledges support from JSPS KAKENHI Grant Number JP21K03634. This work makes use of the following ALMA data: ADS/JAO.ALMA\#2017.1.00161.L and ADS/JAO.ALMA\#2018.1.00162.S. ALMA is a partnership of ESO (representing its member states), NSF (USA) and NINS (Japan), together with NRC (Canada), MOST and ASIAA (Taiwan), and KASI (Republic of Korea), in cooperation with the Republic of Chile. The Joint ALMA Observatory is operated by ESO, AUI/NRAO and NAOJ.

\end{acknowledgements}

%-------------------------------------------------------------------
\bibliographystyle{aa} % style aa.bst
\bibliography{main} % your references Yourfile.bib

\begin{thebibliography}{113}
\expandafter\ifx\csname natexlab\endcsname\relax\def\natexlab#1{#1}\fi

\bibitem[{{Anantharamaiah} \& {Goss}(1996)}]{Anantharamaiah1996}
{Anantharamaiah}, K.~R. \& {Goss}, W.~M. 1996, \apjl, 466, L13

\bibitem[{{Ando} {et~al.}(2017){Ando}, {Nakanishi}, {Kohno}, {Izumi},
  {Mart{\'\i}n}, {Harada}, {Takano}, {Kuno}, {Nakai}, {Sugai}, {Sorai},
  {Tosaki}, {Matsubayashi}, {Nakajima}, {Nishimura}, \& {Tamura}}]{Ando2017}
{Ando}, R., {Nakanishi}, K., {Kohno}, K., {et~al.} 2017, \apj, 849, 81

\bibitem[{{Araya} {et~al.}(2005){Araya}, {Hofner}, {Kurtz}, {Bronfman}, \&
  {DeDeo}}]{Araya2005}
{Araya}, E., {Hofner}, P., {Kurtz}, S., {Bronfman}, L., \& {DeDeo}, S. 2005,
  \apjs, 157, 279

\bibitem[{{Bacon} {et~al.}(2017){Bacon}, {Conseil}, {Mary}, {Brinchmann},
  {Shepherd}, {Akhlaghi}, {Weilbacher}, {Piqueras}, {Wisotzki}, {Lagattuta},
  {Epinat}, {Guerou}, {Inami}, {Cantalupo}, {Courbot}, {Contini}, {Richard},
  {Maseda}, {Bouwens}, {Bouch{\'e}}, {Kollatschny}, {Schaye}, {Marino},
  {Pello}, {Herenz}, {Guiderdoni}, \& {Carollo}}]{Bacon2017}
{Bacon}, R., {Conseil}, S., {Mary}, D., {et~al.} 2017, \aap, 608, A1

\bibitem[{{Batrla} {et~al.}(1987){Batrla}, {Matthews}, {Menten}, \&
  {Walmsley}}]{Batrla1987}
{Batrla}, W., {Matthews}, H.~E., {Menten}, K.~M., \& {Walmsley}, C.~M. 1987,
  \nat, 326, 49

\bibitem[{{Belloche} {et~al.}(2019){Belloche}, {Garrod}, {M{\"u}ller},
  {Menten}, {Medvedev}, {Thomas}, \& {Kisiel}}]{Belloche2019}
{Belloche}, A., {Garrod}, R.~T., {M{\"u}ller}, H.~S.~P., {et~al.} 2019, \aap,
  628, A10

\bibitem[{{Belloche} {et~al.}(2013){Belloche}, {M{\"u}ller}, {Menten},
  {Schilke}, \& {Comito}}]{Belloche2013}
{Belloche}, A., {M{\"u}ller}, H.~S.~P., {Menten}, K.~M., {Schilke}, P., \&
  {Comito}, C. 2013, \aap, 559, A47

\bibitem[{{Bendo} {et~al.}(2015){Bendo}, {Beswick}, {D'Cruze}, {Dickinson},
  {Fuller}, \& {Muxlow}}]{Bendo2015}
{Bendo}, G.~J., {Beswick}, R.~J., {D'Cruze}, M.~J., {et~al.} 2015, \mnras, 450,
  L80

\bibitem[{{Billington} {et~al.}(2020){Billington}, {Urquhart}, {K{\"o}nig},
  {Beuther}, {Breen}, {Menten}, {Campbell-White}, {Ellingsen}, {Thompson},
  {Moore}, {Eden}, {Kim}, \& {Leurini}}]{Billington2020}
{Billington}, S.~J., {Urquhart}, J.~S., {K{\"o}nig}, C., {et~al.} 2020, \mnras,
  499, 2744

\bibitem[{{Breen} {et~al.}(2011){Breen}, {Ellingsen}, {Caswell}, {Green},
  {Fuller}, {Voronkov}, {Quinn}, \& {Avison}}]{Breen2011}
{Breen}, S.~L., {Ellingsen}, S.~P., {Caswell}, J.~L., {et~al.} 2011, \apj, 733,
  80

\bibitem[{{Breen} {et~al.}(2013){Breen}, {Ellingsen}, {Contreras}, {Green},
  {Caswell}, {Stevens}, {Dawson}, \& {Voronkov}}]{Breen2013}
{Breen}, S.~L., {Ellingsen}, S.~P., {Contreras}, Y., {et~al.} 2013, \mnras,
  435, 524

\bibitem[{{Breen} {et~al.}(2015){Breen}, {Fuller}, {Caswell}, {Green},
  {Avison}, {Ellingsen}, {Gray}, {Pestalozzi}, {Quinn}, {Richards}, {Thompson},
  \& {Voronkov}}]{Breen2015}
{Breen}, S.~L., {Fuller}, G.~A., {Caswell}, J.~L., {et~al.} 2015, \mnras, 450,
  4109

\bibitem[{{Chen} {et~al.}(2015){Chen}, {Ellingsen}, {Baan}, {Qiao}, {Li}, {An},
  \& {Breen}}]{Chen2015}
{Chen}, X., {Ellingsen}, S.~P., {Baan}, W.~A., {et~al.} 2015, \apjl, 800, L2

\bibitem[{{Cohen} {et~al.}(2020){Cohen}, {Turner}, \& {Consiglio}}]{Cohen2020}
{Cohen}, D.~P., {Turner}, J.~L., \& {Consiglio}, S.~M. 2020, \mnras, 493, 627

\bibitem[{{Cotton} \& {Yusef-Zadeh}(2016)}]{Cotton2016}
{Cotton}, W.~D. \& {Yusef-Zadeh}, F. 2016, \apjs, 227, 10

\bibitem[{{Cragg} {et~al.}(1992){Cragg}, {Johns}, {Godfrey}, \&
  {Brown}}]{Cragg1992}
{Cragg}, D.~M., {Johns}, K.~P., {Godfrey}, P.~D., \& {Brown}, R.~D. 1992,
  \mnras, 259, 203

\bibitem[{{Cragg} {et~al.}(1993){Cragg}, {Mikhtiev}, {Bettens}, {Godfrey}, \&
  {Brown}}]{Cragg1993}
{Cragg}, D.~M., {Mikhtiev}, M.~A., {Bettens}, R.~P.~A., {Godfrey}, P.~D., \&
  {Brown}, R.~D. 1993, \mnras, 264, 769

\bibitem[{{Cyganowski} {et~al.}(2018){Cyganowski}, {Hannaway}, {Brogan},
  {Hunter}, \& {Zhang}}]{Cyganowski2018}
{Cyganowski}, C.~J., {Hannaway}, D., {Brogan}, C.~L., {Hunter}, T.~R., \&
  {Zhang}, Q. 2018, in Astrophysical Masers: Unlocking the Mysteries of the
  Universe, ed. A.~{Tarchi}, M.~J. {Reid}, \& P.~{Castangia}, Vol. 336,
  281--282

\bibitem[{{Darling} {et~al.}(2003){Darling}, {Goldsmith}, {Li}, \&
  {Giovanelli}}]{Darling2003}
{Darling}, J., {Goldsmith}, P., {Li}, D., \& {Giovanelli}, R. 2003, \aj, 125,
  1177

\bibitem[{{de Vaucouleurs} {et~al.}(1991){de Vaucouleurs}, {de Vaucouleurs},
  {Corwin}, {Buta}, {Paturel}, \& {Fouque}}]{deVaucouleurs1991}
{de Vaucouleurs}, G., {de Vaucouleurs}, A., {Corwin}, Herold~G., J., {et~al.}
  1991, {Third Reference Catalogue of Bright Galaxies}

\bibitem[{{Ellingsen}(2018)}]{Ellingsen2018}
{Ellingsen}, S.~P. 2018, in Astrophysical Masers: Unlocking the Mysteries of
  the Universe, ed. A.~{Tarchi}, M.~J. {Reid}, \& P.~{Castangia}, Vol. 336,
  117--120

\bibitem[{{Ellingsen} {et~al.}(2010){Ellingsen}, {Breen}, {Caswell}, {Quinn},
  \& {Fuller}}]{Ellingsen2010}
{Ellingsen}, S.~P., {Breen}, S.~L., {Caswell}, J.~L., {Quinn}, L.~J., \&
  {Fuller}, G.~A. 2010, \mnras, 404, 779

\bibitem[{{Ellingsen} {et~al.}(2017){Ellingsen}, {Chen}, {Breen}, \&
  {Qiao}}]{Ellingsen2017}
{Ellingsen}, S.~P., {Chen}, X., {Breen}, S.~L., \& {Qiao}, H.~H. 2017, \mnras,
  472, 604

\bibitem[{{Ellingsen} {et~al.}(2014){Ellingsen}, {Chen}, {Qiao}, {Baan}, {An},
  {Li}, \& {Breen}}]{Ellingsen2014}
{Ellingsen}, S.~P., {Chen}, X., {Qiao}, H.-H., {et~al.} 2014, \apjl, 790, L28

\bibitem[{{Ellingsen} {et~al.}(1994){Ellingsen}, {Norris}, {Whiteoak}, {Vaile},
  {McCulloch}, \& {Price}}]{Ellingsen1994}
{Ellingsen}, S.~P., {Norris}, R.~P., {Whiteoak}, J.~B., {et~al.} 1994, \mnras,
  267, 510

\bibitem[{{Fuente} {et~al.}(1993){Fuente}, {Martin-Pintado}, {Cernicharo}, \&
  {Bachiller}}]{Fuente1993}
{Fuente}, A., {Martin-Pintado}, J., {Cernicharo}, J., \& {Bachiller}, R. 1993,
  \aap, 276, 473

\bibitem[{{Garc{\'\i}a-Burillo} {et~al.}(2000){Garc{\'\i}a-Burillo},
  {Mart{\'\i}n-Pintado}, {Fuente}, \& {Neri}}]{GarciaBurillo2000}
{Garc{\'\i}a-Burillo}, S., {Mart{\'\i}n-Pintado}, J., {Fuente}, A., \& {Neri},
  R. 2000, \aap, 355, 499

\bibitem[{{Goldsmith} \& {Langer}(1999)}]{Goldsmith1999}
{Goldsmith}, P.~F. \& {Langer}, W.~D. 1999, \apj, 517, 209

\bibitem[{{Gorski} {et~al.}(2017){Gorski}, {Ott}, {Rand}, {Meier}, {Momjian},
  \& {Schinnerer}}]{Gorski2017}
{Gorski}, M., {Ott}, J., {Rand}, R., {et~al.} 2017, \apj, 842, 124

\bibitem[{{Gorski} {et~al.}(2018){Gorski}, {Ott}, {Rand}, {Meier}, {Momjian},
  \& {Schinnerer}}]{Gorski2018}
{Gorski}, M., {Ott}, J., {Rand}, R., {et~al.} 2018, \apj, 856, 134

\bibitem[{{Gorski} {et~al.}(2019){Gorski}, {Ott}, {Rand}, {Meier}, {Momjian},
  {Schinnerer}, \& {Ellingsen}}]{Gorski2019}
{Gorski}, M.~D., {Ott}, J., {Rand}, R., {et~al.} 2019, \mnras, 483, 5434

\bibitem[{{Green} {et~al.}(2017){Green}, {Breen}, {Fuller},
  {McClure-Griffiths}, {Ellingsen}, {Voronkov}, {Avison}, {Brooks}, {Burton},
  {Chrysostomou}, {Cox}, {Diamond}, {Gray}, {Hoare}, {Masheder}, {Pestalozzi},
  {Phillips}, {Quinn}, {Richards}, {Thompson}, {Walsh}, {Ward-Thompson},
  {Wong-McSweeney}, \& {Yates}}]{Green2017}
{Green}, J.~A., {Breen}, S.~L., {Fuller}, G.~A., {et~al.} 2017, \mnras, 469,
  1383

\bibitem[{{Green} {et~al.}(2008){Green}, {Caswell}, {Fuller}, {Breen},
  {Brooks}, {Burton}, {Chrysostomou}, {Cox}, {Diamond}, {Ellingsen}, {Gray},
  {Hoare}, {Masheder}, {McClure-Griffiths}, {Pestalozzi}, {Phillips}, {Quinn},
  {Thompson}, {Voronkov}, {Walsh}, {Ward-Thompson}, {Wong-McSweeney}, {Yates},
  \& {Cohen}}]{Green2008}
{Green}, J.~A., {Caswell}, J.~L., {Fuller}, G.~A., {et~al.} 2008, \mnras, 385,
  948

\bibitem[{{Harada} {et~al.}(2021){Harada}, {Mart{\'\i}n}, {Mangum}, {Sakamoto},
  {Muller}, {Tanaka}, {Nakanishi}, {Herrero-Illana}, {Yoshimura}, {M{\"u}hle},
  {Aladro}, {Colzi}, {Rivilla}, {Aalto}, {Behrens}, {Henkel}, {Holdship},
  {Humire}, {Meier}, {Nishimura}, {van der Werf}, \& {Viti}}]{Harada2021}
{Harada}, N., {Mart{\'\i}n}, S., {Mangum}, J.~G., {et~al.} 2021, \apj, 923, 24

\bibitem[{{Harada} {et~al.}(2019){Harada}, {Sakamoto}, {Mart{\'\i}n},
  {Watanabe}, {Aladro}, {Riquelme}, \& {Hirota}}]{Harada2019}
{Harada}, N., {Sakamoto}, K., {Mart{\'\i}n}, S., {et~al.} 2019, \apj, 884, 100

\bibitem[{{Hartquist} {et~al.}(1995){Hartquist}, {Menten}, {Lepp}, \&
  {Dalgarno}}]{Hartquist1995}
{Hartquist}, T.~W., {Menten}, K.~M., {Lepp}, S., \& {Dalgarno}, A. 1995,
  \mnras, 272, 184

\bibitem[{{Haschick} \& {Baan}(1993)}]{HaschickBaan1993}
{Haschick}, A.~D. \& {Baan}, W.~A. 1993, \apj, 410, 663

\bibitem[{{Hastings}(1970)}]{Hastings1970}
{Hastings}, W.~K. 1970, Biometrika, 57, 97

\bibitem[{{Holdship} {et~al.}(2021){Holdship}, {Viti}, {Mart{\'\i}n}, {Harada},
  {Mangum}, {Sakamoto}, {Muller}, {Tanaka}, {Yoshimura}, {Nakanishi},
  {Herrero-Illana}, {M{\"u}hle}, {Aladro}, {Colzi}, {Emig},
  {Garc{\'\i}a-Burillo}, {Henkel}, {Humire}, {Meier}, {Rivilla}, \& {van der
  Werf}}]{Holdship2021}
{Holdship}, J., {Viti}, S., {Mart{\'\i}n}, S., {et~al.} 2021, \aap, 654, A55

\bibitem[{{Humire} {et~al.}(2020){Humire}, {Henkel}, {Gong}, {Leurini},
  {Mauersberger}, {Levshakov}, {Winkel}, {Tarchi}, {Castangia}, {Malawi},
  {Asiri}, {Ellingsen}, {McCarthy}, {Chen}, \& {Tang}}]{Humire2020}
{Humire}, P.~K., {Henkel}, C., {Gong}, Y., {et~al.} 2020, \aap, 633, A106

\bibitem[{{Impellizzeri} {et~al.}(2008){Impellizzeri}, {Henkel}, {Roy}, \&
  {Menten}}]{Impellizzeri2008}
{Impellizzeri}, C.~M.~V., {Henkel}, C., {Roy}, A.~L., \& {Menten}, K.~M. 2008,
  \aap, 484, L43

\bibitem[{{Iodice} {et~al.}(2014){Iodice}, {Arnaboldi}, {Rejkuba}, {Neeser},
  {Greggio}, {Gonzalez}, {Irwin}, \& {Emerson}}]{Iodice2014}
{Iodice}, E., {Arnaboldi}, M., {Rejkuba}, M., {et~al.} 2014, \aap, 567, A86

\bibitem[{{Kalenskii} {et~al.}(2010){Kalenskii}, {Johansson}, {Bergman},
  {Kurtz}, {Hofner}, {Walmsley}, \& {Slysh}}]{Kalenskii2010}
{Kalenskii}, S.~V., {Johansson}, L.~E.~B., {Bergman}, P., {et~al.} 2010,
  \mnras, 405, 613

\bibitem[{{Kalenskii} {et~al.}(2006){Kalenskii}, {Promyslov}, {Slysh},
  {Bergman}, \& {Winnberg}}]{Kalenskii2006}
{Kalenskii}, S.~V., {Promyslov}, V.~G., {Slysh}, V.~I., {Bergman}, P., \&
  {Winnberg}, A. 2006, Astronomy Reports, 50, 289

\bibitem[{{Kalenskii} {et~al.}(2002){Kalenskii}, {Slysh}, \&
  {Val'tts}}]{Kalenskii2002}
{Kalenskii}, S.~V., {Slysh}, V.~I., \& {Val'tts}, I.~E. 2002, in Cosmic Masers:
  From Proto-Stars to Black Holes, ed. V.~{Migenes} \& M.~J. {Reid}, Vol. 206,
  191

\bibitem[{{Kang} {et~al.}(2015){Kang}, {Kim}, {Byun}, {Lee}, \&
  {Park}}]{Kang2015}
{Kang}, H., {Kim}, K.-T., {Byun}, D.-Y., {Lee}, S., \& {Park}, Y.-S. 2015,
  \apjs, 221, 6

\bibitem[{{Kelly} {et~al.}(2017){Kelly}, {Viti}, {Garc{\'\i}a-Burillo},
  {Fuente}, {Usero}, {Krips}, \& {Neri}}]{Kelly2017}
{Kelly}, G., {Viti}, S., {Garc{\'\i}a-Burillo}, S., {et~al.} 2017, \aap, 597,
  A11

\bibitem[{{Kim} {et~al.}(2020){Kim}, {Wyrowski}, {Urquhart},
  {P{\'e}rez-Beaupuits}, {Pillai}, {Tiwari}, \& {Menten}}]{Kim2020}
{Kim}, W.~J., {Wyrowski}, F., {Urquhart}, J.~S., {et~al.} 2020, \aap, 644, A160

\bibitem[{{Kim} {et~al.}(2012){Kim}, {Seo}, \& {Kim}}]{Kim2012}
{Kim}, W.-T., {Seo}, W.-Y., \& {Kim}, Y. 2012, \apj, 758, 14

\bibitem[{{Ladeyschikov} {et~al.}(2019){Ladeyschikov}, {Bayandina}, \&
  {Sobolev}}]{Ladeyschikov2019}
{Ladeyschikov}, D.~A., {Bayandina}, O.~S., \& {Sobolev}, A.~M. 2019, \aj, 158,
  233

\bibitem[{{Lees}(1973)}]{Lees1973}
{Lees}, R.~M. 1973, \apj, 184, 763

\bibitem[{{Lees} \& {Baker}(1968)}]{Lees1968}
{Lees}, R.~M. \& {Baker}, J.~G. 1968, \jcp, 48, 5299

\bibitem[{{Leroy} {et~al.}(2015){Leroy}, {Bolatto}, {Ostriker}, {Rosolowsky},
  {Walter}, {Warren}, {Donovan Meyer}, {Hodge}, {Meier}, {Ott}, {Sandstrom},
  {Schruba}, {Veilleux}, \& {Zwaan}}]{Leroy2015}
{Leroy}, A.~K., {Bolatto}, A.~D., {Ostriker}, E.~C., {et~al.} 2015, \apj, 801,
  25

\bibitem[{{Leroy} {et~al.}(2018){Leroy}, {Bolatto}, {Ostriker}, {Walter},
  {Gorski}, {Ginsburg}, {Krieger}, {Levy}, {Meier}, {Mills}, {Ott},
  {Rosolowsky}, {Thompson}, {Veilleux}, \& {Zschaechner}}]{Leroy2018}
{Leroy}, A.~K., {Bolatto}, A.~D., {Ostriker}, E.~C., {et~al.} 2018, \apj, 869,
  126

\bibitem[{{Leurini} \& {Menten}(2018)}]{Leurini2018}
{Leurini}, S. \& {Menten}, K.~M. 2018, in Astrophysical Masers: Unlocking the
  Mysteries of the Universe, ed. A.~{Tarchi}, M.~J. {Reid}, \& P.~{Castangia},
  Vol. 336, 17--22

\bibitem[{{Leurini} {et~al.}(2016){Leurini}, {Menten}, \&
  {Walmsley}}]{Leurini2016}
{Leurini}, S., {Menten}, K.~M., \& {Walmsley}, C.~M. 2016, \aap, 592, 31

\bibitem[{{Li} {et~al.}(2015){Li}, {Shen}, \& {Kim}}]{Li2015}
{Li}, Z., {Shen}, J., \& {Kim}, W.-T. 2015, \apj, 806, 150

\bibitem[{{Liechti} \& {Wilson}(1996)}]{LiechtiWilson1996}
{Liechti}, S. \& {Wilson}, T.~L. 1996, \aap, 314, 615

\bibitem[{{Lo}(2005)}]{Lo2005}
{Lo}, K.~Y. 2005, \araa, 43, 625

\bibitem[{{Ma} {et~al.}(1998){Ma}, {Arias}, {Eubanks}, {Fey}, {Gontier},
  {Jacobs}, {Sovers}, {Archinal}, \& {Charlot}}]{Ma1998}
{Ma}, C., {Arias}, E.~F., {Eubanks}, T.~M., {et~al.} 1998, \aj, 116, 516

\bibitem[{{Mangum} \& {Shirley}(2015)}]{Mangum2015}
{Mangum}, J.~G. \& {Shirley}, Y.~L. 2015, \pasp, 127, 266

\bibitem[{{Mart{\'\i}n} {et~al.}(2021){Mart{\'\i}n}, {Mangum}, {Harada},
  {Costagliola}, {Sakamoto}, {Muller}, {Aladro}, {Tanaka}, {Yoshimura},
  {Nakanishi}, {Herrero-Illana}, {M{\"u}hle}, {Aalto}, {Behrens}, {Colzi},
  {Emig}, {Fuller}, {Garc{\'\i}a-Burillo}, {Greve}, {Henkel}, {Holdship},
  {Humire}, {Hunt}, {Izumi}, {Kohno}, {K{\"o}nig}, {Meier}, {Nakajima},
  {Nishimura}, {Padovani}, {Rivilla}, {Takano}, {van der Werf}, {Viti}, \&
  {Yan}}]{Martin2021}
{Mart{\'\i}n}, S., {Mangum}, J.~G., {Harada}, N., {et~al.} 2021, \aap, 656, A46

\bibitem[{{Mart{\'\i}n} {et~al.}(2019){Mart{\'\i}n}, {Mart{\'\i}n-Pintado},
  {Blanco-S{\'a}nchez}, {Rivilla}, {Rodr{\'\i}guez-Franco}, \&
  {Rico-Villas}}]{Martin2019b}
{Mart{\'\i}n}, S., {Mart{\'\i}n-Pintado}, J., {Blanco-S{\'a}nchez}, C.,
  {et~al.} 2019, \aap, 631, A159

\bibitem[{{Mart{\'\i}n} {et~al.}(2006){Mart{\'\i}n}, {Mauersberger},
  {Mart{\'\i}n-Pintado}, {Henkel}, \& {Garc{\'\i}a-Burillo}}]{Martin2006}
{Mart{\'\i}n}, S., {Mauersberger}, R., {Mart{\'\i}n-Pintado}, J., {Henkel}, C.,
  \& {Garc{\'\i}a-Burillo}, S. 2006, \apjs, 164, 450

\bibitem[{{Matsumoto} {et~al.}(2014){Matsumoto}, {Hirota}, {Sugiyama}, {Kim},
  {Kim}, {Byun}, {Jung}, {Chibueze}, {Honma}, {Kameya}, {Kim}, {Lyo}, {Motogi},
  {Oh}, {Shino}, {Sunada}, {Bae}, {Chung}, {Chung}, {Cho}, {Han}, {Han},
  {Hwang}, {Je}, {Jike}, {Jung}, {Jung}, {Kang}, {Kang}, {Kang}, {Kan-ya},
  {Kawaguchi}, {Kim}, {Kim}, {Ryoung Kim}, {Kim}, {Kobayashi}, {Kono},
  {Kurayama}, {Lee}, {Lee}, {Lee}, {Lee}, {Lee}, {Lee}, {Minh}, {Miyazaki},
  {Oh}, {Oyama}, {Park}, {Roh}, {Sasao}, {Sawada-Satoh}, {Shibata}, {Sohn},
  {Song}, {Tamura}, {Wajima}, {Wi}, {Yeom}, \& {Yun}}]{Matsumoto2014}
{Matsumoto}, N., {Hirota}, T., {Sugiyama}, K., {et~al.} 2014, \apjl, 789, L1

\bibitem[{{McCarthy} {et~al.}(2018){McCarthy}, {Ellingsen}, {Breen},
  {Voronkov}, \& {Chen}}]{McCarthy2018}
{McCarthy}, T.~P., {Ellingsen}, S.~P., {Breen}, S.~L., {Voronkov}, M.~A., \&
  {Chen}, X. 2018, \apjl, 867, L4

\bibitem[{{McCarthy} {et~al.}(2020){McCarthy}, {Ellingsen}, {Breen},
  {Voronkov}, {Chen}, \& {Qiao}}]{McCarthy2020}
{McCarthy}, T.~P., {Ellingsen}, S.~P., {Breen}, S.~L., {et~al.} 2020, \mnras,
  491, 4642

\bibitem[{{McCarthy} {et~al.}(2017){McCarthy}, {Ellingsen}, {Chen}, {Breen},
  {Voronkov}, \& {Qiao}}]{McCarthy2017}
{McCarthy}, T.~P., {Ellingsen}, S.~P., {Chen}, X., {et~al.} 2017, \apj, 846,
  156

\bibitem[{{Meier} {et~al.}(2015){Meier}, {Walter}, {Bolatto}, {Leroy}, {Ott},
  {Rosolowsky}, {Veilleux}, {Warren}, {Wei{\ss}}, {Zwaan}, \&
  {Zschaechner}}]{Meier2015}
{Meier}, D.~S., {Walter}, F., {Bolatto}, A.~D., {et~al.} 2015, \apj, 801, 63

\bibitem[{{Menten}(1991)}]{Menten1991b}
{Menten}, K.~M. 1991, \apj, 380, L75

\bibitem[{{Menten}(2012)}]{Menten2012}
{Menten}, K.~M. 2012, in Cosmic Masers - from OH to H0, ed. R.~S. {Booth},
  W.~H.~T. {Vlemmings}, \& E.~M.~L. {Humphreys}, Vol. 287, 506--515

\bibitem[{{Meyer} {et~al.}(2004){Meyer}, {Zwaan}, {Webster}, {Staveley-Smith},
  {Ryan-Weber}, {Drinkwater}, {Barnes}, {Howlett}, {Kilborn}, {Stevens},
  {Waugh}, {Pierce}, {Bhathal}, {de Blok}, {Disney}, {Ekers}, {Freeman},
  {Garcia}, {Gibson}, {Harnett}, {Henning}, {Jerjen}, {Kesteven}, {Knezek},
  {Koribalski}, {Mader}, {Marquarding}, {Minchin}, {O'Brien}, {Oosterloo},
  {Price}, {Putman}, {Ryder}, {Sadler}, {Stewart}, {Stootman}, \&
  {Wright}}]{Meyer2004}
{Meyer}, M.~J., {Zwaan}, M.~A., {Webster}, R.~L., {et~al.} 2004, \mnras, 350,
  1195

\bibitem[{{Morris} \& {Serabyn}(1996)}]{Morris1996}
{Morris}, M. \& {Serabyn}, E. 1996, \araa, 34, 645

\bibitem[{{Moscadelli} {et~al.}(2003){Moscadelli}, {Menten}, {Walmsley}, \&
  {Reid}}]{Moscadelli2003}
{Moscadelli}, L., {Menten}, K.~M., {Walmsley}, C.~M., \& {Reid}, M.~J. 2003,
  \apj, 583, 776

\bibitem[{{M{\"u}ller} {et~al.}(2005){M{\"u}ller}, {Schl{\"o}der}, {Stutzki},
  \& {Winnewisser}}]{Mueller2005}
{M{\"u}ller}, H. S.~P., {Schl{\"o}der}, F., {Stutzki}, J., \& {Winnewisser}, G.
  2005, Journal of Molecular Structure, 742, 215

\bibitem[{{M{\"u}ller-S{\'a}nchez} {et~al.}(2010){M{\"u}ller-S{\'a}nchez},
  {Gonz{\'a}lez-Mart{\'\i}n}, {Fern{\'a}ndez-Ontiveros}, {Acosta-Pulido}, \&
  {Prieto}}]{MullerSanchez2010}
{M{\"u}ller-S{\'a}nchez}, F., {Gonz{\'a}lez-Mart{\'\i}n}, O.,
  {Fern{\'a}ndez-Ontiveros}, J.~A., {Acosta-Pulido}, J.~A., \& {Prieto}, M.~A.
  2010, \apj, 716, 1166

\bibitem[{{Oike} {et~al.}(2004){Oike}, {Kawaguchi}, {Takano}, \&
  {Nakai}}]{Oike2004}
{Oike}, T., {Kawaguchi}, K., {Takano}, S., \& {Nakai}, N. 2004, \pasj, 56, 431

\bibitem[{{Pence}(1980)}]{Pence1980}
{Pence}, W.~D. 1980, \apj, 239, 54

\bibitem[{{Pickett} {et~al.}(1998){Pickett}, {Poynter}, {Cohen}, {Delitsky},
  {Pearson}, \& {M{\"u}ller}}]{Pickett1998}
{Pickett}, H.~M., {Poynter}, R.~L., {Cohen}, E.~A., {et~al.} 1998, \jqsrt, 60,
  883

\bibitem[{{Pihlstr{\"o}m} {et~al.}(2014){Pihlstr{\"o}m}, {Sjouwerman}, {Frail},
  {Claussen}, {Mesler}, \& {McEwen}}]{Pihlstrom2014}
{Pihlstr{\"o}m}, Y.~M., {Sjouwerman}, L.~O., {Frail}, D.~A., {et~al.} 2014,
  \aj, 147, 73

\bibitem[{{Plambeck} \& {Menten}(1990)}]{Plambeck1990}
{Plambeck}, R.~L. \& {Menten}, K.~M. 1990, \apj, 364, 555

\bibitem[{{Rabli} \& {Flower}(2010)}]{Rabli2010}
{Rabli}, D. \& {Flower}, D.~R. 2010, \mnras, 406, 95

\bibitem[{{Regan} \& {Teuben}(2003)}]{Regan2003}
{Regan}, M.~W. \& {Teuben}, P. 2003, \apj, 582, 723

\bibitem[{{Rekola} {et~al.}(2005){Rekola}, {Richer}, {McCall}, {Valtonen},
  {Kotilainen}, \& {Flynn}}]{Rekola2005}
{Rekola}, R., {Richer}, M.~G., {McCall}, M.~L., {et~al.} 2005, \mnras, 361, 330

\bibitem[{{Rodr{\'\i}guez-Garza} {et~al.}(2017){Rodr{\'\i}guez-Garza}, {Kurtz},
  {G{\'o}mez-Ruiz}, {Hofner}, {Araya}, \& {Kalenskii}}]{Rodriguez2017}
{Rodr{\'\i}guez-Garza}, C.~B., {Kurtz}, S.~E., {G{\'o}mez-Ruiz}, A.~I.,
  {et~al.} 2017, \apjs, 233, 4

\bibitem[{{Rosolowsky} \& {Leroy}(2006)}]{Rosolowsky2006}
{Rosolowsky}, E. \& {Leroy}, A. 2006, \pasp, 118, 590

\bibitem[{{Sakamoto} {et~al.}(2006){Sakamoto}, {Ho}, {Iono}, {Keto}, {Mao},
  {Matsushita}, {Peck}, {Wiedner}, {Wilner}, \& {Zhao}}]{Sakamoto2006}
{Sakamoto}, K., {Ho}, P. T.~P., {Iono}, D., {et~al.} 2006, \apj, 636, 685

\bibitem[{{Salii} {et~al.}(2002){Salii}, {Sobolev}, \& {Kalinina}}]{Salii2002}
{Salii}, S.~V., {Sobolev}, A.~M., \& {Kalinina}, N.~D. 2002, Astronomy Reports,
  46, 955

\bibitem[{{Salter} {et~al.}(2008){Salter}, {Ghosh}, {Catinella}, {Lebron},
  {Lerner}, {Minchin}, \& {Momjian}}]{Salter2008}
{Salter}, C.~J., {Ghosh}, T., {Catinella}, B., {et~al.} 2008, \aj, 136, 389

\bibitem[{{S{\'a}nchez-Monge} {et~al.}(2018){S{\'a}nchez-Monge}, {Schilke},
  {Ginsburg}, {Cesaroni}, \& {Schmiedeke}}]{SanchezMonge2018}
{S{\'a}nchez-Monge}, {\'A}., {Schilke}, P., {Ginsburg}, A., {Cesaroni}, R., \&
  {Schmiedeke}, A. 2018, \aap, 609, A101

\bibitem[{{Schmidt} {et~al.}(2019){Schmidt}, {Mast}, {D{\'\i}az}, {Ag{\"u}ero},
  {G{\"u}nthardt}, {Gimeno}, {Oio}, \& {Gaspar}}]{Schmidt2019}
{Schmidt}, E.~O., {Mast}, D., {D{\'\i}az}, R.~J., {et~al.} 2019, \aj, 158, 60

\bibitem[{{Sinclair} {et~al.}(1992){Sinclair}, {Carrad}, {Caswell}, {Norris},
  \& {Whiteoak}}]{Sinclair1992}
{Sinclair}, M.~W., {Carrad}, G.~J., {Caswell}, J.~L., {Norris}, R.~P., \&
  {Whiteoak}, J.~B. 1992, \mnras, 256, 33P

\bibitem[{{Sjouwerman} {et~al.}(2010){Sjouwerman}, {Murray}, {Pihlstr{\"o}m},
  {Fish}, \& {Araya}}]{Sjouwerman2010}
{Sjouwerman}, L.~O., {Murray}, C.~E., {Pihlstr{\"o}m}, Y.~M., {Fish}, V.~L., \&
  {Araya}, E.~D. 2010, \apjl, 724, L158

\bibitem[{{Sobolev}(1993)}]{Sobolev1993}
{Sobolev}, A.~M. 1993, Astronomy Letters, 19, 293

\bibitem[{{Strickland} {et~al.}(2000){Strickland}, {Heckman}, {Weaver}, \&
  {Dahlem}}]{Strickland2000}
{Strickland}, D.~K., {Heckman}, T.~M., {Weaver}, K.~A., \& {Dahlem}, M. 2000,
  \aj, 120, 2965

\bibitem[{{Szczepanski} {et~al.}(1989){Szczepanski}, {Ho}, {Haschick}, \&
  {Baan}}]{Szczepanski1989}
{Szczepanski}, J.~C., {Ho}, P.~T.~P., {Haschick}, A.~D., \& {Baan}, W.~A. 1989,
  in The Center of the Galaxy, ed. M.~{Morris}, Vol. 136, 383

\bibitem[{{Thomas} \& {Fuller}(2008)}]{Thomas2008}
{Thomas}, H.~S. \& {Fuller}, G.~A. 2008, \aap, 479, 751

\bibitem[{{Turner}(1985)}]{Turner1985b}
{Turner}, B.~E. 1985, \apj, 299, 312

\bibitem[{{Val'tts} {et~al.}(1995){Val'tts}, {Dzura}, {Kalenskii}, {Slysh},
  {Booth}, \& {Winnberg}}]{Valtts1995}
{Val'tts}, I.~E., {Dzura}, A.~M., {Kalenskii}, S.~V., {et~al.} 1995, \aap, 294,
  825

\bibitem[{{van der Tak} {et~al.}(2007){van der Tak}, {Black}, {Sch{\"o}ier},
  {Jansen}, \& {van Dishoeck}}]{vanderTak2007}
{van der Tak}, F.~F.~S., {Black}, J.~H., {Sch{\"o}ier}, F.~L., {Jansen}, D.~J.,
  \& {van Dishoeck}, E.~F. 2007, \aap, 468, 627

\bibitem[{{Voronkov} {et~al.}(2012){Voronkov}, {Caswell}, {Ellingsen}, {Breen},
  {Britton}, {Green}, {Sobolev}, \& {Walsh}}]{Voronkov2012}
{Voronkov}, M.~A., {Caswell}, J.~L., {Ellingsen}, S.~P., {et~al.} 2012, in
  Cosmic Masers - from OH to H0, ed. R.~S. {Booth}, W.~H.~T. {Vlemmings}, \&
  E.~M.~L. {Humphreys}, Vol. 287, 433--440

\bibitem[{{Voronkov} {et~al.}(2014){Voronkov}, {Caswell}, {Ellingsen}, {Green},
  \& {Breen}}]{Voronkov2014}
{Voronkov}, M.~A., {Caswell}, J.~L., {Ellingsen}, S.~P., {Green}, J.~A., \&
  {Breen}, S.~L. 2014, \mnras, 439, 2584

\bibitem[{{Voronkov} {et~al.}(2010){Voronkov}, {Caswell}, {Ellingsen}, \&
  {Sobolev}}]{Voronkov2010}
{Voronkov}, M.~A., {Caswell}, J.~L., {Ellingsen}, S.~P., \& {Sobolev}, A.~M.
  2010, \mnras, 405, 2471

\bibitem[{{Walmsley} {et~al.}(1988){Walmsley}, {Batrla}, {Matthews}, \&
  {Menten}}]{Walmsley1988}
{Walmsley}, C.~M., {Batrla}, W., {Matthews}, H.~E., \& {Menten}, K.~M. 1988,
  \aap, 197, 271

\bibitem[{{Walsh} {et~al.}(2001){Walsh}, {Bertoldi}, {Burton}, \&
  {Nikola}}]{Walsh2001}
{Walsh}, A.~J., {Bertoldi}, F., {Burton}, M.~G., \& {Nikola}, T. 2001, \mnras,
  326, 36

\bibitem[{{Wang} {et~al.}(2014){Wang}, {Zhang}, {Gao}, {Zhang}, {Li}, {Fang},
  \& {Shi}}]{Wang2014}
{Wang}, J., {Zhang}, J., {Gao}, Y., {et~al.} 2014, Nature Communications, 5,
  5449

\bibitem[{{Wilson}(2009)}]{Wilson2009}
{Wilson}, T.~L. 2009, arXiv e-prints, arXiv:0903.0562

\bibitem[{{Yang} {et~al.}(2019){Yang}, {Chen}, {Shen}, {Li}, {Wang}, {Jiang},
  {Li}, {Dong}, {Wu}, \& {Qiao}}]{Yang2019}
{Yang}, K., {Chen}, X., {Shen}, Z.-Q., {et~al.} 2019, \apjs, 241, 18

\bibitem[{Yang {et~al.}(2017)Yang, Chen, Shen, Li, Wang, Jiang, Li, Dong, Wu,
  Qiao, \& Ren}]{Yang2017}
Yang, K., Chen, X., Shen, Z.-Q., {et~al.} 2017, The Astrophysical Journal, 846,
  160

\bibitem[{{Yang} {et~al.}(2020){Yang}, {Xu}, {Choi}, {Ellingsen}, {Sobolev},
  {Chen}, {Li}, \& {Lu}}]{Yang2020}
{Yang}, W., {Xu}, Y., {Choi}, Y.~K., {et~al.} 2020, \apjs, 248, 18

\bibitem[{{Yu} {et~al.}(2018){Yu}, {Xu}, \& {Wang}}]{Yu2018}
{Yu}, N.-P., {Xu}, J.-L., \& {Wang}, J.-J. 2018, Research in Astronomy and
  Astrophysics, 18, 015

\bibitem[{{Yusef-Zadeh} {et~al.}(2013){Yusef-Zadeh}, {Cotton}, {Viti},
  {Wardle}, \& {Royster}}]{Yusef-Zadeh2013}
{Yusef-Zadeh}, F., {Cotton}, W., {Viti}, S., {Wardle}, M., \& {Royster}, M.
  2013, \apjl, 764, L19

\bibitem[{{Zinchenko} {et~al.}(2017){Zinchenko}, {Liu}, {Su}, \&
  {Sobolev}}]{Zinchenko2017}
{Zinchenko}, I., {Liu}, S.~Y., {Su}, Y.~N., \& {Sobolev}, A.~M. 2017, \aap,
  606, L6

\end{thebibliography}

\begin{appendix}

\section{Rotation diagrams}
\label{sect.apen.rot_diag}

We make use of rotation diagrams as our first method to unveil the presence of methanol maser lines. This procedure assumes the Rayleigh-Jeans (RJ) approximation\footnote{see e.g. the derivation presented in \citet{Araya2005}, between Equations A9 and A10.}, valid when the following condition is fulfilled: $\nu$[GHz]$\ll$20.84\,[GHz/K]\,$T_{\rm ex}$[K] (see e.g. \citealt{Wilson2009}). The rotation diagram method also assumes a negligible background continuum \citep[see e.g.][their Chapter 4.4]{Belloche2019}, and relates the total column density $N_{\rm tot}$ of a given species with the excitation temperature $T_{\rm ex}$\footnote{Under LTE, the underlying assumption for the use of the rotation diagram method, the excitation temperature of any transition is equal to the rotation temperature that describes the populations of the rotational levels} and the integrated intensity of the line profiles as

%\footnote{For our data, the conditions for the RJ approximation are near the limit, but it still applies in most of our regions with less than $\sim$30\% of induced uncertainty. On average, considering the available regions, we have a $T_{\rm ex}$ of $\sim$14--25\,K (Sect.\,\ref{Sec.rot_diags}, E-- and A--CH$_{3}$OH, respectively) and $\sim$19-25\,K (Sect.\,\ref{sec.models}, E-- and A--CH$_{3}$OH, respectively). The RJ approximation is fulfilled when $\nu$[GHz]$\ll$20.84\,[GHz/K]\,$T_{\rm ex}$[K] (see e.g. \citealt{Wilson2009}). For an average frequency of 229\,GHz in ALCHEMI data (Sect.\,\ref{Sec.Observations}), this condition translates into 229\,GHz$<$292--521\,GHz (14--25$\times$20.84) and 229\,GHz$<$395--521\,GHz (19--25$\times$20.84), taking into account the mentioned $T_{\rm ex}$ obtained from the rotation diagrams and the LTE modelling, respectively.}

\begin{equation}
    \frac{N_{\rm tot}}{N_{\rm up}} = \frac{Q_{\rm rot(\textit{T}_{\rm ex})}}{g_{\rm up}}\exp{\Bigg(\frac{E_{\rm up}}{kT_{\rm ex}}\Bigg)},
\label{eq_rotd}    
\end{equation}

\noindent
where $Q_{\rm rot}(T)$ is the partition function of the species, which is the multiplication factor needed to go from a column density in a single state to the entire column density of a molecular species summed over all states \citep{Mangum2015}. $g_{\rm up}$ is the statistical weight of the upper level, $E_{\rm up}$/k is the energy of the upper level above the ground state, and $k$ is the Boltzmann constant. In the optically thin case, the beam-averaged $N_{\rm up}$ is given by

\begin{equation}
\label{eq.N_up}
    N_{\rm up} = \frac{8\pi k \nu^2}{hc^3 A_{\rm ul}} \int T_{\rm mb} dv,
\end{equation}

\noindent
where $\nu$ is the frequency, $h$ is the Planck constant, $c$ is the speed of light, $A_{\rm ul}$ is the Einstein coefficient for spontaneous emission, $T_{\rm mb}$ is the intensity in Kelvin units and $v$ denotes the radial velocity over which the integral is calculated, covering the entire observed range of a specific line emission. Thus, the term $\int T_{\rm mb} dv$ corresponds to the integrated intensity of the line, assuming that the emitting region completely covers the beam. When plotting equation \ref{eq_rotd} with $N_{\rm up}$ on a logarithmic scale, we obtain a straight line, with a slope proportional to the negative inverse value of $T_{\rm ex}$.

%Generated by https://www.tablesgenerator.com/
\begin{table*}[!htp]
\caption{Selected methanol transitions.} 
\label{tab.methanol_lines}
\scriptsize
\renewcommand{\tabcolsep}{0.16cm}
\begin{center}
\begin{tabular}{llllllllllllcl}
\hline \hline
Transition                 & $\nu$ [GHz]& A/E      & R1         & R2         & R3         & R4         & R5         & R6         & R7         & R8         & R9         & R10  & potential blending lines\\ \hline \\
5$_{-1}\rightarrow$4$_{0}$ & 84.521172  & E        & \checkmark & \checkmark & \checkmark & \checkmark & --         & \checkmark & \checkmark & \checkmark & \checkmark & \checkmark & -- \\
2$_{1}\rightarrow$1$_{1}$  & 95.914310  & A$^{+}$        & \checkmark & \checkmark & \checkmark & \checkmark & --         & \checkmark & \checkmark & \checkmark & \checkmark & \checkmark & C$_{2}$H$_{5}$OH\\
2$_{1}\rightarrow$1$_{1}$  & 97.582798  & A$^{-}$        & \checkmark & \checkmark & \checkmark & \checkmark & --         & \checkmark & \checkmark & \checkmark & \checkmark & --         & CH$_{3}$COCH$_{3}$\\
3$_{1}\rightarrow$4$_{0}$  & 107.013831 & A$^{+}$        & --         & --         & \checkmark & \checkmark & --         & \checkmark & \checkmark & --         & --         & --         & CH$_{3}$COCH$_{3}$\\
0$_{0}\rightarrow$1$_{-1}$ & 108.893945 & E        & \checkmark & \checkmark & \checkmark & \checkmark & --         & \checkmark & \checkmark & \checkmark & \checkmark & \checkmark & -- \\
6$_{-1}\rightarrow$5$_{0}$ & 132.890759 & E        & \checkmark & \checkmark & \checkmark & \checkmark & --         & \checkmark & \checkmark & \checkmark & \checkmark & \checkmark & CH$_{3}$OCHO\\
3$_{1}\rightarrow$2$_{1}$  & 143.865795 & A$^{+}$        & *          & \checkmark & \checkmark & \checkmark & --         & \checkmark & \checkmark & \checkmark & \checkmark & --         & C$_{2}$H$_{5}$OH\\
3$_{1}\rightarrow$2$_{1}$  & 146.368328 & A$^{-}$        & *          & \checkmark & \checkmark & \checkmark & --         & \checkmark & \checkmark & \checkmark & \checkmark & bN         & cis--DCOOH, n--C$_{3}$H$_{7}$CN\\
9$_{0}\rightarrow$8$_{1}$  & 146.618697 & A$^{+}$        & \checkmark & --         & --         & \checkmark & --         & --         & *          & \checkmark & \checkmark & --         & SO$_{2}$\\
8$_{0}\rightarrow$8$_{-1}$ & 156.488902 & E        & --         & --         & --         & \checkmark & --         & --         & --         & --         & --         & --         & C$_{2}$H$_{3}$CN\\
2$_{1}\rightarrow$3$_{0}$  & 156.602395 & A$^{+}$        & --         & N          & \checkmark & \checkmark & --         & \checkmark & \checkmark & --         & \checkmark & --         & C$_{2}$H$_{5}$C--13--N\\
7$_{0}\rightarrow$7$_{-1}$ & 156.828517 & E        & --         & --         & --         & \checkmark & --         & \checkmark & \checkmark & --         & --         & --        & C$_{2}$H$_{5}$CN, NH$_{2}$CHO\\
6$_{0}\rightarrow$6$_{-1}$ & 157.048617 & E        & --         & --         & --         & \checkmark & --         & \checkmark & \checkmark & --         & --         & --         & HCOOH\\
5$_{0}\rightarrow$5$_{-1}$ & 157.178987 & E        & --         & --         & --         & \checkmark & --         & \checkmark & \checkmark & --         & --         & --         & CH$_{3}$COCH$_{3}$, $^{33}$SO\\
5$_{1}\rightarrow$5$_{0}$  & 165.369341 & E        & \checkmark & \checkmark & \checkmark & \checkmark & --         & \checkmark & \checkmark & \checkmark & \checkmark & *          & C$_{2}$H$_{5}$CN\\
4$_{1}\rightarrow$3$_{1}$  & 191.810503 & A$^{+}$ & --         & *          & \checkmark & \checkmark & --         & \checkmark & \checkmark & bN         & \checkmark & *           & HCOOC$_{2}$H$_{5}$, n--C$_{3}$H$_{7}$CN\\
4$_{1}\rightarrow$3$_{1}$  & 195.146790 & A$^{-}$ & *          & *          & \checkmark & \checkmark & --         & \checkmark & \checkmark & \checkmark & \checkmark & N         & CH$_{3}$OCHO, CH$_{3}$OCH$_{3}$\\
10$_{0}\rightarrow$9$_{1}$ & 198.403067 & A$^{+}$ & *          & *          & \checkmark & \checkmark & --         & \checkmark & \checkmark & N          & bN         & *          & -- \\
1$_{1}\rightarrow$2$_{0}$  & 205.791270 & A$^{+}$ & *          & \checkmark & \checkmark & \checkmark & --         & \checkmark & \checkmark & \checkmark & \checkmark & bN         & C$_{2}$H$_{5}$OH, CH$_{3}$OCHO\\
1$_{1}\rightarrow$0$_{0}$  & 213.427061 & E        & \checkmark & \checkmark & \checkmark & \checkmark & --         & \checkmark & \checkmark & \checkmark & \checkmark & \checkmark & \textit{c}--C$_{3}$H\\
5$_{1}\rightarrow$4$_{2}$  & 216.945521 & E        & --         & --         & --         & \checkmark & --         & \checkmark & \checkmark & --         & --         & --         & -- \\
4$_{2}\rightarrow$3$_{1}$  & 218.440063 & E        & *          & \checkmark & \checkmark & \checkmark & --         & \checkmark & *          & \checkmark & \checkmark & \checkmark & \textit{c}--C$_{3}$H$_{2}$\\
8$_{-1}\rightarrow$7$_{0}$ & 229.758756 & E        & \checkmark & --         & --         & \checkmark & --         & \checkmark & \checkmark & \checkmark & \checkmark & --         & CH$_{3}$CHO\\
5$_{1}\rightarrow$4$_{1}$  & 239.746219 & A$^{-}$  & \checkmark & --         & \checkmark & \checkmark & --         & \checkmark & \checkmark & \checkmark & \checkmark & bN         & HCOOD\\
5$_{1}\rightarrow$4$_{1}$  & 243.915788 & A$^{+}$  & *          & --         & \checkmark & \checkmark & --         & \checkmark & \checkmark & *          & \checkmark & bN         & D$_{2}$ $^{34}$S, $^{34}$SO$_{2}$\\
4$_{2}\rightarrow$5$_{1}$  & 247.228587 & A        & --         & --         & bN         & \checkmark & --         & --         & \checkmark & --         & --         & --         & C$_{2}$H$_{5}$OH\\
2$_{0}\rightarrow$1$_{-1}$ & 254.015377 & E        & \checkmark & \checkmark & \checkmark & \checkmark & --         & \checkmark & \checkmark & \checkmark & \checkmark & \checkmark & \textit{c}--C$_{3}$D$_{2}$, C$_{2}$H$_{5}$OH\\
2$_{1}\rightarrow$1$_{0}$  & 261.805675 & E        & \checkmark & \checkmark & \checkmark & \checkmark & --         & \checkmark & \checkmark & \checkmark & \checkmark & \checkmark & C$_{2}$H, c--C$_{3}$H$_{2}$ \\
6$_{1}\rightarrow$5$_{1}$  & 287.670767 & A$^{+}$        & \checkmark & --         & \checkmark & \checkmark & --         & \checkmark & \checkmark & \checkmark & \checkmark & --         & CCS\\
6$_{0}\rightarrow$5$_{0}$  & 289.939377 & E        & --         & --         & \checkmark & \checkmark & --         & \checkmark & *          & \checkmark & \checkmark & --         & S$^{17}$O, HNC$^{18}$O\\
6$_{1}\rightarrow$5$_{1}$  & 292.672889 & A$^{-}$        & \checkmark & --         & \checkmark & \checkmark & --         & \checkmark & \checkmark & *          & \checkmark & --         & CH$_{3}$COCH$_{3}$, CH$_{3}$CHO\\
3$_{0}\rightarrow$2$_{-1}$ & 302.369753 & E        & \checkmark & \checkmark & \checkmark & \checkmark & --         & \checkmark & \checkmark & \checkmark & \checkmark & \checkmark & \textit{c}--HCOOH, CH$_{3}$OCHO\\
1$_{1}\rightarrow$1$_{0}$  & 303.366921 & A$^{-+}$        & \checkmark & \checkmark & \checkmark & \checkmark & --         & \checkmark & \checkmark & *          & \checkmark & \checkmark & H$_{2}$CN\\
2$_{1}\rightarrow$2$_{0}$  & 304.208348 & A$^{-+}$        & \checkmark & \checkmark & \checkmark & \checkmark & --         & \checkmark & \checkmark & \checkmark & \checkmark & \checkmark & NH$_{2}$CHO, C$_{2}$H$_{5}$OH\\
3$_{1}\rightarrow$3$_{0}$  & 305.473491 & A$^{-+}$        & \checkmark & \checkmark & \checkmark & \checkmark & --         & \checkmark & \checkmark & \checkmark & \checkmark & bN         & C$_{2}$H$_{5}$CN, CH$_{3}$COCH$_{3}$\\
4$_{1}\rightarrow$4$_{0}$  & 307.165924 & A$^{-+}$        & \checkmark & \checkmark & \checkmark & \checkmark & --         & \checkmark & \checkmark & \checkmark & \checkmark & \checkmark & CH$_{3}$OCHO\\
5$_{1}\rightarrow$5$_{0}$  & 309.290360 & A$^{-+}$        & *          & \checkmark & \checkmark & \checkmark & --         & \checkmark & \checkmark & \checkmark & \checkmark & bN         & $^{33}$SO\\
3$_{1}\rightarrow$2$_{0}$  & 310.192994 & E        & \checkmark & \checkmark & \checkmark & \checkmark & --         & \checkmark & \checkmark & \checkmark & \checkmark & \checkmark & CH$_{3}$OCHO, CH$_{3}$COCH$_{3}$\\
6$_{1}\rightarrow$6$_{0}$  & 311.852612 & A$^{-+}$        & *          & bN         & \checkmark & \checkmark & --         & \checkmark & \checkmark & \checkmark & \checkmark & \checkmark & \textit{c}--C$_{3}$H$_{2}$, CH$_{3}$COCH$_{3}$\\
7$_{1}\rightarrow$7$_{0}$  & 314.859528 & A$^{-+}$        & *          & --         & \checkmark & \checkmark & --         & \checkmark & \checkmark & *          & \checkmark & --         & C$_{2}$H$_{3}$CN\\
6$_{2}\rightarrow$5$_{1}$  & 315.266861 & E        & \checkmark & --         & \checkmark & \checkmark & --         & \checkmark & \checkmark & *          & *          & --         & C$_{2}$H$_{3}$CN\\
8$_{1}\rightarrow$8$_{0}$  & 318.318919 & A$^{-+}$        & --         & --         & \checkmark & \checkmark & --         & \checkmark & \checkmark & bN         & bN         & --         & c--C$_{3}$H$_{2}$, C$_{2}$H$_{5}$OH\\
7$_{1}\rightarrow$6$_{1}$  & 335.582017 & A$^{+}$        & \checkmark & --         & \checkmark & \checkmark & --         & \checkmark & \checkmark & --         & --         & --         & n--C$_{3}$H$_{7}$CN, l--C$_{3}$H$_{2}$\\
7$_{0}\rightarrow$6$_{0}$  & 338.124488 & E        & --         & \checkmark & \checkmark & \checkmark & --         & \checkmark & \checkmark & \checkmark & \checkmark & --         & CH$_{3}$OCHO\\
7$_{1}\rightarrow$6$_{1}$  & 341.415615 & A$^{-}$     & --         & --         & \checkmark & \checkmark & --         & \checkmark & \checkmark & --         & --         & --         & NH$_{2}$CHO, CH$_{3}$OCHO\\
4$_{0}\rightarrow$3$_{-1}$ & 350.687662 & E        & \checkmark & \checkmark & \checkmark & \checkmark & --         & \checkmark & \checkmark & \checkmark & \checkmark & \checkmark & NO\\
1$_{1}\rightarrow$0$_{0}$  & 350.905100 & A$^{+}$     & \checkmark & \checkmark & \checkmark & \checkmark & --         & \checkmark & \checkmark & \checkmark & \checkmark & \checkmark & CH$_{3}$COCH$_{3}$, $^{33}$SO$_{2}$\\
4$_{1}\rightarrow$3$_{0}$  & 358.605799 & E        & \checkmark & *          & \checkmark & \checkmark & --         & \checkmark & \checkmark & \checkmark & \checkmark & --         & CH$_{3}$OCHO, n--C$_{3}$H$_{7}$CN\\
7$_{2}\rightarrow$6$_{1}$  & 363.739868 & E        & --         & --         & \checkmark & \checkmark & --         & \checkmark & \checkmark & --         & --         & --         & --\\ 
TNoL                       &            &          & 25(35)     & 23(29)     & 41(42)     & 49         & --         & 46         & 45(48)     & 31(38)     & 35(39)     & 16(26)     &\\
%SFR                        &            &          & 7.4        & 9.4        & 26.0       & 33.5       & 25.7       & 32.5       & 29.8       & 12.2       & 10.9       & 3.1        &\\
H$_{\alpha}$               &            &          & --          & 1147       & 58303      & 58579      & 17814      & 36998      & 36291      & 8150       & 7073       & 2603       &\\ \hline
\end{tabular}
\tablefoot{Selected methanol lines to perform our rotation diagrams. Possible contamination with other molecules is mentioned in the rightmost column. The symbol * indicates bad-shaped lines. The letter N indicates lines embedded in a noisy continuum but with intensities above the 3.5$\sigma$ level. bN denotes lines tentatively detected (2--3$\sigma$). $\checkmark$ indicates well-distinguished/confidently-detected lines, and the dash (--) indicates a non-detection. The third last row provides the total number of lines (TNoL) detected toward each region, considering lines confidently observed (check-mark). The TNoL considering the ones not clearly detected (*, N, bN), are given in parenthesis. The last row presents the H$_{\alpha}$ mean intensity for each region considering an angular extent of 1\farcs6 (the beam size), obtained from MUSE archival data (ID:0102.B-0078(A), PI:Laura Zschaechner) in units of 10$^{-20}$ erg/s/cm$^{2}$/\si{\angstrom}. For an alternative approach on line blending, see Tables\,\ref{tab.methanol_lines_model_E} and \ref{tab.methanol_lines_model_A}.}

\end{center}
\end{table*}

%%% 16 April, 2022:
Given the relatively broad frequency range covered by ALCHEMI, it can be expected that the RJ approximation will not be correct for all transitions and temperatures measured in the CMZ of NGC\,253. It is impossible to know a priori whether the $T_{\rm ex}$ will be low enough to cause a significant distortion between the Planck function and the RJ approximation. Given the results summarized in our Table A.2, this approximation has an appreciable bias and we have decided to calculate a correction factor ($CF$) meant to be multiplied by the upper level column density of each methanol transition in the rotation diagrams. This $CF$ is equal to the Planck function over the RJ approximation:

\begin{equation}
\label{eq.CF}
   CF = \frac{h\nu}{kT_{\rm ex}} \frac{1}{\exp(\frac{h\nu}{kT_{\rm ex}})-1}
\end{equation}

Considering methanol transitions with frequencies of 200 and 300\,GHz inside a region with a temperature of 10--20\,K, we obtained correction factors in the range of 0.6--0.8 and 0.45--0.7, respectively. After applying this correction to our rotation diagrams (assuming RJ) the resulting $T_{\rm ex}$ obtained for each region vary mostly by less than 1\,K. The total column densities, on the other hand, can decrease by $\sim$50\% in the coldest regions, such as regions 2 and 10 for E--CH$_{3}$OH (see Table\,\ref{tab.apen.RD_params}), and by $\sim$20\% in warmer environments. The lowest values obtained in this way are included in the uncertainties shown in Table\,\ref{tab.apen.RD_params}. 

We could do an iterative process from the temperature obtained from RJ to the one obtained after applying the correction factor until reaching a certain $T_{\rm ex}$ asymptotically. However, since $T_{\rm  ex}$ hardly varies, we decided to include the new $N_{\rm{up}}$ values inside the uncertainties after a single correction is applied. Indeed, the change of the $T_{\rm ex}$ assumed when applying $CF$ will not have an important effect on the slope derived in the rotation diagrams. It might slightly increase the dispersion of the individual $N_{\rm{up}}$ values with respect to the LTE slope (solid black lines in Fig.\,\ref{fig.rotationdiagrams}).

This dispersion is inversely proportional to $T_{\rm ex}$, and $T_{\rm ex}$ is higher for the A-CH$_{3}$OH species with respect to the E-CH$_{3}$OH species (see Table\,\ref{tab.apen.RD_params}). One might suspect that an increased dispersion may discard the 10$_{0}\rightarrow 9_{1} - A^{+}$ transition at 198.4\,GHz as a maser candidate because of its departure from the LTE slope of a factor of 3.6$\pm$1.2 in region\,6, as indicated in Sect.\,\ref{subsec.Amasers}, given our threshold of 3.3 (3$\sigma$; Sect.\,\ref{Sec.rot_diags}). However, this region has a $T_{\rm ex}$ of 24.3\,K indicating that RJ is a good approximation for the Planck formula. All the other methanol maser candidates depart from LTE by a factor higher than four times the LTE expected value.

The results from our rotation diagrams fitting to the LTE lines are listed in Table\,\ref{tab.apen.RD_params}. The effect of having a strong continuum might affect our synthetic spectra, potentially leading to an overestimation of the emission \citep[e.g.][]{Belloche2019}. However, comparing our data with the models, this can be ruled out. Indeed, we noticed that considering our deviation from the Planck function in the total column density uncertainties, the rotation diagram results are in better agreement with the modelling results (see Table\,\ref{tab.LTE_model_params}). This is especially critical for the colder regions. It can also be appreciated that sometimes the lower limits are placed much below the ones derived by the radiative transfer models, as in the case of region\,4, where the temperatures derived from the rotation diagram method seem to be underestimated by $\sim$5\,K with respect to the LTE models, generating an artificial scatter and lower total column densities. The latter support our predicament of including the newly derived upper-level column density values obtained after applying the $CF$ inside an error bar instead of replacing the values obtained from RJ. Accounting for the excitation temperatures, they are always higher in the modelling as compared to the rotation diagrams for E-methanol the nuclear regions (3 to 6) in A-methanol, and the opposite is observed at the outskirts of the NGC\,253's CMZ for the A-methanol case (regions\,1 and 8 to 10).

%Another reason to keep the original $N_{\rm{up}}$ values is the good fit (low dispersion) we observe when doing the linear regression to the diagrams.}
%%%%%%

\begin{table}[!h]
\caption{Fit parameters from our rotation diagrams.} \label{tab.apen.RD_params}
\begin{center}
\centering\scriptsize
%\begin{tabular}{lllll}
\renewcommand{\tabcolsep}{0.11cm}
\begin{tabular}{cll|ll}
\hline \hline
Region & $N$(Sp) & \multicolumn{1}{c}{$T_{\rm ex}$} & $N$(Sp) & \multicolumn{1}{c}{T$_{ex}$} \\
       & \multicolumn{1}{c}{[$\times$10$^{13}$ cm$^{-2}$]}  & \multicolumn{1}{c}{[K]}     & \multicolumn{1}{c}{[$\times$10$^{13}$ cm$^{-2}$]}  & \multicolumn{1}{c}{[K]}  \\
       & \multicolumn{1}{c}{E--CH$_{3}$OH} & & \multicolumn{1}{c}{A--CH$_{3}$OH} \\ \hline \\

R1      & 67.3$^{+10.1}_{-38.4}$   & 9.7$^{+1.0}_{-2.1}$  & 9.9$^{+1.5}_{-3.8}$     & 25.0$^{+3.1}_{-2.9}$ \\
R2      & 78.1$^{+11.7}_{-30.6}$   & 8.4$^{+0.8}_{-1.9}$  & 10.0$^{+1.5}_{-5.0}$    & 14.5$^{+1.9}_{-2.0}$ \\
R3      & 263.0$^{+39.4}_{-105.0}$ & 19.0$^{+1.6}_{-2.7}$ & 84.2$^{+12.6}_{-32.1}$  & 23.1$^{+3.1}_{-3.3}$ \\
R4      & 91.2$^{+13.7}_{-28.8}$   & 27.5$^{+3.9}_{-4.0}$ & 76.2$^{+11.4}_{-26.5}$  & 27.7$^{+5.3}_{-5.5}$ \\
R6      & 253.7$^{+38.1}_{-92.1}$  & 20.3$^{+2.9}_{-1.9}$ & 115.4$^{+17.3}_{-36.6}$ & 24.3$^{+3.1}_{-4.8}$ \\  %Tex(E):lower limit is the same(w/CF) , thanks to the temperature around 20K. Higher limit increased.
R7      & 275.2$^{+41.3}_{-134.3}$ & 12.3$\pm$1.3         & 58.4$^{+8.8}_{-21.1}$   & 25.7$^{+4.6}_{-5.2}$ \\ %Tex(E):same values, the dispersion increased.
R8      & 77.8$^{+11.7}_{-44.4}$   & 9.9$^{+0.9}_{-1.9}$  & 12.6$^{+1.9}_{-4.1}$    & 30.8$^{+7.3}_{-7.1}$ \\
R9      & 103.3$^{+11.7}_{-57.2}$  & 10.4$^{+1.0}_{-1.9}$ & 17.9$^{+2.7}_{-5.6}$    & 29.3$^{+7.7}_{-7.6}$ \\
R10     & 11.6$^{+1.74}_{-6.8}$    & 9.6$^{+0.8}_{-2.7}$  & 3.0$^{+0.4}_{-1.1}$     & 23.8$^{+14.5}_{-13.5}$ \\
\hline \\
\end{tabular}
\tablefoot{Total column densities ($N$(Sp)) and excitation temperatures ($T_{\rm ex}$) for each methanol symmetric type derived from the best LTE fit to the rotation diagrams (see Fig.\,\ref{fig.rotationdiagrams}). Uncertainties correspond to the standard deviation (1$\sigma$).}
\end{center}
\end{table}

\section{Rotation diagram: non-masing outliers}
\label{sec.apen.rotdiag_outliers}

Making use of the rotation diagram method (Sec.\,\ref{Sec.rot_diags}), we have found a variety of methanol lines that do not follow LTE conditions but that, at the same time, do not correspond to masers. These lines are placed beyond a 3$\sigma$ scatter, surpassing by more than a factor of 3.3 the expected $N_{\rm up}$ value from LTE conditions. We list them in Table\,\ref{tab.RD_outliers} and provide a description below, in frequency order.

We realized that the five A--CH$_{3}$OH lines, at 95.9\,GHz, 97.6, 143.9, 146.4, and 156.6\,GHz, that usually depart from the LTE conditions of A-type methanol, actually follow the LTE trend of the E--CH$_{3}$OH species. Since the conditions between A- and E-methanol types differ towards the edge of the CMZ, these five transitions are clearly apart from LTE conditions in regions 1, 2, and 7 to 9. Therefore we decided to not include them in the LTE fitting in any region.

With the exception of the line at 156.6\,GHz, the other four can be considered as members of the same family, although varying between symmetric labels (A$^{+}$ and A$^{-}$), corresponding to the $(J+1)_{1}\rightarrow J_{1}-A$ series. In general, the mentioned five transitions have Einstein coefficients for spontaneous emission (A$_{ij}$) below 1.78$\times$10$^{-5}$\,s$^{-1}$, unlike the other A--CH$_{3}$OH lines in LTE. 

As can be seen in Fig\,\ref{fig.rotationdiagrams}, these five A--CH$_{3}$OH transitions with low A$_{ij}$ also have 60--600 times lower critical densities than transitions at similar $E_{\rm up}$/k and this is not significantly changed if $T_{\rm ex}$ is moderately (by less than a factor of five) lower than $T_{\rm kin}$. This means that those transitions require a particularly small critical density to emerge, pointing out that they are tracing conditions of cooler gas located in layers farther away from the GMC cores as compared to the rest of the A-methanol lines. The latter is also based on the overall higher temperatures in A-type methanol compared to E-type methanol in the nine analyzed regions (see fourth panel in Fig.\,\ref{fig.delta}). 

The line at 261.8\,GHz is located beyond LTE in region\,6 only. This can be caused by line blending with C$_{2}$H and c--C$_{3}$H$_{2}$, molecules previously observed in the central regions of NGC\,253 \citep{Oike2004,Martin2006,Holdship2021}.

In regions\,1 and 8, the $6_{2}\rightarrow5_{1}-E$ line at 315.3\,GHz ($E_{\rm up}$/k$=$63.1\,K) seems to act as a maser. However, this line is not clearly observed as an outlier in the other 7 regions. Additionally, if methanol is tracing two different temperatures, the line at 315.2\,GHz is related to the temperature better described by A--CH$_{3}$OH and is therefore likely not a maser. 

The line at 318.3\,GHz ($E_{\rm up}$/k$=98.8\,K$) strongly departs from LTE in the inner regions (regions\,3 to 6). Similar to the case of the line at 261.8\,GHz, this departure is likely due to the contribution of c--C$_{3}$H$_{2}$.

The line at 338.1\,GHz ($E_{\rm up}$/k$=78.8$\,K) falls beyond LTE in region\,8. In regions\,6 to 9 this transition is placed above the LTE trend described by E-type methanol, being closer to the conditions described by A-type methanol instead. In this case we can not argue for a higher critical density for this behavior, contrary to other E--CH$_{3}$OH transitions (see below).

The line at 350.7\,GHz ($E_{\rm up}$/k$=36.3$\,K) appears to be out of LTE in region\,6. Performing a simple LTE model for NO in region\,6, we find that NO contributes to 70\% to the measured blended feature. We also find that NO is stronger in region\,6 than in any other studied region (by doing a moment\,0 map of this line, not shown), explaining why we do not see the 350.7\,GHz methanol line seemingly departing from LTE in any other region.

The E--CH$_{3}$OH transitions at 358.605 ($E_{\rm up}$/k$=36.4$\,K) and 363.739\,GHz ($E_{\rm up}$/k$=79.4$\,K) are emerging at the highest densities, tracing conditions of A--CH$_{3}$OH. They clearly depart from LTE in region\,7, and serve as a complementary case with respect to the five A--CH$_{3}$OH lines mentioned above.

\subsection{lines blended with maser candidates}

\textbf{The} $\mathbf{5_{-1}\rightarrow 4_{0}-E}$ line at 84.52\,GHz occupies a similar frequency window as the $12_{10}\rightarrow12_{11}-E$ methanol line at 84.53\,GHz in all selected regions except region\,10, where the line profiles are narrow enough to be disentangled. Line blending, however, is unlikely since the potential companion line connects levels 1085\,K above the ground state and has an Einstein coefficient $\sim$2.6 times lower than the masing line (see Table\,\ref{tab.RD_outliers}). In addition to that, there is no observed emission from the line at 84.53\,GHz in region\,10 (only the maser line is observed). Furthermore, as mentioned in Sections\,\ref{Sec.rot_diags} and \ref{sec.models}, in general there is no methanol emission involving energy levels $>$150\,K above the ground state. 

\textbf{The} $\mathbf{7_{-1}\rightarrow 6_{0}-E}$ line at 181.295\,GHz which, to our knowledge, was not yet reported as a (Galactic) maser line \citep[e.g.][]{Leurini2016}, is dramatically contaminated by the $J=$2--1 HNC line at 181.324\,GHz, impeding a proper line inspection. This methanol transition is also affected by its proximity to telluric contamination (H$_{2}$O at $\sim$183\,GHz) that decreases the quality of our data near this frequency.

\textbf{The} $\mathbf{9_{-1}\rightarrow 8_{0}-E}$ transition at 278.305\,GHz is the highest excited line ($E_{\rm up}$/k$\sim$102\,K) covered by our observations in the $J_{-1}\rightarrow (J-$ 1)$_{0}-E$ series. It is strongly contaminated by the 2$_{-2}\rightarrow 3_{-1}-E$ line at 278.342\,GHz. Due to this blending, this methanol line is not considered in the rotation diagrams. However, as presented in Sect.\,\ref{sec.models}, a maser line profile emerges in regions\,8, 9 and 10.

%At first sight, the line at 146.6\,GHz seems to be masing in region\,4 given that its departure from LTE is a factor of 3.3$\pm$1.2 times.

\textbf{The} $\mathbf{9_{0}\rightarrow 8_{1}-A^{+}}$ line at 146.618\,GHz appears to be masing in region\,4, with a departure from LTE of 3.3$\pm$1.2 times. However, this line is considerably contaminated by SO$_{2}$ in region\,5 ($\sim$80\% of the total integrated intensity of the blended feature, Mart\'in et al. in prep.) and therefore, due to the proximity to this region, we expect an important contamination in region\,4 and also some contribution in region\,7 (after performing a moment\,0 map for SO$_{2}$, not shown). The above influences its departure from LTE in the mentioned regions, producing a larger displacement than for the lines at 84.5\,GHz and 132.9\,GHz, which are not contaminated. The $9_{0}\rightarrow 8_{1}-A^{+}$ line at 146.618\,GHz is also blended with the 14$_{1}\rightarrow 13_{2}-A^{+}$ methanol line at 146.617\,GHz, which requires twice the $E_{\rm up}$/k (256\,K, see Table\,\ref{tab.RD_outliers}) to emerge. However, the contribution of this companion line at 146.617\,GHz to the maser line at 146.618\,GHz is not higher than 2\% (Mart\'in et al. in prep.).

\section{Detailed model analysis}
\label{sec.apen.detailed_modelling}

In Sect.\,\ref{sec.models}, we limited our LTE and non-LTE modelling to a single component for each methanol symmetric type and then we preferred to continue with the LTE model only. This is enough to characterize our selected regions for the scope of this study, resulting in synthetic spectra that adjust to most of the observed methanol emission lines, as can be seen in our Tables\,\ref{tab.methanol_lines_model_E} and \ref{tab.methanol_lines_model_A}. The comparison between the observed and the resulting synthetic spectra clearly unveil lines with intensities far above the LTE trend. 

\begin{table*}[!htp]
\caption{E-type methanol lines used as input for our LTE model.} \label{tab.methanol_lines_model_E}
\scriptsize
\begin{center}
\renewcommand{\tabcolsep}{0.14cm}
\begin{tabular}{lllllllllllll}
\hline \hline
Transition                  &$\nu$ [GHz]  & R1          & R2         & R3         & R4         & R5         & R6         & R7         & R8         & R9        & R10      & Blending lines\\ \hline \\
6$_{-2}\rightarrow$7$_{-1}$  & 85.568131 & NL          & NL         & F          & F          & N          & N          & $\sim$     & NL         & NL        & NL        & -- \\
%0$_{0}\rightarrow$1$_{-1}$  & 108.893945 & $\sim$      & NL         & F          & F          & $\sim$     & F          & F          & F          & F          & N        & --\\
5$_{-2}\rightarrow$6$_{-1}$ & 133.605439 & $\sim$      & NL         & F          & F          & $\sim$     & F          & F          & $\sim$     & F          & N        & --\\
8$_{0}\rightarrow$8$_{-1}$  & 156.488902 & $\sim$      & N          & F          & F          & F          & F          & F          & N          & $\sim$     & NL       & C$_{2}$H$_{5}$OH\\
7$_{0}\rightarrow$7$_{-1}$  & 156.828517 & F           & NL         & F          & F          & F          & F          & F          & N          & $\sim$     & NL       & --\\
6$_{0}\rightarrow$6$_{-1}$  & 157.048617 & $\sim$      & NL         & F          & F          & F          & F          & F          & F          & F          & N        & C$_{3}$H\\
5$_{0}\rightarrow$5$_{-1}$  & 157.178987 & F           & F          & F          & F          & F          & F          & F          & F          & F          & $\sim$   & --\\
4$_{0}\rightarrow$4$_{-1}$  & 157.246062 & F           & F          & F          & $\sim$     & F          & F          & F          & F          & F          & F        & --\\
1$_{0}\rightarrow$1$_{-1}$  & 157.270832 & F           & F          & $\sim$     & F          & F          & $\sim$     & F          & $\sim$     & F          & F        & --\\
3$_{0}\rightarrow$3$_{-1}$  & 157.272338 & F           & F          & $\sim$     & F          & F          & $\sim$     & F          & F          & F          & F        & --\\
2$_{0}\rightarrow$2$_{-1}$  & 157.276019 & F           & F          & $\sim$     & F          & F          & $\sim$     & F          & $\sim$     & F          & F        & --\\
5$_{1}\rightarrow$5$_{0}$   & 165.369341 & F           & $\sim$     & $\sim$     & $\sim$     & F          & $\sim$     & F          & $\sim$     & F          & N        & --\\
5$_{1}\rightarrow$4$_{2}$   & 216.945521 & N           & $\sim$     & F          & F          & F          & F          & F          & F          & F          & $\sim$   & C$_{2}$H$_{3}$CN, CH$_{3}$OCHO\\
8$_{0}\rightarrow$7$_{1}$   & 220.078561 & NL          & NL         & F          & F          & N          & F          & F          & NL         & F          & NL       & HC$_{3}$N \\
3$_{-2}\rightarrow$4$_{-1}$ & 230.027047 & F           & F          & $\sim$     & ~          & F          & N          & F          & F          & F          & F        & CH$_{3}$NH$_{2}$\\
5$_{0}\rightarrow$4$_{0}$   & 241.700159 & $\sim$      & F          & F          & F          & F          & F          & F          & F          & F          & F        & --\\
5$_{2}\rightarrow$4$_{2}$   & 241.904643 & F           & $\sim$     & F          & F          & F          & F          & F          & F          & F          & $\sim$   & --\\
6$_{1}\rightarrow$5$_{2}$   & 265.289562 & F           & $\sim$     & F          & F          & F          & F          & F          & F          & F          & F        & --\\
5$_{2}\rightarrow$4$_{1}$   & 266.838148 & F           & $\sim$     & F          & F          & F          & $\sim$     & F          & $\sim$     & $\sim$     & N        & --\\
6$_{0}\rightarrow$5$_{0}$   & 289.939377 & F           & F          & F          & F          & F          & F          & F          & F          & F          & F        & --\\
6$_{2}\rightarrow$5$_{2}$   & 290307.738 & F           & N          & F          & F          & F          & F          & F          & F          & F          & N        & --\\
3$_{0}\rightarrow$2$_{-1}$  & 302.369773 & F           & F          & F          & $\sim$     & F          & $\sim$     & F          & F          & F          & F        & --\\
3$_{1}\rightarrow$2$_{0}$   & 310.192994 & $\sim$      & F          & F          & F          & F          & F          & F          & F          & F          & F        & --\\
6$_{2}\rightarrow$5$_{1}$   & 315.266861 & F           & N          & F          & F          & F          & F          & F          & $\sim$     & $\sim$     & N        & C$_{2}$H$_{3}$CN\\
7$_{0}\rightarrow$6$_{0}$   & 338.124488 & F           & N          & F          & F          & F          & F          & F          & $\sim$     & $\sim$     & F        & --\\
7$_{1}\rightarrow$6$_{1}$   & 338.614936 & F           & $\sim$     & F          & $\sim$     & F          & $\sim$     & F          & $\sim$     & F          & N        & CH$_{3}$NH$_{2}$, HC$_{3}$N\\
7$_{2}\rightarrow$6$_{2}$   & 338.721693 & F           & $\sim$     & F          & F          & F          & F          & F          & $\sim$     & F          & N        & --\\
7$_{-2}\rightarrow$6$_{-2}$ & 338.722898 & F           & $\sim$     & F          & F          & F          & F          & F          & $\sim$     & F          & N        & --\\
4$_{0}\rightarrow$3$_{-1}$  & 350.687662 & F           & F          & F          & N          & N          & N          & F          & F          & F          & F        & NO\\
7$_{2}\rightarrow$6$_{1}$   & 363.739868 & $\sim$      & N          & $\sim$     & F          & $\sim$     & $\sim$     & F          & $\sim$     & $\sim$     & N        & --\\ \hline
\end{tabular}
\tablefoot{The upper energy level was limited to 150\,K and the catalog used inside CASSIS was VASTEL (see Sect.\,\ref{Sec.rot_diags}). The frequency ranges approximately from 84 to 373\,GHz. We indicate with letter N the lines that were not fitted (reproduced by less than 40\% or over-predicted by more than 40\% of its intensity). NL means that the corresponding transition was not observed and therefore not fitted. All the fitted transitions are labelled with the letter F. The symbol $\sim$ is representing lines that were fitted at an important percentage, around 40--70\%, and usually without confusion with other lines. Blending lines, in terms of integrated intensity contamination percentage, based on preliminary results from synthetic spectra in region\,5 (Mart\'in et al. in prep.) are shown in the last column when surpassing 5\%.}

\end{center}
\end{table*}

\begin{table*}[!htp]
\caption{A-type methanol lines used as input for our LTE model.} \label{tab.methanol_lines_model_A}
\scriptsize
\begin{center}
\renewcommand{\tabcolsep}{0.15cm}
\begin{tabular}{llllllllllllll}
\hline \hline
Transition                 & $\nu$ [GHz] & Type    & R1          & R2         & R3         & R4         & R5     & R6      & R7         & R8         & R9         & R10       & Blending lines \\ \hline \\
2$_{1}\rightarrow$1$_{1}$  & 95.914310  & A$^{+}$ & N           & N     & N          & F          & F      & $\sim$       & N          & N          & N          & $\sim$       & C$_{2}$H$_{5}$OH\\
2$_{1}\rightarrow$1$_{1}$  & 97.582798  & A$^{+}$ & N           & N     & N          & F          & F      & $\sim$       & N          & N          & N          & F         & --    \\
%7$_{2}\rightarrow$8$_{1}$  & 111.289453  & A$^{+}$ & NL           & NL     & F          & F          & F      & F       & N          & F          & N          & F         & c--C$_{2}$H$_{2}$O$_{2}$(85.7\%)    \\
%6$_{2}\rightarrow$7$_{1}$  & 132.621824  & A$^{+}$ & N           & $\sim$     & F          & F          & F      & F       & N          & F          & N          & F         & C$_{5}$H$_{2}$O(9.7\%), t--HOCO(50.7\%), syn--C$_{2}$H$_{3}$OH(72.4\%), C$_{3}$H$_{6}$O$_{2}$(34.7\%)\\

4$_{1}\rightarrow$3$_{1}$  & 191.810503  & A$^{+}$ & N           & $\sim$     & F          & F          & F      & F       & N          & F          & N          & F         & --    \\
4$_{1}\rightarrow$3$_{1}$  & 195.146790  & A$^{-}$ & N           & $\sim$     & $\sim$     & $\sim$     & F      & $\sim$  & N          & N          & N          & N        & CH$_{3}$OCHO\\
%10$_{0}\rightarrow$9$_{1}$ & 198.403067  & A$^{+}$ & N           & NL         & N          & $\sim$     & F      & N        & N          & NL         & N          & NL        & n--C$_{3}$H$_{7}$CN(9.1\%)\\
1$_{1}\rightarrow$2$_{0}$  & 205.791270  & A$^{+}$ & $\sim$      & F          & F          & F          & $\sim$      & F    & F          & F          & F          & $\sim$    & C$_{2}$H$_{5}$OH\\
4$_{2}\rightarrow$5$_{1}$  & 234.683370  & A$^{-}$ & $\sim$      & F          & F          & F          & $\sim$ & F       & F          & $\sim$     & $\sim$     & $\sim$   & C$_{2}$H$_{5}$OH, H$_{2}$CS\\
5$_{1}\rightarrow$4$_{1}$  & 239.746219  & A$^{+}$ & F           & F          & F          & F          & F & F       & N          & $\sim$     & $\sim$     & $\sim$    & --\\
4$_{3}\rightarrow$4$_{2}$  & 251.866524  & A$^{-+}$ & F           & F          & F          & $\sim$     & N & F       & F          & F          & F          & N        & --\\
5$_{3}\rightarrow$5$_{2}$  & 251.890886  & A$^{+-}$ & F           & F          & F          & $\sim$     & N & F       & F          & F          & F          & F        & --\\
6$_{3}\rightarrow$6$_{2}$  & 251.895728  & A$^{+-}$ & F           & F          & F          & $\sim$     & N & F       & F          & F          & F          & F        & --\\
4$_{3}\rightarrow$4$_{2}$  & 251.900452  & A$^{+-}$ & F           & F          & F          & $\sim$     & N & F       & F          & F          & F          & F        & --\\
3$_{3}\rightarrow$3$_{2}$  & 251.905729  & A$^{-+}$ & F           & F          & F          & $\sim$     & N & F       & F          & F          & F          & F        & --\\
3$_{3}\rightarrow$3$_{2}$  & 251.917065  & A$^{+-}$ & F           & F          & F          & $\sim$     & N & F       & F          & F          & F          & F        & --\\
7$_{3}\rightarrow$7$_{2}$  & 251.923701  & A$^{+-}$ & F           & F          & F          & $\sim$     & N & F       & F          & F          & F          & F        & --\\
8$_{3}\rightarrow$8$_{2}$  & 251.984837  & A$^{+-}$ & F           & F          & F          & $\sim$     & N & F       & F          & F          & F          & F        & \\
6$_{1}\rightarrow$5$_{1}$  & 287.670767  & A$^{+}$ & F           & $\sim$     & F          & F          & F      & $\sim$  & $\sim$     & F          & F          & F         & --\\
1$_{1}\rightarrow$1$_{0}$  & 303.366921  & A$^{-+}$ & F           & $\sim$     & $\sim$     & $\sim$     & --      & F       & F          & $\sim$     & $\sim$     & $\sim$   & H$_{2}$CN\\
2$_{1}\rightarrow$2$_{0}$  & 304.208348  & A$^{-+}$ & $\sim$      & $\sim$     & F          & $\sim$     & --      & F       & F          & $\sim$     & $\sim$     & $\sim$   & C$_{2}$H$_{5}$OH\\
3$_{1}\rightarrow$3$_{0}$  & 305.473491  & A$^{-+}$ & $\sim$      & $\sim$     & $\sim$     & $\sim$     & --      & $\sim$  & F          & $\sim$     & $\sim$     & F        & c--C$_{3}$H$_{2}$, CH$_{3}$SH\\
4$_{1}\rightarrow$4$_{0}$  & 307.165924  & A$^{-+}$ & F           & $\sim$     & $\sim$     & N          & --     & $\sim$  & F          & $\sim$     & F          & F        & CH$_{3}$OCHO\\
5$_{1}\rightarrow$5$_{0}$  & 309.290360  & A$^{-+}$ & F           & $\sim$     & $\sim$     & $\sim$     & --      & $\sim$  & F          & F          & F          & F        & --\\
6$_{1}\rightarrow$6$_{0}$  & 311.852612  & A$^{-+}$ & F           & F          & F          & $\sim$     & $\sim$      & F       & F          & F          & F          & $\sim$   & --\\
7$_{1}\rightarrow$7$_{0}$  & 314.859528  & A$^{-+}$ & $\sim$      & F          & F          & $\sim$     & $\sim$      & F       & $\sim$     & $\sim$     & N          & $\sim$   & --\\
2$_{2}\rightarrow$3$_{1}$  & 335.133570  & A$^{-}$ & N           & $\sim$     & F          & F          & F      & F       & F          & F          & F          & $\sim$   & CH$_{3}$COOH\\
7$_{1}\rightarrow$6$_{1}$  & 335.582017  & A$^{+}$ & $\sim$      & $\sim$     & $\sim$     & $\sim$     & F & $\sim$  & F          & F          & $\sim$     & $\sim$    & --\\
7$_{1}\rightarrow$6$_{1}$  & 341.415615  & A$^{-}$ & F           & N          & $\sim$     & F          & F      & F       & F          & F          & F          & $\sim$   & --\\
1$_{1}\rightarrow$0$_{0}$  & 350.905100  & A$^{+}$ & F           & F          & F          & F          & F & F       & $\sim$     & F          & F          & F         & --\\ \hline
\end{tabular}
\tablefoot{Same as for Table\,\ref{tab.methanol_lines_model_E}, but for A-type methanol. Lines denoted with the symbol -- were not attempted to fit in Region\,5 (see Section.\,\ref{sec.LTE_outliers}).}

\end{center}
\end{table*}

In this Appendix, we present a couple of more sophisticated models, aiming to characterize in a more refined way both regions well characterized by LTE conditions and regions with strong maser emission. The former are being located toward the nucleus of NGC\,253, while the latter occupy the outskirts of the CMZ of NGC\,253, in a bi-symmetric configuration. 

We performed a model (hereafter model (1)) to fit the spectrum of region\,4 (see Table\,\ref{tab.positions}), which shows the best agreement between E- and A-type methanol in our rotation diagrams (see Table\,\ref{tab.apen.RD_params}), both in terms of excitation temperatures and column densities. In this model we used two physical components (both with slab geometry) for each methanol type. 

In CASSIS it is possible to use the MCMC method to explore the space of parameters and fit the line profiles of both methanol types together by adjusting the ISO parameter, which corresponds to the column density ratio of E--CH$_{3}$OH over A--CH$_{3}$OH. Unfortunately, this two-component approach is constrained to a fixed single excitation temperature per component. Thus, our region with the lowest difference in $T_{\rm ex}$ (see Fig.\,\ref{fig.delta}) between methanol types, i.e. region\,4, is the best candidate to perform such analysis (model (1)).

The other model (2), addressing region\,8, presents the highest difference between A- and E-methanol excitation temperatures (Table\,\ref{tab.apen.RD_params}), in addition of having all maser lines but the one at 198.4\,GHz observed through the rotation diagram method (see Table\,\ref{tab.RD_outliers}). For model (2) we found better results by fitting both methanol types together, as for our model (1), with two components each. This is not in contradiction with the rotation diagrams, as they only accounted for one gaseous component while here we are using two of them. It is then possible to have one component related to each slope in Fig.\,\ref{fig.rotationdiagrams}.

\begin{figure*}[!htp]
\includegraphics[width=1.107\linewidth, trim={2.5cm 0 0 0},clip]{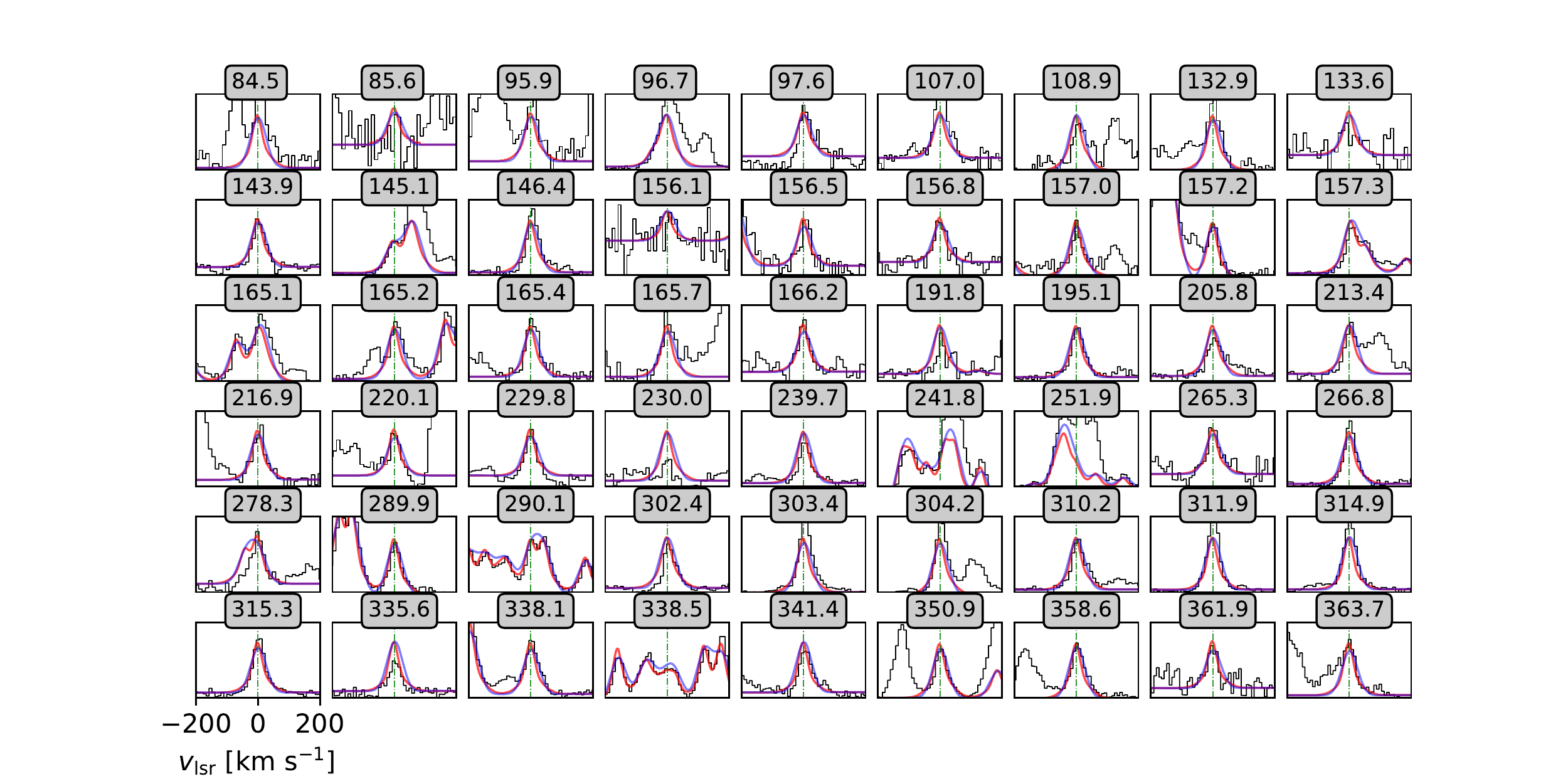}
\caption{Two-component LTE model for region\,4 (in red) over-plotted to the rest-frame spectra (in black). The common velocity range (in km\,s$^{-1}$, after applying the radio convention and subtracting the $v_{\rm LSR}$ velocity of the region, as in Fig.\,\ref{fig.abs_lines}, namely 252\,km\,s$^{-1}$, see Table\,\ref{tab.positions}) is depicted at the bottom left panel while line frequencies are indicated above each panel in GHz and by green dashed lines inside the panels. Previous single-component LTE modelling (see Fig.\,\ref{fig.fitted_lines}) is shown in blue for comparison.}
\label{fig.LTE_R4}
\end{figure*}

The resulting parameters of the two models are shown in Table\,\ref{tab.apen.sophisticated_models_params} after 600 iterations for each model. While model (1) describes the full LTE conditions, model (2) is mostly non-LTE: the component with the highest column density shows gas densities of $\sim$6$\times$10$^{5}$\,cm$^{-3}$, far below the critical densities of methanol lines used in Sect.\,\ref{Sec.rot_diags}. The LTE synthetic spectrum obtained for region\,4 is shown in Fig.\,\ref{fig.LTE_R4} for selected lines covering the entire ALCHEMI frequency range. We did not find important over-fits to the observed spectra in this two-component LTE modelling. Although some lines are still underpredicted at frequencies below $\sim$150\,GHz, in general there is a better agreement in comparison to the previous model (single-component) results in terms of line shape (in blue in Fig.\,\ref{fig.LTE_R4}). 

Summing over column densities of each component in model (1) we obtain column densities of 1.86$\pm 0.024\times$10$^{15}$ and 8.29$\pm0.17\times$10$^{14}$cm$^{-2}$ for E--CH$_{3}$OH and A--CH$_{3}$OH, respectively (model uncertainties only, the calibration uncertainty (15\%, see Sect.\,\ref{Sec.Observations}) should be added). These values are slightly smaller than in our single-component approach (see Table\,\ref{tab.LTE_model_params}). That is probably due to the avoidance of over-fitting in some line transitions when the spectrum becomes crowded due to nearby methanol lines, like our panels centered at 157.3, 241.8, 290.2 and 338.5\,GHz presented in Fig.\,\ref{fig.LTE_R4}.

\begin{figure*}[!htp]
\includegraphics[width=1.107\linewidth, trim={2.5cm 0 0 0},clip]{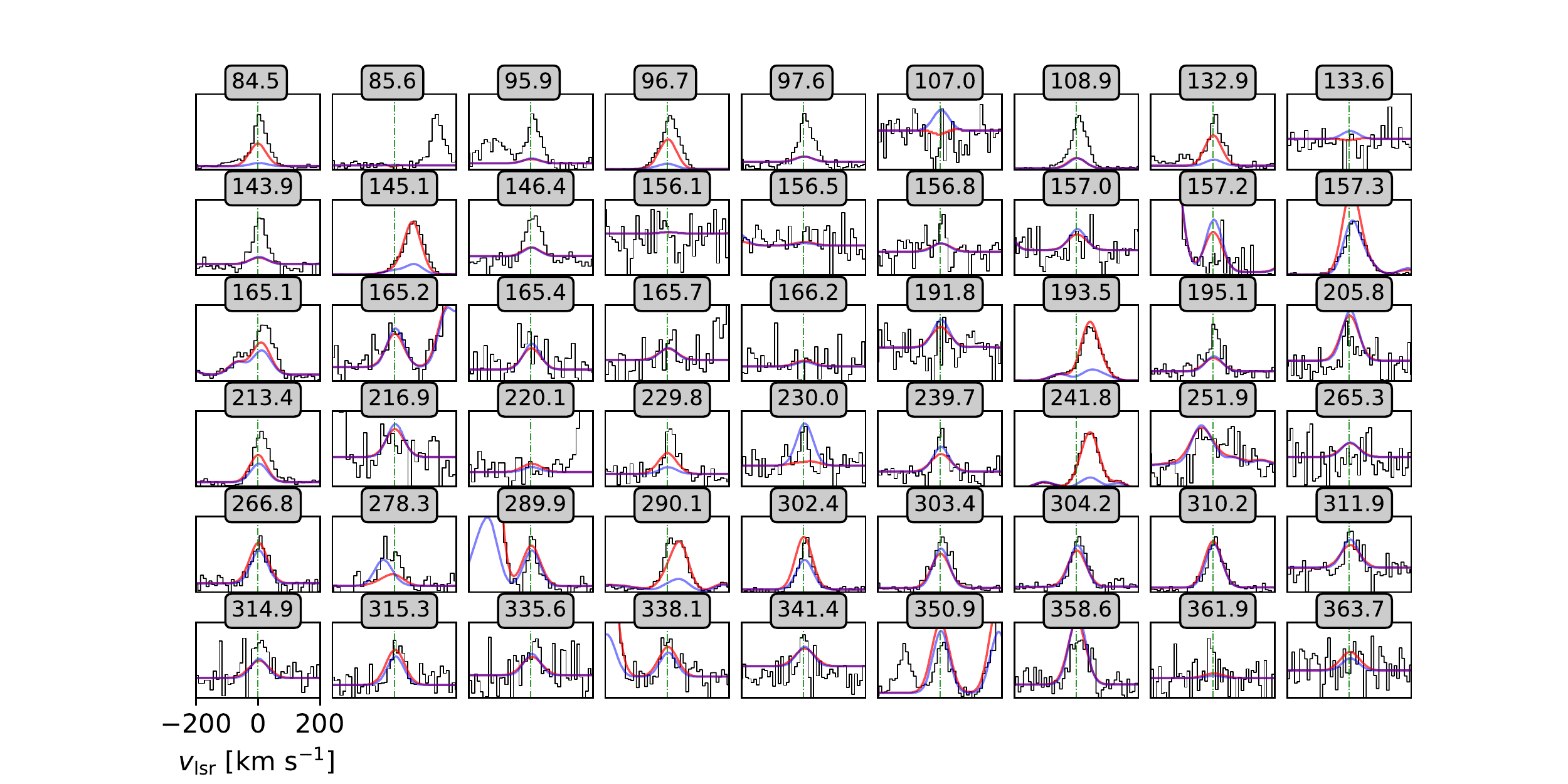}
\caption{Two-component RADEX non-LTE model for region\,8 (in red) over-plotted on the rest-frame spectra (in black). The common velocity range is depicted at the bottom left panel and was obtained in the same manner as in Fig.\,\ref{fig.LTE_R4} but this time subtracting a $v_{\rm LSR}$ velocity of 205\,km\,s$^{-1}$ (see Table\,\ref{tab.positions}). Line frequencies are indicated above each panel in GHz and by green dashed lines inside the panels. Previous single-component LTE modelling (see Fig.\,\ref{fig.fitted_lines}) is shown in blue for comparison.}
\label{fig.apen.RADEX_R8}
\end{figure*}

Our non-LTE modelling allows us to obtain better results than the LTE model for region\,8. This is remarkable in the case of the $J_{K}\rightarrow (J-1)_{K}$ transitions (for  both methanol symmetric types). Namely, the $2_{K}\rightarrow 1_{K}$ line series ($K$=-1 ... 1) centered around 96.7\,GHz, the $3_{K}\rightarrow 2_{K}$ series ($K=$-2 ... 2) at $\sim$145.1\,GHz; the $4_{K}\rightarrow 3_{K}$ series ($K=$-3 ... 3) at $\sim$193.5\,GHz, the $5_{K}\rightarrow 4_{K}$ series ($K=$-4 ... 4) at $\sim$241.8\,GHz, and the $6_{K}\rightarrow 5_{K}$ series ($K=$-5 ... 5) at $\sim$290.1\,GHz, as can be seen in Fig.\,\ref{fig.apen.RADEX_R8}. Our only concern regarding this non-LTE approach is the overestimation of the $J_{0}\rightarrow J_{-1}-E$ line series at about 157.3\,GHz, with $J$ from 1 to 3. From this model, it can also be observed that the line at 108\,GHz is underpredicted by far, similar to the case of the maser candidates, although the optical depth for this transition is positive.

From model (2) we have obtained the optical depths ($\tau$) for each methanol transition, and we use them to discriminate between maser line candidates and pure outliers in both our rotation diagrams and radiative transfer models (Tables\,\ref{tab.RD_outliers} and \ref{tab.methanol_maser_candidates_LTE_model}, respectively). In this way, methanol maser candidates can be limited to transitions with negative $\tau$ only. The general departure from LTE observed in the outliers of the CMZ by the two methods outlined in Sects.\ref{Sec.rot_diags} and \ref{sec.models} can be explained by blending lines, non-LTE conditions, or an extra colder and more extended component, when they have positive $\tau$.

%\setlength\tabcolsep{0.3pt}  % default value: 6pt
%%%TABLA MOFIFICADA CON VALORES PARA REGION 7 , 8 9 Y 10
\begin{table*}[!t]
\caption{Best E-type methanol fit parameters from our detailed two components modelling.} \label{tab.apen.sophisticated_models_params}
\scriptsize
\begin{center}
\setlength\tabcolsep{4pt}  % default value: 6pt
%\begin{tabular}{llllllll}
\begin{tabular}{cccccccc}

\hline \hline
Region & Component & $N$(Sp)                                         & $T_{\rm kin}$                & FWHM                               & $V_{\rm LSR}$                      & ISO & $n_{\rm H_2}$\\
&       &  \multicolumn{1}{c}{[cm$^{-2}$]}              & \multicolumn{1}{c}{[K]} & \multicolumn{1}{c}{[km\,s$^{-1}$]} & \multicolumn{1}{c}{[km\,s$^{-1}$]}  &     & \multicolumn{1}{c}{[cm$^{-3}$]}\\ \hline \\
4      & 1         & 7.67$\times$10$^{14}\pm$1.20$\times$10$^{13}$ & 26.68$\pm$0.21          & 84.78$\pm$0.32                     &252                      & 1.516$\pm$0.025 & -- \\
       & 2         & 4.19$\times$10$^{14}\pm$1.19$\times$10$^{13}$ & 32.32$\pm$0.39          & 30.66$\pm$0.25                     &248                      & 1.295$\pm$0.061 & -- \\
8      & 1         & 1.35$\times$10$^{14}\pm$1.29$\times$10$^{13}$ & 24.76$\pm$0.80          & 75.59$\pm$0.37                     &205                      & 1.984$\pm$0.121 & 2.777($\pm$0.101)$\times$10$^{8}$\\
       & 2         & 1.34$\times$10$^{15}\pm$5.00$\times$10$^{13}$ & 14.84$\pm$0.32          & 62.42$\pm$0.68                     &200                      & 3.141$\pm$0.578 & 5.949($\pm$0.279)$\times$10$^{5}$\\
\hline \\
\end{tabular}
\tablefoot{Uncertainties correspond to 3$\sigma$. Velocity uncertainties (not shown) correspond to less than 10\% of the instrumental uncertainty ($\sim$8--9\,km\,s$^{-1}$) 
%and were obtained averaging the model results between the A and E methanol symmetric types #when we use the ISO parameter in CASSIS, both methanol types are fixed in terms of parameters, the only available change between E- and A-type methanol is for N(Sp)
. Column density ($N$(Sp)) values for A-type methanol can be obtained dividing the them by the ISO number (column density ratio of E- over A-type methanol, see Appendix\,\ref{sec.apen.detailed_modelling}), while the rest of the parameters are the same for both methanol symmetric types. The minus symbol (-) indicates that no $n_{\rm H_2}$ was derived as the model is in LTE.}
\end{center}
\end{table*}

% trim={<left> <lower> <right> <upper>}

% trim={<left> <lower> <right> <upper>}

%\section{LTE model parameters}
%\label{sect.apen.model_params}

%In this Section we present our results from the synthetic spectra performed by CASSIS, as described in Sect.\,\ref{sec.models}, they are listed in Table\,\ref{tab.LTE_model_params}.

\section{Outliers identified through LTE modelling}
\label{sec.apen.modelling_outliers}

As mentioned in Sect.\,\ref{sec.LTE_outliers}, there are methanol transitions not reproduced by our synthetic spectra in LTE conditions. Without considering the two maser candidates at 95.2 and 278.3\,GHz detected by our radiative transfer modelling, there are still a number of transitions not reproduced (see Table\,\ref{tab.methanol_maser_candidates_LTE_model}). Some intriguing cases will be described below, ordered by frequency.

\textbf{The} $\mathbf{3_{1} \rightarrow 4_{0}-A^{+}}$ transition line at 107.0\,GHz ($E_{\rm up}/$k$=$28.3\,K) has intensities surpassing the LTE models in regions\,3, 4, 6, and 7. This line is not identified as a maser candidate through the rotation diagram method, presenting only a couple of values slightly above the LTE fit of the rotation diagrams by a factor ($N_{\rm up, 107\,GHz}$ -- $N_{\rm up, LTE}$) of 0.25 to 0.46 (in regions\,6 and 4, respectively), while in other regions it falls below the fit, like in the case of region\,7, where it appears below the expected value by a factor of 1.1. In the LTE modelling its intensity is not strongly underpredicted (see Fig.\,\ref{fig.unknown_maser_lines}), in contrast to the case of the Class\,I methanol masers in the $J_{-1}\rightarrow(J-$ 1)$_{0}-E$ and $J_{0}\rightarrow(J-$ 1)$_{1}-A^{+}$ line series (see Fig.\,\ref{fig.known_maser_lines}).

Methanol transitions, frequently observed as masers in interstellar space have sometimes the property that they show, in case of absence of a proper stimulating environment, anti-inversion. A characteristic case is the 12.1\,GHz $2_{0} \rightarrow3_{-1}-E$ transition, often seen as a Class\,II maser line in massive star forming regions \citep[e.g.][]{Breen2011}. This line has been observed in absorption toward dark clouds, thus indicating a typical excitation temperature below 2.7\,K in this class of objects \citep{Walmsley1988}. Furthermore, from Galactic data, Class\,I and Class\,II maser regions are not well correlated \citep{Kang2015}. The 3$_{1}\rightarrow4_{0}-A^{+}$ transition at 107.0\,GHz may be a related case. The line shows weak absorption in regions\,1 and 9, where class\,I emission is observed. This leads us to relate this line to a Class\,II maser. Another possibility is that this transition comes from colder gas, following an LTE behavior in the rotation diagrams but requiring an extra component to be fitted by synthetic spectra. The presence of an extra component is in line with recent observations performed by \citet{Holdship2021}, based on C$_{2}$H observations from the same ALCHEMI dataset. They found an extended and relatively cold component seen predominantly at lower frequencies (mainly in the $J$=1--0 transition at 87\,GHz), where C$_{2}$H has a gas density ($n_{\rm H2}$) in the range of (0.35--1.7)$\times$10$^{4}$\,cm$^{-3}$.

When this line is not in emission, it shows either absorption or a low signal. This happens in regions\,1, 2, 5, 8, 9, and 10 (Fig.\,\ref{fig.unknown_maser_lines}), the same regions where Class\,I masers are detected in emission in the rotation diagrams (see Sect.\,\ref{sec.maser_line_candidates_RD}), with the only exception of region\,7 and the previously avoided region\,5. Masers related to both classes might arise in region\,7 due to our limited angular resolution, that cannot disentangle individual maser spots. Coming back to the rotation diagram method, within all regions where the 107\,GHz line is detected in emission, the only one where its intensity is lower than expected by the LTE behavior is region\,7. This opens the possibility that this line is suffering absorption not detectable by the LTE models due to a colder gas component that may be contaminating our spectra below 156\,GHz.

We cannot be confident about the nature of this line, but if it is maser emission, that will indicate a first detection of methanol Class\,II maser emission in NGC\,253 and the farthest detection of such a maser (107\,GHz) so far. We note that the strongest Class\,II line in our Galaxy is the 5$_{1} \rightarrow 6_{0}-A^{+}$ line at 6.7\,GHz \citep{Leurini2016}, first reported by \citet{Menten1991b}, which belongs to the same family as the 107.0\,GHz line detected in the present study. Absorption might then be expected for this methanol line in regions populated by Class\,I methanol masers. We highlight, however, that based on a non-LTE analysis of region\,8, this line has positive optical depths (Appendix\,\ref{sec.apen.detailed_modelling}). 

\textbf{The} $\mathbf{0_{0} \rightarrow 1_{-1}-E}$ transition at 108.9\,GHz ($E_{\rm up}/$k$ = 5.2$\,K) was not reproduced by our models, although it follows LTE conditions according to the rotation diagrams, where it is placed slightly above the LTE fit in most regions (see Fig.\,\ref{fig.rotationdiagrams}). Its intensity cannot be reached by our non-LTE approximation either (see Appendix\,\ref{sec.apen.detailed_modelling}) and, interestingly, its line shape is similar to that observed in our maser candidates at 84.5 and 132.9\,GHz. We do not expect negative optical depths for this line (see Table\,\ref{tab.methanol_maser_candidates_LTE_model}) but also no clear blending lines, either from preliminary model results (Mart\'in et al. in prep.), or from our current inspection (see Table\,\ref{tab.methanol_lines}, last column). Masers in the 0$_{0} \rightarrow 1_{-1}-E$ transition are expected to be pumped by radiation of hot dust rather than by ultra-compact {H\,\sc{ii}} regions and to be thermalized faster than other Class\,II masers \citep{Kalenskii2002}.

Figure\,\ref{fig.fitted_lines} shows that there are plenty of confirmed LTE lines spanning a frequency range between 86 and 364\,GHz. As can be noted toward regions around the core of the CMZ (regions\,3 to 6), the $J_{1}\rightarrow J_{0}-A^{+}$ series of lines at 304.2, 305.5, 307.2, and 309.3\,GHz is slightly underestimated by our synthetic spectra and is weakly misaligned from the LTE behavior in the rotation diagrams. Those lines belong to the same family as the line at 318.3\,GHz, previously observed to depart from the rotation diagrams (see Table\,\ref{tab.RD_outliers}). Since the difference between predicted and observed line strengths is mostly not remarkable, we do not account for them as new maser candidates. In addition, based on a non-LTE model performed for region\,8 (Appendix\,\ref{sec.apen.detailed_modelling}), only positive optical depths are calculated for these lines.

\textbf{The} $\mathbf{J_{K}\rightarrow (J-1)_{K}-X}$ line series, with $X$ being either $E$, $A^{+}$, or $A^{-}$ are successfully reproduced in region\,8 by a mixture between LTE and non-LTE models (see Appendix\,\ref{sec.apen.detailed_modelling}), being placed out of LTE but with positive optical depth, indicating a quasi-thermal nature. We have therefore decided to discard them as maser candidates.

\textbf{The} $\mathbf{5_{3}\rightarrow5_{2}-A^{-+}}$ line at 251.81\,GHz is the only among 13 transitions in the $J_{K}\rightarrow J_{K}-A$ line series (251--252\,GHz) which shows a clear departure from our LTE modelling. The intensities of this line are much higher than expected in the central regions of NGC\,253. We performed an LTE model for SO (1000 iterations in region\,6 only, not shown) that can account for $\sim$64\% of the emission at 251.83\,GHz, and therefore we do not expect an important contribution for the methanol line, assuming that other species can be also contaminating. However, as this line is not fully reproduced, we add this as an outlier in our Table\,\ref{tab.methanol_maser_candidates_LTE_model}.

\section{Maser line distribution at 146.6\,GHz}
\label{sec.apen.146_distribution}

As mentioned in Sect\,\ref{subsec.Amasers}, the $9_{0}\rightarrow 8_{1}-A^{+}$ line at 146.6\,GHz is observed to depart from LTE in regions\,1, 7, 8, and 9, with intensities $\geq$8.4$\pm$1.2 stronger than predicted, indicating a maser behaviour according to our threshold of 3.3 for this factor. Following its cousins at lower frequencies in the $J_{0}\rightarrow (J-1)_{1}-A^{+}$ series, at 44.1 ($J=7$) and 95.2\,GHz ($J=8$), it belongs to the Class\,I maser category \citep[e.g.][and references therein]{Leurini2016,Yang2020}.

\begin{figure*}[!htp]
\includegraphics[width=\linewidth, trim={0 0 0 10cm},clip]{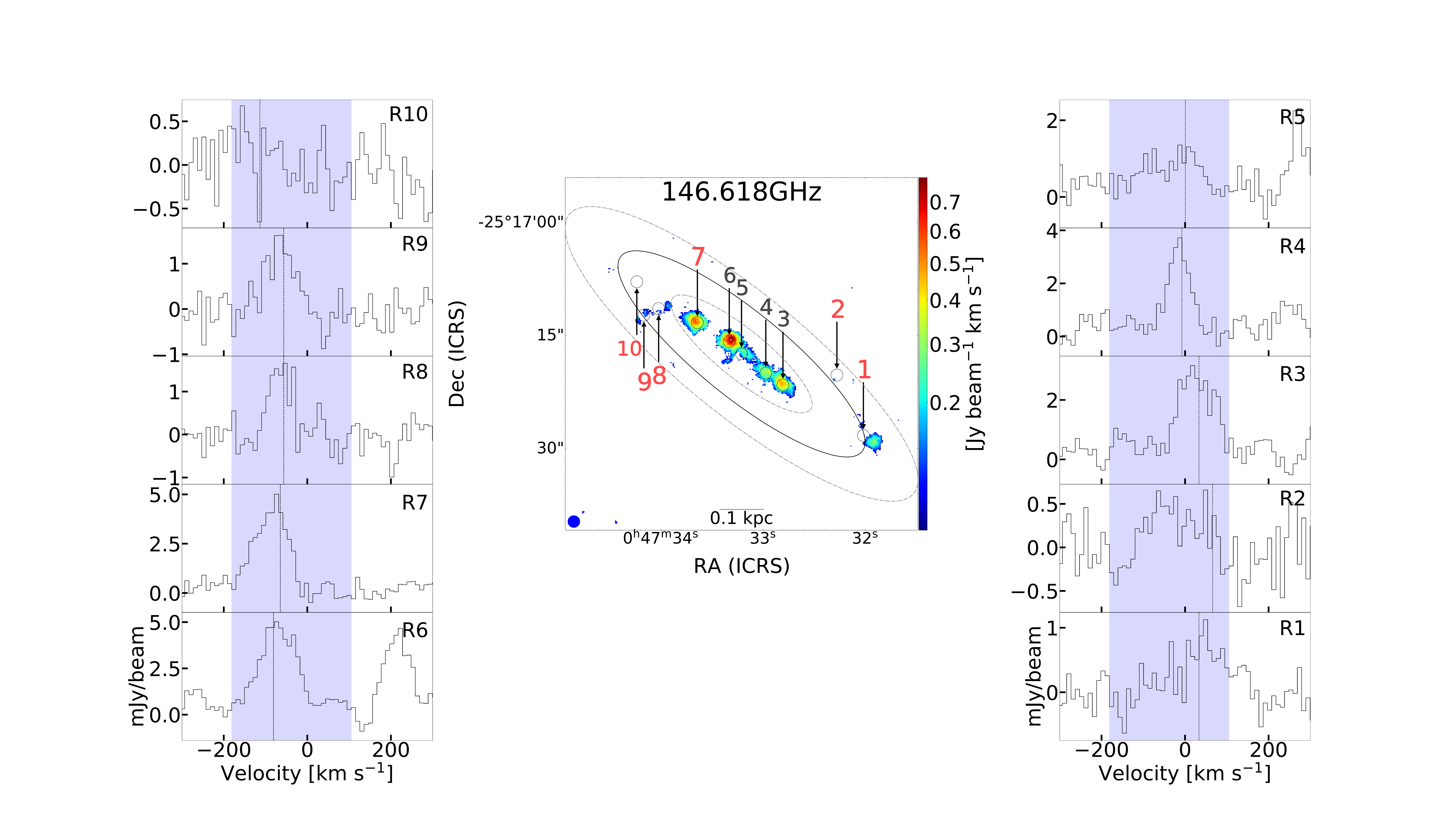}
\caption{Same as for Figure\,\ref{fig.m0_and_spectra_84} but for the $9_{1}\rightarrow8_{0}-A^{+}$ methanol line at 146.6\,GHz.}
\label{fig.apen.m0_and_spectra_146}
\end{figure*}

It can be noted in Fig.\,\ref{fig.apen.m0_and_spectra_146} that, contrary to what is observed for methanol masers in the $J_{-1}\rightarrow(J-1)_{-1}-E$ family of lines (except the last covered transition at 278.3\,GHz), the $9_{1}\rightarrow8_{0}-A^{+}$ methanol line at 146.6\,GHz peaks in region 6. That might be an indication that this line is a better tracer of the highest densities
%and strongest star formation (see Fig.\,\ref{fig.SFR_from_HCN}) 
instead of cloud-cloud collisions, that dominate the outskirts of the CMZ of NGC\,253.

% trim={<left> <lower> <right> <upper>}

\section{Equivalency between intensity and column density ratios}
\label{Sec.apen.equivalency}

\begin{figure}[!htp]
\includegraphics[width=\linewidth, trim={0 0 0 0},clip]{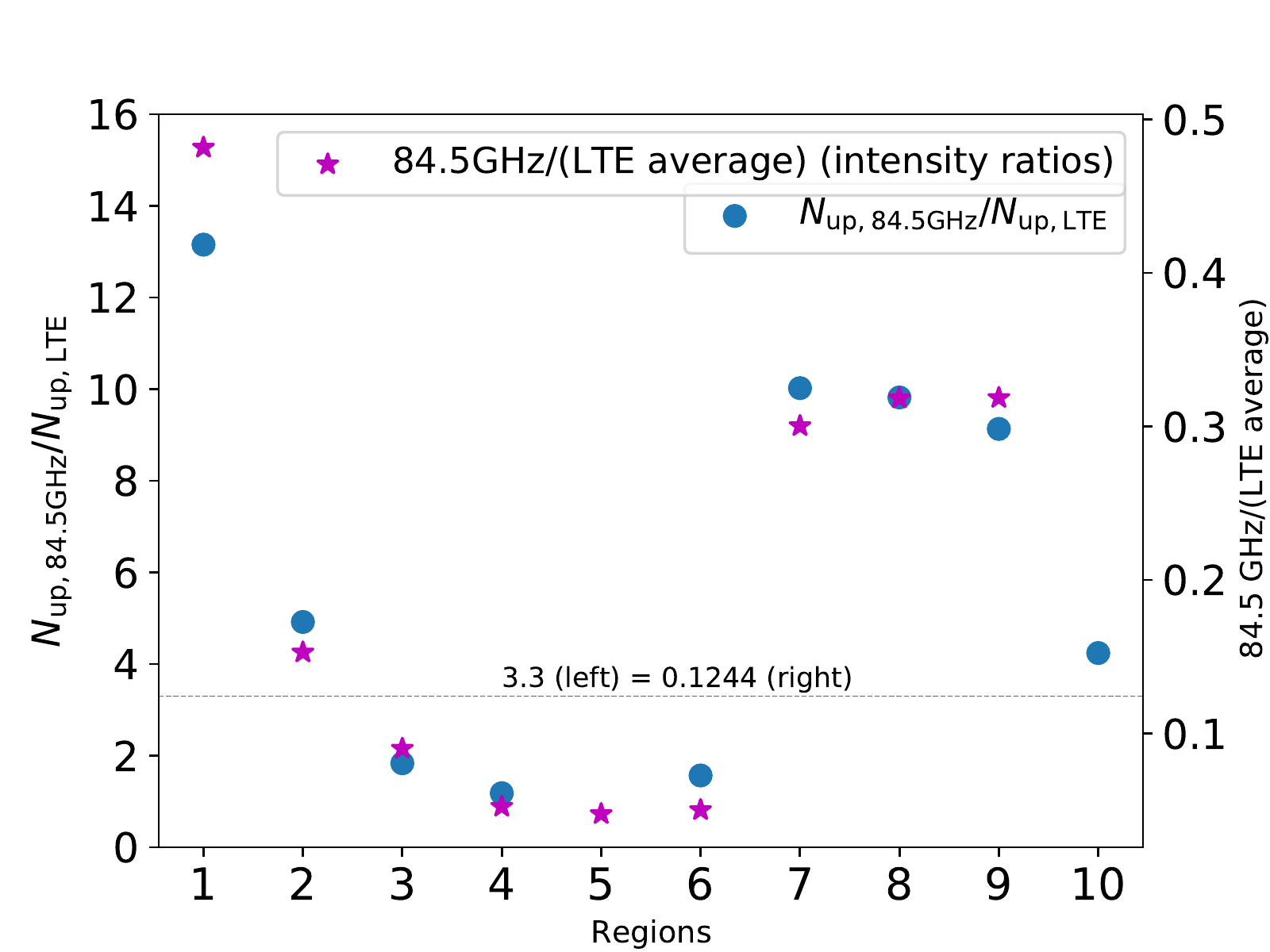}
\caption{Magenta stars are scaled according to the left y-axis, and are equivalent to the magenta stars in the first panel of Fig.\,\ref{fig.delta}. Blue dots are scaled with respect to the right y-axis and correspond to integrated intensities of the 5$_{-1}\rightarrow$4$_{0}-E$ methanol transition at 84.5\,GHz over an LTE approximation (see Sect.\,\ref{sec.maser_emission_distribution}) averaged inside a beam size aperture (1\farcs6 diameter) for each of the studied regions in this work (see Table\,\ref{sec.selected_regions}).}
\label{fig.apen.equivalency}
\end{figure}

In Fig.\,\ref{fig.maser_distribution}  we present line intensity ratios between the 5$_{-1}\rightarrow$4$_{0}-E$ methanol transition at 84.5\,GHz and the average of a couple of E-type methanol lines that are always found to follow LTE conditions (according to our rotation diagrams; see Sect.\,\ref{sec.maser_emission_distribution}). It is important to note a certain equivalency between LTE departure, in terms of $N_{\rm up,maser}$/$N_{\rm up,LTE}$ (as in Fig.\,\ref{fig.delta}) and line intensity ratios between the 84.5\,GHz line and the proposed LTE average. They are not necessarily the same, especially considering optically thick transitions and negative optical depths and excitation temperatures present in masers. Therefore, we reproduce in Fig.\,\ref{fig.apen.equivalency} the first panel of Fig.\,\ref{fig.delta} with the values on the leftmost y-axis along with the intensity ratios averaged inside the beam size (1\farcs6 diameter) presented in Fig.\,\ref{fig.maser_distribution}. The results are in agreement in terms of distribution: regions where the 5$_{-1}\rightarrow$4$_{0}-E$ line is observed in LTE through the rotation diagram method (regions\,3, 4, and 6) present the lowest integrated intensity ratios. We note that region\,5 was avoided when performing the rotation diagrams and that region\,10 is not available for all the thermal lines used to obtain the LTE average. We found that a threshold of 3.3 in terms of upper level column density ratios is equivalent to a factor of 0.1244 in terms of integrated intensity ratios, at least within the studied regions.

% trim={<left> <lower> <right> <upper>}

\end{appendix}

\end{document}